\documentclass[11pt,preprint]{aastex}

\newcommand{\pr}{^\prime}
\newcommand{\e}{\epsilon}

\newcommand{\gp}{\gamma^\prime}
\newcommand{\pp}{p^\prime}
\newcommand{\gpp}{\gamma_p^\prime}
\newcommand{\ep}{\epsilon^\prime}
\newcommand{\Ep}{E^\prime}
\newcommand{\psim}{\lower.5ex\hbox{$\; \buildrel \propto \over\sim \;$}}
\newcommand{\ebf}{}

\shorttitle{Energetic Particle Production in Gamma Ray Bursts}
\shortauthors{Dermer}

\begin{document}

\title{Neutrino, Neutron, and Cosmic Ray Production \\
    in the External Shock Model of Gamma Ray Bursts}

\author{Charles D. Dermer}
\affil{Naval Research Laboratory, Code 7653, Space Science Division, 
Washington, DC 20375-5352}
\email{dermer@gamma.nrl.navy.mil}

\begin{abstract}
The hypothesis that ultra-high energy ($\gtrsim 10^{19}$ eV) cosmic
rays (UHECRs) are accelerated by gamma-ray burst (GRB) blast waves is
assumed to be correct. Implications of this assumption are then
derived for the external shock model of gamma-ray bursts. The evolving
synchrotron radiation spectrum in GRB blast waves provides target
photons for the photomeson production of neutrinos and neutrons. Decay
characteristics and radiative efficiencies of the neutral particles
that escape from the blast wave are calculated. The
diffuse high-energy GRB neutrino background and the distribution of
high-energy GRB neutrino events are calculated  for specific
parameter sets, and a scaling
relation for the photomeson production efficiency in surroundings
with different densities is derived.

GRBs provide an intense flux of high-energy neutrons, with
neutron-production efficiencies exceeding $\sim 1$\% of the total
energy release. The radiative characteristics of the neutron
$\beta$-decay electrons from the GRB ``neutron bomb" are solved in a
special case. Galaxies with GRB activity should be surrounded by
radiation halos of $\sim 100$ kpc extent from the outflowing neutrons,
consisting of a nonthermal optical/X-ray synchrotron component and a
high-energy gamma-ray component from Compton-scattered microwave
background radiation. The peak luminosity emitted by the diffuse
$\beta$-electron halo from a single GRB with $\gtrsim 2\times 10^{53}$
ergs isotropic energy release is $\sim 10^{35}$ ergs s$^{-1}$, with a
potentially much brighter signal from the neutron-decay protons. The
decay halo from a single GRB can persist for $\gtrsim 0.1$-1 Myr.
Stronger neutrino fluxes and neutron decay halos can be produced by
external shocks in  clumpy external media and in scenarios
involving internal shock scenarios, so detection of neutrinos
associated with  smooth-profile GRBs could  rule out an
implusive GRB central engine and an external shock model for the
prompt phase.

The luminosity of sources of GRBs and relativistic outflows in L$^*$
galaxies such as the Milky Way is at the level of $\sim 10^{40\pm 1}$
ergs s$^{-1}$. This is sufficient to account for UHECR generation by
GRBs. We briefly speculate on the possibility that hadronic cosmic
rays originate from the subset of supernovae that collapse to form
relativistic outflows and GRBs.

\end{abstract}

\keywords{cosmic rays---galaxies: halos---gamma rays: theory---
gamma rays: burst}

\section{Introduction}

The distance scale to the sources of gamma-ray bursts (GRBs) with
durations $\gtrsim 1$ s has been established as a consequence of
observations made with the Beppo-SAX satellite \citep{cos97,vpa97}.
The Beppo-SAX discovery of decaying X-ray afterglows permits follow-on
optical observations that give redshift determinations from absorption
and emission lines in optical transient counterparts or from
directionally coincident host galaxies. Nearly 20 GRB sources have
measured redshifts (for a recent review, see \citet{vpa00}), with a
mean redshift $z \sim 1$ for the sample. The distribution of redshifts
is as yet poorly established, but ranges from $z = 0.0085$ for GRB
980425 to $z = 4.50$ for GRB 000131.  The redshift of GRB 980425 is
based upon its temporal and spatial coincidence with SN 1998bw
\citep{gal98,kul98,pia00}, and points to a relationship between GRB
sources and supernovae (SNe). A GRB/SN relationship is strengthened by
the detection of highly reddened excesses in the optical afterglows of
several GRBs, which would arise if supernova ejecta, powered by the
decay of radioactive $^{56}$Ni \citep{blo99,rei99,gal00}, are formed
in GRB explosions. Measured  apparent isotropic
 $\gamma$-ray energy releases range from
$E_\gamma\sim 10^{48}$ ergs for GRB 980425 to $\sim 2.4\times 10^{54}$ ergs
for GRB 990123 at $z = 1.60$, with $E_\gamma \gtrsim
3\times 10^{51}$ 
ergs in all cases except GRB 980425 \citep{fra01}. 
Achromatic temporal breaks in the
optical light curves of GRB 990123 \citep{kul99} and GRB 990510
\citep{har99} suggest, however, that the most luminous GRBs might be
beamed, so that only directional energy releases are actually
measured. In the case of GRB 990123  \citep{bri99}, the
directional $\gamma$-ray power and $\gamma$-ray energy release reach
peak values $\partial L_\gamma /\partial \Omega \sim 3\times 10^{51}$
ergs s$^{-1}$ sr$^{-1}$ and $\partial E/\partial \Omega \sim 2\times
10^{53}$ ergs sr$^{-1}$, respectively.

Considerable evidence linking the sources of GRBs with star-forming
regions in galaxies has recently been obtained \citep{lam99,djo01}. 
Optical
transients associated with GRBs are superposed on the stellar fields
of associated host galaxies in essentially all 14 cases of GRBs with
deep follow-up optical observations \citep{vpa00,fru99,blo99a,ode98},
rather than far outside the galaxies' disks, as might be expected in a
scenario of merging neutron stars and black holes \citep{npp92}. Host
galaxies that are directionally coincident with optical transients
discovered within the field of GRB X-ray afterglows have blue colors,
consistent with galaxy types that are undergoing active star formation
\citep{fru99,cl99a,cl99b}. The host galaxy luminosities are consistent
with a Schechter luminosity function \citep{sch00}, and span a wide
range of extinction-corrected $R$ magnitudes from $R \sim 13$ for the
host galaxy of GRB 980425 associated with SN1998bw to $R > 27.1$ for
GRB 980326 \citep{sch00,hf99}. Lack of optical counterparts in some
GRBs such as GRB 970828 and GRB 991226, which have associated radio
counterparts \citep{fra99}, could be due to extreme reddening from
large quantities of gas and dust in the host galaxy \citep{owe98}.
X-ray evidence \citep{piro00} for Fe K$_\alpha$-line signatures in GRB
991216, requiring large masses and column densities of nearby gas
\citep{boe00}, also indicates that GRBs originate in regions with
active star formation.

Knowledge of the distance scale to GRBs makes it possible to determine
their effects on the surrounding environment. Some of the claimed
effects of GRB explosions are the formation of HI shells and stellar
arcs \citep{eeh98,lp98}, the melting of dust grains by GRB UV
radiation to produce flash-heated chondrules in the early Solar system
\citep{mh99}, and the formation of sites of enhanced annihilation
radiation in the interstellar medium (ISM) originating from large
numbers of mildly relativistic positrons produced by a GRB
\citep{db00,fl02}. UV and X-rays from nearby GRBs could also
have produced biologically significant dosages on Earth in the past
\citep{sw02}.

Another effect of GRBs, proposed prior to the Beppo-SAX discovery, is
that GRB sources accelerate the highest energy cosmic
rays. \citet{mil95} argued for this connection on the basis of a
directional association of two $> 10^{20}$ eV air shower events with
earlier BATSE GRBs. \citet{wax96} pointed out, however, that the
intergalactic field must disperse the arrival time of the cosmic rays
by $\gtrsim 50$ yrs to be consistent with the detection rate of
GRBs. \cite{vie95} noted that the isotropy of the UHECR arrival
direction was consistent with the isotropic distribution of GRB
sources, and that the extreme energies of UHECRs could be explained
through first-order Fermi acceleration by a relativistic blast wave
with Lorentz factor $\Gamma$. At each shock crossing, a particle would
increase its energy by a factor $\sim 4 \Gamma^2 \sim 4\times
10^{5}(\Gamma/300)^2$, so that only a few such cycles would suffice to
produce UHECRs starting from low-energy particles.  The efficiency to
accelerate low-energy particles to ultra-high energies through
relativistic shock acceleration has since been shown to be infeasible
\citep{gal99,gak99}. Following the first shock crossing, the blast
wave intercepts the particle before its angular deflection from the
shock normal is much larger than $1/\Gamma$; thus subsequent cycles
lead to energy increases by only factors of $\sim 2$. Second-order
Fermi acceleration, for example, due to magnetohydrodynamic turbulence
generated by charged dust or irregularities in the external medium
\citep{wax95,sd00,dh01}, or by first-order Fermi acceleration
involving putative shocks in a relativistic wind \citep{wax95} could,
however, accelerate UHECRs in GRB blast waves.

Both \citet{vie95} and \citet{wax95} pointed out a remarkable
coincidence between the energy density of the highest energy cosmic
rays and the power of GRB sources within the Greisen-Zatsepin-Kuzmin
(GZK) radius, outside of which UHECRs are degraded by photomeson
production on the cosmic microwave background.  If GRB sources convert
a comparable amount of energy into UHECRs as is detected in the form
of $\gamma$ rays, then these sources can account for the observed
intensity of UHECRs.  The comparisons of \citet{vie95} and
\citet{wax95} made use of statistical studies where the most distant
GRBs detected with BATSE were assumed to be at $z \sim 1$.  Redshift
measurements of GRB sources now permit more refined studies of GRB
statistics, yielding the comoving space density of GRB sources and the
volume-averaged energy injection rate of GRB sources into the ISM.
This coincidence can therefore be more carefully tested.

In this paper, it is assumed that the sources of UHECRs are GRBs. We
then examine the implications that follow from this assumption.
Theoretical problems with accelerating particles to ultra-high
energies are not dealt with here (see, e.g.,
\citet{rm98,vie98b,der01,dh01}). In Section 2, we summarize a recent
statistical study employing the external shock model for GRBs
\citep{bd00} and compare it with other statistical studies of GRBs and
the constant-energy-reservoir result of \citet{fra01}. An external
shock model  is more energetically efficient than internal shocks to
generate $\gamma$ rays in the prompt phase of a GRB, so this study
yields a lower limit to the energy production rate of GRB sources per
comoving volume.  We show that an external shock model is consistent
with the UHECR/GRB hypothesis, so that the coincidence originally
identified by \citet{vie95} and \citet{wax95} holds.

In Section 3, the evolving temporal and spectral  behavior
of synchrotron radiation in GRB blast waves is characterized.  This
radiation provides a target photon source for high-energy protons,
and we calculate neutron and neutrino production from photopion
processes in GRB blast waves. Neutral particle production spectra,
integrated over the prompt and afterglow phases of a GRB, are
calculated. The diffuse high-energy neutrino background and the
distribution of neutrino event rates are calculated in Section 4. The
outflowing neutrons decay to form high-energy protons and
electrons. In Section 5, the radiation halos formed through
synchrotron and Thomson processes of neutron $\beta$-decay electrons
are derived in the special case of a power-law distribution of
neutrons that are impulsively released from a GRB source. The
hypothesis that hadronic cosmic rays originate from the subset of
supernovae that collapse to form relativistic outflows and GRBs is
briefly considered in Section 6. Fuller discussions of this hypothesis
can be found elsewhere \citep{der00a,der00b}. Summary and conclusions
are given in Section 7.  Appendix A gives the synchrotron radiation 
limit used to determine the maximum proton energies, and Appendix 
B gives a scaling for the photomeson production efficiency in 
surrounding circumburster media (CBM) with different densities.

\section{Statistics and Energetics of UHECRs and GRBs}

\subsection{GRB Statistics}

The BATSE instrument on the {\it Compton Gamma Ray Observatory}
provides a data base of peak count rates and peak fluxes for several
thousand GRBs with unknown redshifts \citep{pac99}.  Many
attempts have been made to derive the GRB rate density and mean
luminosities by modeling this size distribution.  Even constraining
the implied redshift distribution to be consistent with the
$z$-distribution for the GRBs with measured redshifts, it has not been
possible to derive these quantities unambiguously from the size
distribution alone. Uncertainties in determining the rate density of
GRBs arise from lack of knowledge of the redshift distribution
\citep{tot97}, the luminosity function \citep{mm98,kth98,hf99,sch99},
and the spectral shape \citep{mpp96} of GRBs. A useful simplification
\citep{tot97,wij98,tot99} is to assume that the GRB rate density is
proportional to the star formation rate (SFR) history of the universe
as traced, for example, by faint galaxy data in the Hubble Deep Field
\citep{mpd98} (which may, however, seriously underestimate the
true star formation rate at $z\gtrsim 1$ \citep{bla99}). An
important result is that GRBs are unlikely to be standard candles,
whether or not their birth rate follows the SFR or a range of
reasonable evolutionary models \citep{sch99,hf99,kth98}.

To constrain the models further, \citet{bd00} jointly modeled the
distributions of peak flux, duration, and peak photon energies of the
$\nu F_\nu$ spectra of GRBs using an analytic representation
\citep{dcb99} of temporally evolving GRB spectra in the external shock
model of GRBs.  The assumption that the GRB source density followed
the star formation history of the universe was maintained, and a 
flat $\Lambda$CDM cosmology with $(\Omega_0, \Omega_{\Lambda}) =
(0.3, 0.7)$ and Hubble constant $H_0 = 100h $~km~s$^{-1}$~Mpc$^{-1}$,
with $h = 0.65$, was used. The model flux was folded through the
simulated triggering response of a BATSE detector to determine
detectability. This approach requires that the total energy $E_0$ and
the initial blast wave Lorentz factor $\Gamma_0$ of a GRB source be
specified. The burst luminosity is then calculated through the
standard blast-wave physics that yielded the analytic representation
of the GRB spectrum. The analytic model is degenerate in the quantity
$n_0\Gamma_0^8$, where $n_0$ is the density of the surrounding medium,
which is assumed to be uniform. The photomeson production efficiency
can be scaled from $n_0$, as shown in Appendix B.

\citet{bd00} showed that fixed values of $E_0$ and $\Gamma_0$ could
not explain the observed distributions, and that broad ranges of
values are required.  The comoving differential density distribution
of GRB sources was obtained by assuming that the $E_0$ and $\Gamma_0$
distributions are separable from the redshift distribution and are
adequately described by single power-law distributions. The
rate-density distribution $\dot n_{\rm GRB}(E_0,\Gamma_0; z)$ of GRB
sources that gives a reasonable fit to the size, duration, and peak
photon energy distributions, in units of Gpc$^{-3}$
yr$^{-1}~E_{52}^{-1}~\Gamma_0^{-1}$, is
\begin{equation}
\dot n_{\rm GRB}(E_{52},\Gamma_0; z) = 0.022\;\Sigma(z)\; E_{52}^{-1.52}
\Gamma_0^{-0.25}\; H[E_{52}; 10^{-4}, 10^2]\; H[\Gamma_0; 1,260]\;.
\label{ndot}
\end{equation}
In equation (\ref{ndot}), $E_0 = 10^{52}E_{52}$ ergs, and the
Heaviside function is defined such that $H[x;a,b]= 1$ for $a \leq x
\leq b$, and $H[x;a,b]= 0$ otherwise. The range of $\Gamma_0$ given
here corresponds to a density $n_0 = 10^2$ cm$^{-3}$ although, again, the
model is degenerate in the quantity $n_0\Gamma_0^8$. The analytic
representation of the SFR function, normalized to unity at $z = 0$, is
\begin{equation}
\Sigma (z) = 
 \cases{1 \; ,& for $z \le 0.3$ \cr
	5 \cdot 10^{z - 1} \; , & for $0.3 < z \le 1.1$ \cr
        6.3 \; , & for $1.1 < z \le 2.8$ \cr
        210 \cdot 10^{-0.4 \, (z + 1)} \; , & for $2.8 < z \le 10$. \cr}
\label{SFR}
\end{equation}
(Note that the $z \leq 0.3$ branch of this function was omitted in
equation (12) of \citet{bd00}.)

The burst rate and energy release rate per unit comoving volume by GRB
progenitors can be easily obtained from equation (\ref{ndot}). In the
local universe, we find that
\begin{equation}
\dot n_{GRB}(z = 0) =  \int_{E^{min}_{52}}^\infty dE_{52} \int_1^\infty d\Gamma_0 \;\dot n_{\rm GRB}(E_{52},\Gamma_0;z=0)\; \cong 3.6 (E^{min}_{52})^{-0.52}~{\rm Gpc}^{-3}{\rm~yr}^{-1}\;
\label{dNdVdt}
\end{equation}
for the burst rate density, where $E^{min}_{52} $ is the minimum
energy of GRB sources in units of $10^{52}$ ergs, and the
expression on the right-hand-side of this equation is valid when
$E^{min}_{52} \ll 100$. The fit to the BATSE statistics is not
sensitive to the value of $E^{min}_{52}$ when $E^{min}_{52} \ll
1$. When $E^{min}_{52} = 1$, $\dot n_{GRB}(z = 0) \cong 3.6$
Gpc$^{-3}$ yr$^{-1}$. If GRB 980425 is assumed to be associated with
SN 1998bw, then the statistical model requires that $E^{min}_{52}
\cong 10^{-4}$. In this case, $\dot n_{GRB}(z = 0) \cong 430$
Gpc$^{-3}$ yr$^{-1}$, and most GRBs have low energy and luminosity and
are consequently not observed.  The event rate is  therefore
very sensitive to the
number of faint bursts which is not well-constrained by present
data \citep{mm98}.  Beaming will increase the rate
density of sources by the inverse of the mean beaming fraction
compared to the isotropic value given here.

The local  energy emissivity  of the sources of GRBs, from 
equation (\ref{ndot}), is
\begin{equation}
\dot\epsilon_{GRB}(z=0) \cong  \int_0^\infty dE_{52} E_0\int_1^\infty d\Gamma_0 \; \dot n_{\rm GRB}(E_{52},\Gamma_0;z=0) = 3.6\times 10^{53} {\rm ~ergs}~{\rm~Gpc}^{-3}{\rm~yr}^{-1} .
\label{dEdVdt}
\end{equation}
The average energy release per burst is just the ratio of equations
(\ref{dEdVdt}) and (\ref{dNdVdt}), and is equal to $\cong 2.8\times
10^{52}$ ergs and $8.2\times 10^{50}$ ergs when $E^{min}_{52} = 0.1$
and $10^{-4}$, respectively. This does not correspond to the average
energy release of {\it detected} GRBs, because very energetic bursts
are much more likely to be detected. One-half of the total energy
generated by burst sources comes from events with energies $\cong
2.3\times 10^{53}$ ergs. This is a lower limit to the average energy
of an event, because the use of a single power-law function for $E_0$
in equation (\ref{ndot}) does not accurately model extremely powerful
and very rare events, such as GRB 990123. Consequently, equation
(\ref{dEdVdt}) is a lower limit to the local emissivity determined by
fits to BATSE data. This value is not sensitive to the choice for
$E^{min}_{52}$, which is required to be $\lesssim 0.1$ in the
statistical study. Thus the local volume-averaged GRB energy emissivity is
better known than the local GRB event rate density.

Possible collimation of GRB sources does not alter the energetics
arguments made here  as it does for the event rate calculation,
because a smaller beaming fraction is offset by a larger number
of sources. Neither would beaming affect the efficiency calculations
performed below. If GRBs exhibit a constant energy reservoir
\citep{fra01,pk01}, the implications for GRB source emissivity in the
statistical study of \citet{bd00} therefore remains unchanged, and in
fact implies a distribution of jet opening angles $\theta$ for the
sources of GRBs ($dN_{GRB}/d\cos\theta \propto (1-\cos\theta)^{-1/2}$,
so that $dN_{GRB}/d\theta \propto const$ when $\theta \ll 1)$. Because the rate
density of GRB sources is inversely proportional to the
beaming factor, the radiative signatures from a single GRB would be
changed due to beaming. Throughout this paper, we quote energy
emissivities and event rates for uncollimated GRB sources, but
additionally consider the constant energy reservoir result when
calculating the rate of GRBs in the Milky Way.

The statistical study of \citet{bd00} is seen to be consistent with
other recent GRB statistical studies once one recognizes that
inefficiencies for generating radiation from the GRB event and for
detecting emission in the BATSE range have been explicitly taken into
account in this approach (this point was not considered by
\citet{ste00}). Moreover, most burst events with $\Gamma_0 \lesssim
100$ will not trigger a GRB detector such as BATSE due to the
triggering criteria and design of burst detectors that have been flown
to date \citep{dcb99}. These undetected dirty fireballs may contribute
as much as 50-70\% of the total emissivity.

For example, \citet{sch99} derives a local emissivity of GRBs in the
10-1000 keV band of $1.0\times 10^{52}$ ergs Gpc$^{-3}$ yr$^{-1}$,
which is a factor 36 smaller than the value obtained here. The
efficiency for the external shock model to produce radiation in the
10-1000 keV band is $\sim 5$-15\%, and $\sim 50$\% of the total energy
is released in the form of dirty fireballs with $\Gamma_0 \lesssim
100$ that would not trigger BATSE. (The clean fireballs with $\Gamma_0
\gg 300$ cannot be very numerous.) Insofar as inefficiencies for
generating $\gamma$-ray emission in a colliding shell model are
typically 1\% or less \citep{kum99,psm99,bel00,frr99}, and that the
collision of a relativistic shell with matter at rest allows the
greatest fraction of directed kinetic energy to be dissipated within
the blast wave shell \citep{pir99}, we think that equation
(\ref{dEdVdt}) therefore provides a conservative lower estimate for
the emissivity of progenitor sources of GRBs in the local universe.

To obtain the emissivity of GRB sources into the Milky Way galaxy, we
proceed in two ways. The first, following \citet{wij98}, is to employ
the Schechter luminosity function $\Phi(L)dL $ $=
(\Phi^*/L^*)$$(L/L^*)^\alpha $ $\exp(-L/L^*)dL$, giving the number
density of galaxies with luminosities in the range $L$ to
$L+dL$. Assuming that the burst emissivity per galaxy is proportional
to the luminosity of the galaxy, then $\dot\e_{GRB} = k\int_0^\infty
dL\cdot L \cdot \Phi(L)$, so that $k = \dot\e_{GRB} [\Phi^* L^*
\Gamma(\alpha + 2)]^{-1}$, where $\Gamma(v)$ is the Gamma
function. The energy released by GRB progenitors in a galaxy with
luminosity $L$ is $dE(L)/dt\cong (dE/dVdtdL)/(dN/dVdL) = k L
\Phi(L)/\Phi(L)$, so that
\begin{equation}
{dE(L)\over dt} \cong {\dot\e_{GRB}\cdot L\over \Phi^* L^* \Gamma(\alpha + 2)} \cong 2.5 \times 10^{39} ({L\over L^*}) {\rm~ergs~s}^{-1}.
\label{dE(L)dt}
\end{equation}
In the last term of equation (\ref{dE(L)dt}), we used the results of
equation (\ref{dEdVdt}) with $\Phi^* = 1.6\times 10^{-2} h^3$
Mpc$^{-3}$, $\alpha = -1.07$, and $h=0.65$ \citep{lov92}.  If the
Milky Way is an $L^*$ galaxy, then the power of GRB sources into the
Milky Way is therefore $dE/dt \gtrsim 2.5 \times 10^{39}$ ergs
s$^{-1}$.

\citet{sw02} argue that a better approach is to weight the burst
emissivity by the ratio of the blue luminosity surface density
$\Sigma_L$ of the Milky Way to the volume-averaged blue luminosity
density $J_{gal,B}$ of galaxies in the local universe. Using the
expressions $\Sigma_L = 20 L_\odot$ pc$^{-2}$ and $J_{gal,B} \cong
1.1\times 10^8 h L_\odot$ Mpc$^{-3}$ quoted by \citet{sw02}, we find
$dE/dAdt = 1.0\times 10^{38}$ ergs pc$^{-2}$ yr$^{-1}$ for the burst
power per unit area in the solar neighborhood. For a 15 kpc radius, we
then obtain $dE/dt \cong 2\times 10^{39}$ ergs s$^{-1}$ for the GRB
source power in the Milky Way, which is in good agreement with the
value obtained through the first approach. An advantage of this method
is to highlight the potentially large variations in the emissivity of
GRB sources in different regions of a galaxy.

The power required to supply the galactic cosmic radiation, assuming
that cosmic rays are uniformly distributed throughout the disk of the
Galaxy, is $\sim 5\times 10^{40}$ ergs s$^{-1}$ \citep{gai90}. We
therefore see that GRB sources and the dirty and clean fireballs,
collectively referred to as fireball transients (FTs), supply a power
to the Milky Way that is $\gtrsim 5$\% of the cosmic ray power, and
may therefore make an appreciable contribution to cosmic ray
production in the Galaxy. The relative FT/cosmic-ray power could be
much larger if the contribution of clean and dirty fireballs that are
invisible to GRB detectors \citep{dcb99} is much larger than derived
on the basis of the single power-law representation of the $\Gamma_0$-
and $E_0$-distributions. This fraction would also be larger if the
efficiency for the sources of GRBs to generate $\gamma$ rays is
smaller than calculated in the external shock model used by
\citet{bd00}.

\subsection{Ultrahigh Energy Cosmic Rays}

The energy density of UHECRs follows from the intensity $E^3 dJ/dE =
3.5\times 10^{24}$ eV$^2$ m$^{-2}$ sr$^{-1}$ s$^{-1}$ measured with
the Akeno Giant Air Shower Array \citep{tak98}. This expression is
valid within the experimental error for all cosmic rays with energy
$1.2 \times 10^{19}< E$(eV)$ < 3\times 10^{20}$ eV, except for being
1.5$\sigma$ away from the $E\cong 1.4\times 10^{20}$ eV data point. It
is accurate to within 2$\sigma$ of all data points at $3\times
10^{18}< E$(eV) $< 3\times 10^{20}$. Above $3\times 10^{20}$ eV,
small-number statistics dominate. From this expression it follows that
the energy density of UHECRs with energy between $E$(eV) and $3 \times
10^{20}$ eV is
\begin{equation}
u_{\rm UH}(E)\;({\rm ergs~cm}^{-3})\;\cong {2.4\times 10^{-21}\over 
(E/10^{20} {\rm~ eV})}\;( 1- {E\over 3\times 10^{20}{\rm ~eV}})  ,
\end{equation}
The evidence for a high energy tail above $E\approx 3\times 10^{20}$
eV is unclear due to the small-number statistics. Equation (6) should
be considered an upper limit to the UHECR energy density, in view of
the smaller flux of $\gtrsim 10^{20}$ eV particles measured with the
monocular HiRes  fluorescence air shower experiment \citep{som02}.

Ultra-high energy particles lose energy by adiabatic losses in the
expanding universe, and by photo-hadron and photo-pair production on
the cosmic microwave background.  The mean energy loss length $x_{\rm
loss}(E)$ due to these processes has been recently recalculated by
\citet{sta00}. The loss length for $10^{20}$ eV protons is about 140
Mpc, and this length is also consistent with their calculations of
horizon distance within which 50\% of the protons survive. The values
of $x_{\rm loss}$ at $E \gtrsim 6\times 10^{19}$ eV defines the GZK
radius insofar as the energy losses are dominated by photo-hadronic
processes at these energies. The quantity $x_{\rm loss}(E)/c$ defines
a characteristic survival time for particles with energy $E$. The
volume-averaged rate at which astronomical sources produce $> 10^{20}$
eV particles in the local universe is therefore $\lesssim 2.4\times
10^{-21}$ ergs cm$^{-3}/(140 {\rm ~Mpc/c}) \cong 1.5\times 10^{53}$
ergs Gpc$^{-3}$ yr$^{-1}$, provided that UHECRs traverse roughly
straight-line trajectories through intergalactic space. If UHECRs
diffuse in the intergalactic magnetic field, then the source volume
 contributing to locally observed cosmic rays
could be reduced,  though with no significant 
change in the local UHECR intensity due to the better trapping of UHECRs in
our vicinity.

This value is $\gtrsim 2.5$ times smaller than the emissivity given in
equation (\ref{dEdVdt}), so that in principle there is a sufficient
amount of energy available in the sources of GRBs to power the UHECRs
\citep{vie95,wax95}. The conversion of the initial energy of a
fireball into UHECRs must, however, be very efficient. If nonthermal
power-law distributions of particles are accelerated in the blast
wave, as expected in simple treatments of Fermi acceleration, then
hard spectra with  a nonthermal particle injection index 
$p \lesssim 2$ place a large fraction of the
nonthermal energy in the form of the highest energy particles. A large
fraction of the blast-wave energy can be dissipated as UHECRs even if
$p\gtrsim 2$ if particle acceleration is  sufficiently rapid 
that particles reach ultra-high energies and diffusively escape on 
the deceleration time scale \citep{dh01}.

\section{Photomeson and Neutral Particle Production in GRB Blast Waves}

\subsection{Photopion Cross Section and Production Spectra}

Only the photomeson process is considered in detail in this paper;
photopair and secondary production losses involving nucleon-nucleon
collisions can be shown to much less important in comparison to
photomeson losses for ultra-high energy particles in the blast-wave
environment. The two dominant channels of photomeson production for
proton-photon ($p+\gamma$) interactions are $p+\gamma \rightarrow
p+\pi^0$ and $p+\gamma \rightarrow n+\pi^+$, which occur with roughly
equal cross sections. In the latter case, the neutron decays with a
lifetime $t_n\cong 10^3$ s through the $\beta$-decay reaction $n
\rightarrow p+e^-+\bar \nu_e$. The decay of the charged pion produces
three neutrinos and a positron through the chain $\pi^+ \rightarrow
\mu^+ + \nu_\mu$, followed by the decay $\mu^+ \rightarrow e^+ + \nu_e
+ \bar \nu_\mu$.  Neutrino production from photomeson interactions in
GRB blast waves has been considered earlier
\citep{wb97,vie98a,vie98b,rm98,hh99,wb00,dl01}, but usually in the
context of an internal shock model. \citet{rm98a} also consider
neutron production in the internal shock model.

To treat neutral particle production, we follow the approach of
\citet{ste79} (see also \citet{bd98}). The cross section is treated in
the $\delta$-function approximation. Thus an interaction takes place
if the photon energy in the proton's rest frame equals
$\gamma_p\pr\epsilon\pr(1-\mu\pr )= \e_\Delta \cong 0.35 m_p/m_e \cong
640$, where primes denote quantities in the comoving frame, $\gpp$ is
the proton Lorentz factor, $\e$ represents photon energies in units of
the electron rest mass energy, $\epsilon_\Delta$ is the energy of the
$\Delta$ resonance, and $\mu\pr$ is the cosine of the angle between
the photon and proton directions. The differential cross section for
the photomeson production of neutrons and neutrinos produced with
energy $\Ep$ is approximated as
\begin{equation}
{d\sigma_{p\gamma\rightarrow\nu,n}\over d\Ep } =  \zeta_i \sigma_0 \delta[\mu\pr - (1 - {\epsilon_\Delta\over \gpp\e\pr})]\delta(\Ep - m_i\gpp)\; .
\label{dsigdedE}
\end{equation}
The multiplicity $\zeta_i = 1/2$ for neutrons and $\zeta_i = 3/2$ for
neutrinos, noting that we are only considering the neutrinos formed
from $\pi^+$ decay (and not from the neutron). The photomeson cross
section $\sigma_0 \cong 2\times 10^{-28}$ cm$^2$. Each neutrino
carries away about 5\% of the proton's initial energy, with the
$\pi^+$-decay positron receiving another 5\%. Thus we let $m_i = 0.8
m_p$ for neutrons and $m_i = 0.05 m_p$ for neutrinos, with the units
of the proton rest mass $m_p$ defining the units of $\Ep$.

The neutral particle production spectrum in the comoving frame is therefore
\begin{equation}
\dot N\pr (\Ep ) \cong {c\over 2}\; \int _1^\infty d\gpp N_p\pr (\gpp )
\int_0^\infty d\ep n_{\rm ph}\pr (\ep ) \int_{-1}^1 d\mu\pr (1-\mu\pr )\; 
({d\sigma_{p\gamma\rightarrow\nu,n}\over d\Ep })\;,
\label{dN(E)/dt}
\end{equation}
where $N_p\pr (\gpp )$ gives the nonthermal proton spectrum in the
comoving frame, and $n_{\rm ph}\pr (\ep)d\ep$ is the differential
number density of soft photons, assumed to be isotropically
distributed in the blast-wave fluid frame, with photon energies
between $\ep$ and $\ep + d\ep$. Substituting equation (\ref{dsigdedE})
into equation (\ref{dN(E)/dt}) gives
\begin{equation}
\dot N_i\pr (\Ep ) \cong {\zeta_i c \sigma_0\e_\Delta \over 2\Ep }\; N_p\pr (\Ep/m_i )
\int_{m_i\epsilon_\Delta/2\Ep}^\infty d\ep \; \e^{\prime - 1}\;n_{\rm ph}\pr (\ep )\; .
\label{dN'(E)/dt}
\end{equation}
The production spectrum of neutral particles as measured by an
observer can be approximately obtained by noting that the differential
time element $dt \cong dt\pr/\Gamma$, and particle energy $E \cong
\Gamma\Ep$, where unprimed quantities refer to observed
quantities. Redshift effects are not considered in this section. In
more accurate treatments, a full angular integration over the
production spectrum should be performed, which is especially important
if the outflow is collimated. In the present treatment, it is adequate
to use the simpler relations for $dt$ and $E$. Thus $\dot N_i\pr (\Ep
)= \dot N_i(E)$, and we have
\begin{equation}
\dot N_i(E ) = {\zeta_i c \sigma_0\e_\Delta \Gamma \over 2 E }\; N_p\pr (E/\Gamma m_i )\;I(y)\; ,
\label{N(E)}
\end{equation}
where 
\begin{equation}
I(y) = \int_y^\infty d\ep \; \e^{\prime - 1}\;n_{\rm ph}\pr (\ep )\;, \; {\rm and} \; y \equiv {\Gamma m_i \e_\Delta\over 2E}
\label{I(y)}
\end{equation}

\subsection{Blast Wave Dynamics}

We consider the case of an adiabatic blast wave decelerating in a
uniform surrounding medium with density $n_0$.\footnote{This treatment
is reasonably consistent with the statistical treatment of GRBs by
\citet{bd00} There the radiative regime that provided the best fit to
the GRB statistics is nearly adiabatic, with the blast wave
decelerating as $\Gamma \propto x^{-1.7}$, compared to $\Gamma\propto
x^{-1.5}$ in the fully adiabatic limit.} When $\Gamma \gg 1$, the
blast wave evolves according to the relation
\begin{equation}
\Gamma(x) = {\Gamma_0 \over \sqrt{1 + (x/x_d)^3} }\; 
\label{Gamma(x)}
\end{equation}
\citep{cd99}, where $x$ is the distance of the blast wave from the explosion center, and the deceleration radius
\begin{equation}
x_d \equiv ( {3 E_0\over 8\pi\Gamma_0^2 m_pc^2 n_0})^{1/3} \cong
2.1\times 10^{16} ({E_{52}\over \Gamma_{300}^2 n_0})^{1/3}\;\rm{cm}
\label{x_d}
\end{equation}
\citep{mr93}, where $\Gamma_{300} = \Gamma_0/300$.
The rate at which nonthermal proton kinetic energy is swept-up in the
comoving frame of an uncollimated blast wave is
\begin{equation}
\dot\Ep_{ke} = 4\pi x^2 n_0\beta c (m_pc^2)\Gamma(\Gamma-1)
\label{dotEp}
\end{equation}
\citep{bm76}, where $\beta = \sqrt{1-\Gamma^{-2}}$. Thus the accumulated nonthermal kinetic energy at radius $x$ is 
\begin{equation}
\Ep_{ke}(x) = \int_0^x d\tilde x |{dt\pr\over d\tilde x}|\dot\Ep_{ke} = {E_0\over \Gamma_0} 
\cases{{1\over 2}({x\over x_d})^3\;,  & for $x \ll x_d$ \cr
        ({x\over x_d})^{3/2} \;, & for $x_d \ll x \ll x_d
\Gamma_0^{2/3}$ , \cr}
\label{Ep}
\end{equation}
where the largest value of $x$ in the second asymptote stems from the
$\Gamma \gg 1 $ restriction, and $dx = \beta\Gamma c dt^\prime$.

It is convenient to relate the observer's time $t$ to $x$, and
describe blast-wave evolution in terms of the dimensionless time $\tau
\equiv t/t_d$, where the deceleration timescale (for the observer) is
\begin{equation}
t_d =  (1+z) {x_d\over \Gamma_0^2 c} \cong 7.7 (1+z)({E_{52}\over \Gamma_{300}^8 n_0})^{1/3}\;\rm{s}
\label{t_d}
\end{equation}
\citep{rm92,mr93}. Because $dt \cong dx /\Gamma^2 c$, 
\begin{equation}
{x\over x_d}\cong \cases{ \tau\;,  & for $\tau \ll 1$ \cr
        (4\tau)^{1/4} \;, & for $1 \ll \tau \ll \Gamma_0^{8/3}$  \cr} \simeq \; {\tau\over 1+4^{-1/4}\tau^{3/4}}\; , \; {\rm for}\;\tau \ll \Gamma_0^{8/3}\;.
\label{x(tau)}
\end{equation}
Likewise, 
\begin{equation}
{\Gamma\over \Gamma_0}\cong \cases{ 1\;,  & for $\tau \ll 1$ \cr
        (4\tau)^{-3/8} \;, & for $1 \ll \tau \ll \Gamma_0^{8/3}$  \cr} \cong {1\over \sqrt{1+(4\tau)^{3/4}}}\; ,
\label{Gamma(tau)}
\end{equation}
and
\begin{equation}
\Ep_{ke}(\tau)\cong {E_0\over \Gamma_0}\cases{ {1\over 2}\tau^3\;,  & for $\tau \ll 1$ \cr
        (4\tau)^{3/8} \;, & for $1 \ll \tau \ll \Gamma_0^{8/3}$  \cr} \simeq \; {E_0\over \Gamma_0[2\tau^{-3}+(4\tau)^{-3/8}]}\;  , \; {\rm for}\;\tau \ll \Gamma_0^{8/3}\;.
\label{Ep(tau)}
\end{equation}
The expressions on the right-hand-sides of equations (\ref{x(tau)})-(\ref{Ep(tau)}) accurately bridge the early and late time behaviors of the asymptotes.

\subsection{Comoving Proton, Electron, and Photon Spectra}

A power-law distribution of nonthermal protons with number index $p$
is assumed to be accelerated in the blast-wave. Because protons and
ions are swept up with Lorentz factor $\Gamma$ and are then
subsequently accelerated, we represent the nonthermal proton
distribution by the expression
\begin{equation}
N\pr(\gpp;\tau) = {(p-2) \xi \Ep_{ke}(\tau)\over m_pc^2 (\Gamma^{2-p}
- \gamma_{\rm max}^{\prime 2-p})}\;\gamma_p^{\prime -p}\;, \; {\rm
for}\; \Gamma(\tau )\leq \gpp < \gamma_{\rm max}\pr
\label{N(g,t)}
\end{equation}
\citep{bd98}. The term $\xi$ represents the fraction of swept-up
particle kinetic energy that is transformed into the  energy
in the nonthermal proton
distribution and could, in principle, be as large as $\sim 0.5$. Not
more than $\sim 10$-20\% of the total nonthermal proton energy could,
however, be radiated if the treatment is to remain consistent with the
assumption of an adiabatic blast wave.  The term $\gamma_{\rm
max}^{\prime}$, giving the maximum proton Lorentz factor in the blast
wave frame, must be $\gtrsim 10^{10}$ for GRBs to account for
UHECRs. \citet{rm98} define limits on various acceleration scenarios
that give large values of $\gamma_{\rm max}^{\prime}$. Particle
spectra from gyroresonant acceleration  due to pitch-angle scatterings
and stochastic energy diffusion in particle interactions with plasma
waves can give $\gamma_{\rm
max}^{\prime}\gtrsim 10^{10}$ and can produce nonthermal spectra with
$p \gtrsim 1$, though the exact value depends on the spectrum of the
turbulence \citep{sd00,dh01}.

A nonthermal electron spectrum is also assumed to be accelerated in
the blast wave with the same index $p$ as the nonthermal
protons. Following \citet{spn98} (see also \citet{dbc00}), we
represent the nonthermal electron spectrum by the expression
\begin{equation}
N_e\pr(\gp_e ) \cong (s-1) N_e \gamma_0^{s-1}
\, \cases{ \gamma_e^{\prime -s}\;, & for $\gamma_0\leq\gp_e\leq\gamma_1$\cr 
	 \gamma_1^{p+1-s}\gamma_e^{\prime -(p+1)} \; ,& for $\gamma_1
\leq \gp_e \leq \gamma_2$,\cr}\; ,
\label{N_e}
\end{equation}
where $N_e = 4\pi x^3 n_0/3$ is the total number of swept-up nonthermal
electrons and $\gp_e$ is the electron Lorentz factor. In the slow
cooling limit, $\gamma_0 = \gamma_m$, $\gamma_1 = \gamma_c$, and the
steady-state electron spectral index $s=p$, whereas in the fast
cooling limit $\gamma_0 = \gamma_c$, $\gamma_1 = \gamma_m$, and $s =
2$. Here the minimum electron Lorentz factor $\gamma_m \cong e_e
(p-2)\Gamma m_p/[(p-1) m_e]$ and the cooling electron Lorentz factor
$\gamma_c = 3m_e/(16 m_p e_B n_0 c \sigma_{\rm T} \Gamma^3 t)$, where
$e_e$ and $e_B$ are parameters describing the swept-up kinetic energy
transferred to the electrons and the magnetic field, respectively
\citep{spn98}. The magnetic field $B$ is defined through the
expression
\begin{equation}
B = \sqrt{32\pi n_0 m_p c^2 e_B \Gamma (\Gamma-1)} \cong 0.39 \sqrt{e_B n_0} \;\Gamma \;{\rm G}\; .
\label{B}
\end{equation} 
We let $\gamma_2 \cong 4\times 10^7 e_{\rm max}/[B{\rm (G)}]^{1/2}$
\citep{cd99} and take $e_{\rm max} = 1$ in this paper.

We consider only nonthermal synchrotron emission here. Synchrotron
self-absorption and synchrotron self-Compton (SSC) processes are
treated by \citet{dbc00}, including a comparison of the analytic
results to detailed numerical simulations. Given the parameters used
here, the neglect of synchrotron self-absorption is not important for
photomeson production, but the inclusion of Compton processes could,
however, depress the intensity of the low-energy photon spectrum when
$e_B \ll e_e$.

In the $\delta$-function approximation for the synchrotron emissivity,
the photon production spectrum
\begin{equation}
\dot N_{ph}\pr(\e\pr ) = {3\over 8} \nu_0 \e^{\prime -1/2} \e_H^{-3/2} N_e\pr ( \sqrt{{\e\pr\over\e_H}})\; ,
\label{dotN(e)}
\end{equation}
where $\nu_0 =  \nu_0(B) = 4 c\sigma_{\rm T} u_B/3$,
 $u_B = B^2/(8 \pi m_ec^2)$ is the
magnetic-field energy density  in units of $m_ec^2$ cm$^{-3}$, and $\e_H =
B/B_{cr} = B/4.413\times 10^{13}$ G. The magnetic field is assumed to
be randomly oriented. This formula is accurate to better than a
factor-of-2 except near the endpoints of the distribution (see Fig.\ 2
in \citet{dbc00}).  By substituting equation (\ref{N_e}) into equation
(\ref{dotN(e)}), we obtain the comoving photon density
\begin{equation}
n_{ph}\pr (\ep ) = K \cases{\gamma_0^{-2/3}(\ep/\e_H)^{-2/3} \; , &
 for $\ep/ \e_H \le \gamma_0^2$ \cr
 \gamma_0^{s-1}(\ep/\e_H)^{-(s+1)/2} \; , & for $\gamma_0^2 < \ep/
 \e_H \le \gamma_1^2$ \cr \gamma_0^{s-1}\gamma_1^{p+1 -s}
 (\ep/\e_H)^{-(p+2)/2} \; , & for $\gamma_1^2 < \ep/ \e_H \le
 \gamma_2^2$ \cr 0 \; , & for $\gamma_2^2 \leq \ep/ \e_H $ , \cr}
\label{nph(e)}
\end{equation}
where 
\begin{equation}
K \equiv {3\over 8}{\nu_0 (s-1) N_e\over 4\pi x^2 c\e_H^2} = {B_{cr}^2 \sigma_T (s-1) x n_0\over 48\pi m_ec^2}\; .
\label{K}
\end{equation}

Substituting equation (\ref{nph(e)}) into equation (\ref{I(y)}),
performing the integrals, and defining $\gamma_{n,i} \equiv
\Gamma\epsilon_\Delta/(2\epsilon_H \gamma_i^2$) for $i = 0, 1,$ and 2,
we obtain
\begin{equation}
I(\gamma) = 2K 
\cases{0\; , & for $\gamma \leq \gamma_{n,2} $  \cr
(p+2)^{-1}\gamma_0^{s-1}\gamma_1^{p+1 -s}
[({\Gamma\epsilon_\Delta\over
2\e_H\gamma})^{-(p+2)/2}-\gamma_2^{-p-2}]\; , & for $\gamma_{n,2} \leq
\gamma < \gamma_{n,1}$ \cr (p+2)^{-1}\gamma_0^{s-1}\gamma_1^{p+1 -s}
(\gamma_1^{-p-2}-\gamma_2^{-p-2}) & $~$ \cr
\;\;\;\;\;\;\;\;\;\;
+(s+1)^{-1}\gamma_0^{s-1} [({\Gamma\epsilon_\Delta\over 2\e_H\gamma})^{-(s+1)/2}-\gamma_1^{-s-1}]
\; , & for $\gamma_{n,1} \leq \gamma < \gamma_{n,0}$ \cr
(p+2)^{-1}\gamma_0^{s-1}\gamma_1^{p+1 -s} (\gamma_1^{-p-2}-\gamma_2^{-p-2})  + 
(s+1)^{-1}\gamma_0^{s-1}
&  $~$ \cr
\;\;\;\;\;\;\;\;\;\;
 \times (\gamma_0^{-s-1}-\gamma_1^{-s-1}) + {3\over
4}\;\gamma_0^{-2/3} [({\Gamma\epsilon_\Delta\over
2\e_H\gamma})^{-2/3}-\gamma_0^{-4/3}]
\; , & for $\gamma \geq \gamma_{n,0}$ ,\cr}
\label{I(gamma)}
\end{equation}
where $\gamma \equiv E/m_i$.   The $(\Gamma\epsilon_\Delta /2\e_H\gamma)$ term dominates each of the branches of equation (\ref{I(gamma)}). A good approximation to $I(\gamma)$ is therefore 
\begin{equation}
I_{ap}(\gamma) = {2K\over \gamma_0^2}\; 
\cases{0\; , & for $\gamma \leq \gamma_{n,2} $  \cr
(p+2)^{-1}({\gamma_0\over \gamma_1})^{s+1}\;({\gamma\over \gamma_{n,1}})^{(p+2)/2}
\; , & for $\gamma_{n,2} \leq \gamma < \gamma_{n,1}$ \cr
(s+1)^{-1}\;({\gamma\over \gamma_{n,0}})^{(s+1)/2}
\; , & for $\gamma_{n,1} \leq \gamma < \gamma_{n,0}$ \cr
 {3\over 4}\;({\gamma\over \gamma_{n,0}})^{2/3}
\; , & for $\gamma \geq \gamma_{n,0}$ .\cr}
\label{Iap(gamma)}
\end{equation}
The production spectrum $\dot N_i(E)$ of neutral particles formed
through photomeson production is therefore given by equation
(\ref{N(E)}), but with $I(y)$ replaced by either $I(\gamma)$ or
$I_{ap}(\gamma)$ given by equations (\ref{I(gamma)}) or
(\ref{Iap(gamma)}), respectively.

\subsection{Energy-Loss Timescales for High Energy Protons}

Energy-loss timescales are derived in the comoving frame for protons
that would have energies $E$ as measured in the observer frame. These
timescales are compared with the comoving time $t\pr$ passing since
the initial explosion event; clearly if the energy-loss timescale is
long compared with the available comoving time, then only a small
fraction of the particle energy can be extracted through that
process. From the relation $dx = \beta \Gamma c dt\pr$, we obtain the
comoving time
\begin{equation}
t\pr \cong \Gamma_0 t_d \cases{ \tau\;,  & for $\tau \ll 1$ \cr
        {2\over 5}(4\tau)^{5/8} \;, & for $1 \ll \tau \ll \Gamma_0^{8/3}$  \cr} \simeq \; {\Gamma_0 t_d\tau\over 1+\tau^{3/8}}\; , \; {\rm for}\;\tau \ll \Gamma_0^{8/3}\;.
\label{tpr(tau)}
\end{equation}

The photopion  energy-loss rate is
\begin{equation}
t_{p\gamma}^{\prime -1} \cong {1\over 5}\;{c\over 2}\;
\int_0^\infty d\ep n_{\rm ph}\pr (\ep ) \int_{-1}^1 d\mu\pr (1-\mu\pr )\;\sigma_{p\gamma\rightarrow\nu,n}(\e\pr,\mu\pr)\;,
\label{tpr_pg}
\end{equation}
(compare eq.\ [\ref{dN(E)/dt}]),  where the factor $1/5$
takes into account that $\approx 5$ interactions are 
required for a high-energy proton to lose a significant
amount of its energy. Here we consider both the
$p\gamma \rightarrow \pi^+n$ and $p\gamma\rightarrow \pi^0p$ chains,
because both will compete against other energy-loss processes. Thus
$\sigma_{p\gamma\rightarrow\nu,n}(\e\pr,\mu\pr) \cong \sigma_0
\delta[\mu\pr - (1 - \epsilon_\Delta/\gpp\e\pr)]$, giving
\begin{equation}
t_{p\gamma}^{\prime -1} \cong {c\sigma_0 \epsilon_\Delta \over 
 10\gp_p }\; I(\gamma )\;,
\label{tpr_pg1}
\end{equation}
where $\gamma = E/m_p = \Gamma \gpp$. Hence
\begin{equation}
-({d\gp_p\over dt\pr })_{p\gamma} \cong {c\sigma_0\e_\Delta\over 
 10}\;I(\Gamma\gp_p)\;.
\label{dg_pgdt}
\end{equation}
The energy-loss rate through photopair ($p + \gamma \rightarrow p +
e^+ + e^-$) production is small compared to the photomeson energy-loss
rate at very high energies because of the greater energy loss per
scattering event in photomeson production. Although photopair
production could dominate the energy-loss rate for protons with $\gpp
\ll 10^8$, it is not important for the highest energy protons and is
not treated here.

The proton synchrotron loss rate is given by
\begin{equation}
t_{p,syn}^{\prime - 1} \cong {\nu_0\gpp \over (m_p/m_e)^3} = 
{e_B n_0({\rm cm}^{-3}) \Gamma^2 \gpp \over 3.2\times 10^{19} {\rm ~s}}\; .
\label{t_psyn}
\end{equation}
The importance of this process for producing high-energy $\gamma$ rays
from GRB blast waves has been considered by \citet{vie97} and
\citet{bd98}.

 The secondary production rate is $t_{pp}^{\prime -1} = n\pr
\sigma_{pp}c$, where $n\pr = 4\Gamma n_0$ from the shock 
jump conditions, and $\sigma_{pp} \cong 30$ mb. The secondary
production efficiency $\eta_{pp} = t/t_{pp} = \hat\eta_{pp}
\tau/[1+(4\tau)^{3/4}]$, where $\hat\eta_{pp} = 8n_0\sigma_{pp} x_d$
$\cong 5\times 10^{-9} E_{52}^{1/3} (n_0/\Gamma_{300})^{2/3}$.
The secondary production efficiency increases through the 
afterglow phase, and reaches a value of $\eta_{pp} \cong 
10^{-7} E_{52}^{1/3}n_0^{2/3}$ at $\tau = \Gamma_0^{8/3}$.
Unless $n_0 \gg 10^{8}$ cm$^{-3}$, the efficiency 
for this process will be low (see \cite{ps00} and \citet{sps02}
for treatments of this process in GRBs and blazars, respectively).

Fig.\ 1 shows results of calculations of the ratio $\eta$ of the
comoving time to timescales for photomeson production (open circles)
and proton synchrotron radiation (filled circles). The timescales are
calculated at different observer times as a function of observed
proton energy $E$, up to the maximum proton energy defined by the
synchrotron radiation limit given in Appendix A. The chosen parameters
in Figs.\ 1a and 1b are typical of those used to fit GRBs in the
prompt and afterglow phase, respectively, and are listed in Table
1. In both cases, we use a total energy release $E_0 = 2\times
10^{53}$ ergs, which is near the mean value of the energy release
distribution (see Section 2.1). The value $p = 2.2$ is similar to that
deduced in fits to afterglow GRB spectra of GRB 990510 \citep{har99}
and GRB 970508 \citep{wg99}; a value of $p$ much steeper than $\sim
2.2$ will make the energetics of UHECR production problematic. 

Other than $E_0$ and $p$, Fig.\ 1a employs the parameter set in Fig.\
1 of \citet{dcm00} which was shown to give good fits to burst spectra
during the $\gamma$-ray luminous phase of GRBs (\citet{cd99}; there we
used $E_0 = 10^{54}$ ergs and $p = 2.5$). The remaining parameters
used in Fig.\ 1a are $\Gamma_0= 300$, $n_0 = 100$ cm$^{-3}$, $e_B =
10^{-4}$, and $e_e = 0.5$. Even with such a large value of $e_e$, the
blast wave evolves in the adiabatic limit because the electrons are in
the weakly cooling regime.  We also take $\xi = 0.5$. The dotted lines
show the photopion timescales obtained using the approximate
expression for $I_{ap}(\gamma)$ in equation (\ref{Iap(gamma)}). Fig.\
1b uses parameters that are typical of those used to model the
afterglow spectra of GRBs \citep{har99,wg99}, and are the same as
Fig.\ 1a except that $e_B = 0.1$ and $e_e = 0.1$. The latter choice
ensures that the GRB blast-wave evolution is nearly (though not quite;
see \citet{bd00a}) adiabatic, given that the electrons are strongly
cooled during the prompt phase and much of the afterglow phase for
this larger value of $e_B$. The major difference between the fits
derived to GRB spectra during the prompt and afterglow phases is thus
the stronger value of field at later times. Other arguments that the
magnetic field evolves to its equipartition value following the prompt
phase are given by
\citet{dcm00}.

As can be seen from Fig.\ 1, the relative timescales for photomeson
production in the external shock model usually dominates the other
processes, and approaches or exceeds unity for the highest energy
protons during the afterglow phase.  Thus a large fraction of the
energy contained in the highest energy protons is converted into an
internal electromagnetic cascade and lost as photomeson neutral
secondaries.  The largest proton energies are constrained by the
synchrotron limit given in Appendix A, but still exceeds $10^{20}$ eV,
in accord with the hypothesis that UHECRs are accelerated by GRBs.
Protons with observed energies $\gtrsim 10^{18}$ eV therefore lose a
significant fraction of their kinetic energy through photomeson
production which is transformed into neutrons, neutrinos, and high
energy leptons. The leptons generate high energy gamma rays during an
electromagnetic cascade in the blast wave \citep{bd98}. When $\e_B
\gtrsim 0.1$, proton synchrotron losses can dominate photomeson losses
during certain phases of the evolution.   Although secondary
production can be the dominant proton energy loss process at $E
\lesssim 10^{16}$ eV, its importance is negligible unless the CBM
density is very high.

When the relative timescales exceed unity, a large fraction of the
proton energy is radiated away during the comoving time $t\pr$, and
the proton distribution will strongly evolve through radiative
cooling. When this occurs, a thick-target calculation is required to
calculate total neutrino and neutron emissivity. This regime begins to
be encountered here, but a complete treatment of photopion production
will require solving a transport equation that is beyond the scope of
this paper. We also note that  the efficiency ratio $\eta = 
t\pr/t\pr(process)$ is
actually shorter when the injection index $p$ becomes larger, because
the energy density of the soft photons is then concentrated into a
narrower bandwidth and is therefore more intense. Nevertheless, much
less energy of the total GRB energy is radiated through photomeson
production when $p\gg 2$, because the total GRB energy carried by the
highest energy protons is much smaller.

\subsection{Instantaneous and Time-Integrated Production Spectra}

It is simple to derive the characteristic spectral behavior of
neutrons or neutrinos produced in the external shock model.  Using
equations (\ref{Iap(gamma)}) and (\ref{N(g,t)}) in equation
(\ref{N(E)}), we find that the instantaneous production spectra,
multiplied by $E^2$, follow the behavior
\begin{equation}
E^2 \dot N_i(E) \propto 
\cases{0\; , & for $E/m_i\leq \max(\Gamma,\gamma_{n,2}) $  \cr
E^{(4-p)/2}\; , & for $\max(\Gamma,\gamma_{n,2})\leq E/m_i < \gamma_{n,1}$ \cr
E^{(3+s-2p)/2}\; , & for $\gamma_{n,1} \leq E/m_i < \gamma_{n,0}$ \cr
 E^{-p + 5/3}\; , & for $\gamma_{n,0} \leq E/m_i < \gamma_{{\rm L},max}$ \cr
0 \;, & for $E/m_i > \gamma_{{\rm L},max}$	,\cr}
\label{E^2N(E)}
\end{equation}
where $\gamma_{{\rm L},max}$ is given by the synchrotron radiation
limit, equation (\ref{gLmax}).  These spectral indices are two units
larger than  particle injection number indices. It is also
assumed in these expressions that $\Gamma < \gamma_{n,1}$ and
$\gamma_{{\rm L},max} >
\gamma_{n,0}$, but it is simple to generalize the results when this is
not the case. The instantaneous production spectra are very hard at
low energies, with $\dot N_i(E) \propto E^{-p/2} \psim E^{-1}$ when $
p \sim 2$. Irrespective of whether we are in the fast cooling ($s =
2$) or slow cooling ($s=p$) regime, the spectra soften to $\dot N_i(E)
\psim E^{-3/2}$ for $p\sim 2$, although the spectra still rise in an
$E^2 \dot N_i(E) $ representation. At energies $E \gtrsim
m_i\gamma_{n_0}$, $\dot N_i(E) \propto E^{-p-1/3} \psim E^{-7/3}$,
where the $-7/3$ behavior holds when $p\sim 2$. The $E^2 \dot N_i(E)$
peak energy is carried primarily by particles with energy
\begin{equation}
E_{pk} \cong m_i \gamma_{n,0} = {m_i \Gamma \e_\Delta \over 2 \e_H \gamma_0^2} = {m_i \Gamma^2 \e_\Delta \over 2\e_{br}}\simeq {(\Gamma/300)^2\over (\e_{br}/0.1)} 
\cases{2\times 10^{17}{\rm ~eV}\; , & for ${\rm neutrons}$  \cr
10^{16} {\rm ~eV}\;, & for ${\rm neutrinos}$ .	\cr}
\label{Epk}
\end{equation}
In this expression, the break energy $\e_{br}$ is the photon energy
separating the $\e^{-2/3}$ portion of the synchrotron emissivity
spectrum produced by an electron distribution with a low-energy cutoff
from the higher-energy portion of the synchrotron spectrum. As is well
known, this often occurs at energies $\sim 50$ keV - several MeV
during the prompt phase of GRBs \citep{coh97}. Equation (\ref{Epk})
also follows from elementary considerations.

Figs.\ 2a and 2b show instantaneous production spectra at different
observing times for neutrons and neutrinos, respectively. Fig.\ 2a
employs the parameter set (A) for the prompt phase of GRBs and Fig.\
2b uses the parameter set (B) that better represents afterglow data
(see Table 1).  The peak of the $E^2 \dot N_i(E)$ spectrum at
$E_{pk}$, above which $\dot N_i(E)\propto E^{-2.53}$, occurs at early
times in the instantaneous spectra of Fig.\ 2a.  The transition to the
$E^{-p-1/3}$ portion of the spectrum is not seen after $\sim 10^4$ s
is Fig.\ 2a nor in the Fig.\ 2b spectra. This is because $\e_{br}$
reaches such low energies that $E_{pk}$ would occur above the maximum
energy defined by equation (\ref{gLmax}). The maximum energies of the
neutrons reach or exceed $\sim 10^{21}$ eV, but the maximum neutrino
energies only reach $\lesssim 10^{20}$ eV due to the smaller amount of
energy transferred to each neutrino in the photomeson production
process (compare cross section (\ref{dsigdedE})). The production
spectrum breaks from $\dot N_i(E) \propto E^{-0.9 }$ at low energies
to $\dot N_i(E) \propto E^{-1.6}$ at intermediate energies in Fig.\
2a, because the electrons distribution starts to evolve in the
uncooled regime. In contrast, the spectrum above the break in Fig.\ 2b
is slightly softer with $\dot N_i(E) \propto E^{-1.7}$, because the
electron distribution evolves in the strongly cooled regime.

We also show the time-integrated production spectra of both the
neutrons and neutrinos for parameter sets (A) and (B) in Figs.\ 2a and
2b, respectively. Here we integrate the instantaneous production
spectra over all times until the blast wave reaches $x =
x_d\Gamma_0^{2/3}$, where it has decelerated to mildly relativistic
speeds. The time-integrated spectra retains its $\dot N_i(E) \propto
{-p/2}$ behavior at $10^{12} \lesssim E$(eV) $\ll 10^{17}$ eV. For the
prompt-phase parameter set (A), the time-integrated spectra steepens
to a $\dot N_i(E) \psim E^{-1.8}$ behavior above the value of $E_{pk}$
evaluated at $t= t_d$, and then cuts off at a maximum energy
determined by equation (\ref{gLmax}) at $\tau = 1$. For parameter set
(B), the time-integrated spectrum remains very hard, with $\dot N_i(E)
\psim E^{-p/2}$, up to nearly the maximum energy defined by
$m_i\gamma_{{\rm L},max}$.

The time-integrated spectra in Fig.\ 2 imply both the total energy
release and the energies of the produced neutrons and neutrinos that
carry the bulk of this energy.  Neutrons with energies between $\sim
10^{18}$ and $\sim 10^{21}$ eV carry $\sim 10^{51}$ ergs of energy for
the chosen parameters. Neutrinos carry $\sim 3/20$ as much total
energy as the neutrons in an energy range that is $\sim 20$ times
smaller than that of the neutrons. The ratio of the energy carried by
either neutrons or neutrinos to the total explosion energy $E_0$, here
called the production efficiency, is therefore $\sim 1$\% for neutrons
and $\sim 0.1$\% for neutrinos. Fig.\ 3 shows calculations for the
neutron and neutrino production efficiencies as a function of
$E_{54}$. The neutron production efficiency increases with increasing
$E_0$ and reaches a few per cent when $E_{54} = 1$.  Parameter set (A)
gives better efficiency at large values of $E_0$ than set (B), but
poorer efficiencies when $E_{54} \lesssim 0.2$. The production
efficiency is only weakly dependent upon $\Gamma_0$, but depends
strongly upon $p$, as outlined earlier and shown in the inset. The
maximum efficiency occurs when $p\sim 2.1$. According to the
statistical treatment of the external shock model described in Section
2.1, $\sim 50$\% of the total GRB energy is radiated by explosions
with $E_0 \gtrsim 2\times 10^{53}$ ergs. Thus we find that $\gtrsim
1$\% and $\gtrsim$ 0.2\% of this energy is converted into high-energy
neutrons and neutrinos, respectively, if the UHECR/GRB hypothesis is
correct.  This will have the observable consequences described in
Sections 4 and 5.

 The efficiencies for neutral particle production correspond to
GRBs with smooth profiles.  In the external shock model result,
smooth-profile GRBs result from blast wave deceleration in a uniform
surrounding medium. We have chosen $n_0\cong 100$ cm$^{-3}$.  Because $\eta
\propto \sqrt{n_0}$ (Appendix B), smooth profile GRBs could produce 
neutrinos and neutrons with even smaller efficiencies if $n_0 \lesssim
1$ cm$^{-3}$. Interactions with an inhomogeneous and clumpy CBM are
thought to produce the short timescale variability observed in rapidly
variable GRBs in the external shock model
\citep{dm99,db02}. Under these circumstances, neutrino production 
could be considerably enhanced. Thus the production calculations only
apply to GRBs which display smooth profiles.

\subsection{Temporal Behavior of Production Spectra}

The temporal indices of the particles formed through photomeson
production can be obtained by examining equations (\ref{N(E)}),
(\ref{N(g,t)}) and (\ref{Iap(gamma)}), noting the temporal dependences
of the various terms. Writing equation (\ref{N(E)}) in more detail, we
have
\begin{equation}
\dot N_i(E ) = {\zeta_i c \sigma_0\e_\Delta \Gamma \over 2 E }\; {(p-2) \xi \Ep_{ke}(\tau)\over m_pc^2 (\Gamma^{2-p} - \gamma_{\rm max}^{\prime 2-p})}\;({E\over \Gamma m_i})^{-p}\;K \; [ I_{ap}(E/m_i)/ K] \; .
\label{N_i(E)}
\end{equation}
provided $\Gamma \leq E/m_i < \gamma\pr_{\rm max}$. The coefficient $K
$ has been extracted from the $I_{ap}(\gamma)$ term, and varies
according to $K({\tau})\propto x$ (equation (\ref{K})), so that it has
the time dependence given by equation (\ref{x(tau)}). The time
dependences of $\Gamma$ and $\Ep_{ke}$ are given by equations
(\ref{Gamma(tau)}) and (\ref{Ep(tau)}), respectively. It then becomes
necessary to determine the time dependences of $\gamma_{n,i} \equiv
\Gamma\e_\Delta/2\e_H\gamma_0^2$ and therefore of the $\gamma_i$ that
enter into equation (\ref{Iap(gamma)}), noting that $\e_H \propto B
\propto \Gamma$.

The temporal behavior of the neutron and neutrino production time
profiles, or ``light curves," depends on whether the electrons are in
the slow or fast cooling regimes \citep{spn98}.  Because of the
progressive weakening of the magnetic field in the standard blast-wave
model, the fast cooling regime will exist only if the cooling electron
Lorentz factor $\gamma_c$ is less than the minimum electron injection
Lorentz factor $\gamma_m$ at $\tau \approx 1$. Using the expressions
for $\gamma_c$ and $\gamma_m$ following equation (\ref{N_e}), we
therefore find that the nonthermal electrons will evolve in the fast
cooling regime at least during some stage of the blast-wave evolution
if
\begin{equation}
\Gamma_0 \gtrsim \bar \Gamma_0 = {0.16\over n^{1/2} E_{52}^{1/4}} \; [{1\over e_e e_B} \cdot({ p-1\over p-2 })]^{3/4}\; ,
\label{G0}
\end{equation}
using equation (\ref{t_d}) for $t_d$. When equation (\ref{G0}) does
not hold, the system is always in the slow cooling regime. For
example, if we vary only $\Gamma_0$ in parameter set (A), there will
be some evolution in the fast cooling regime when $\Gamma_0 \gtrsim
70$. There will be evolution in the fast cooling regime for
essentially all values of $\Gamma_0 \gg 1$ with parameter set
(B). Note that the baryon-loading factor $\bar\Gamma_0$ separating the
different cooling regimes is quite sensitive to $n_0$, with $\bar
\Gamma_0 \propto n_0^{-1/2}$.

Fig.\ 4 is a sketch of the temporal indices of particles produced with
different energies as a function of dimensionless time $\tau$. First
consider the outside boundaries of the temporal-index plane. Particles
will only be produced if $E/m_i \gtrsim \Gamma$. This defines the
lower region bordered by the short dashed lines. Due to threshold
effects, another limit to low-energy particle production arises from
photomeson threshold effects due to the upper cutoff of the highest
energy photons at $\e\pr \sim \gamma_2^2 \e_H$. Neutrons and neutrinos
will not be produced with energies $m_i \gamma_{n,2} < m_i
\Gamma\epsilon_\Delta/(2\gamma_2^2 \e_H$) because of this cutoff. For
the synchrotron radiation limit given by equation (\ref{gLmax}) ---
but now for electrons --- $\gamma_2 \propto B^{1/2}$, so $\gamma_{n,2}
\propto \Gamma^{-1}$ as sketched by the dashed-triple-dotted
lines. The synchrotron radiation limit for protons defines the upper
boundary for the highest energy particles that are produced. It is
$\propto \Gamma^{1/2}$, and is shown by the dot-dashed
lines. Different Fermi acceleration models could give different
maximum particle energies, but all would likely be bounded from above
by this limit due to the competition between the synchrotron loss and
acceleration rates.

The blast-wave system, as illustrated in Fig.\ 4, passes through a
fast cooling regime. Thus there are two dimensionless times $\tau_<$
and $\tau_>$ defined by the relation $\gamma_c = \gamma_m$ that bound
the period when the blast wave is in this regime. The blast wave is in
the slow cooling regime when either equation (\ref{G0}) fails to hold
or, if not, when $\tau < \tau_<$ and $\tau > \tau_>$. We define
$\gamma_{n,m} \equiv \Gamma\epsilon_\Delta/(2\e_H\gamma_m^2)\propto
\Gamma^{-2}$ and $\gamma_{n,c}\equiv
\Gamma\epsilon_\Delta/(2\e_H\gamma_c^2)\propto \Gamma^6\tau^2$. Hence
$\gamma_{n,m} \propto 1$ and $\gamma_{n,c} \propto \tau^2$ for $\tau
\lesssim 1$, and $\gamma_{n,m} \propto \tau^{3/4}$ and $\gamma_{n,c}
\propto \tau^{-1/4}$ when $1 \lesssim \tau \lesssim \Gamma_0^{8/3}$.
The behaviors of $\gamma_{n,m}$ and $\gamma_{n,c}$ are indicated by
the thick lines and the double lines, respectively, in Fig.\ 4.

In the slow-cooling regime, $\gamma_0 = \gamma_m$ and $\gamma_1 =
\gamma_c$. In the fast-cooling regime, $\gamma_0 = \gamma_c$ and
$\gamma_1 = \gamma_m$.  It is straightforward though tedious to derive
the temporal indices $\chi$ displayed in Fig.\ 4 using the above
relations and equation (\ref{Iap(gamma)}) in equation
(\ref{N_i(E)}). The important point to notice is how hard the values
of $\chi$ are in the afterglow phase. The highest energy particles
with $\gamma_{n,0} < \gamma < \Gamma \gamma\pr_{\rm max}$ are due to
interactions with the $\e^{-2/3}$ part of the soft photon spectrum. In
the uncooled regime, $\chi \approx -0.25$ when $p\sim 2$, so that the
bulk of the energy is radiated at late times. Because of the rapid
decay of $\e_{br}$ with time, however, this phase does not persist
very long. Nevertheless, the temporal indices in the lower and
intermediate energy regimes are $\chi \cong -0.75$ and $\chi \cong
-0.875$, respectively, in the afterglow phase when $p \sim 2$. Thus
the bulk of the energy is still radiated at late times. Although this
temporal behavior will be difficult to detect from neutrinos and
neutrons from GRBs, they are relevant to the high-energy gamma-ray
spectrum observed from GRBs. Charged pions will decay into leptons,
which will scatter soft photons to high energies to generate a
cascade, and neutral pions from $p + \gamma \rightarrow p + \pi^0$
will decay to form $\gamma$ rays that can pair produce until the
photons are at sufficiently low energies to escape. As noted by
\citet{bd98}, the temporal decay of the high-energy emission from
hadrons is much slower than the synchrotron decay. Thus high-quality
GeV observations of GRBs could reveal the presence of a high-energy
hadronic component, though it must be carefully distinguished from the
SSC component, for example, by its spectral characteristics. The {\it
Gamma ray Large Area Space Telescope} (GLAST)
mission\footnote{http://glast.gsfc.nasa.gov} will be well-suited to
measure the $\gamma$-ray afterglow of GRBs and thus test for an
energetic hadronic component in GRB blast waves.

Fig.\ 5 shows calculations of the neutron and neutrino production time
profiles at $10^{12}$, $10^{15}$, $10^{18}$, and $10^{20}$ eV for
parameter sets (A) and (B), as described in the figure caption. Here
we have multiplied the $E^2 \dot N(E)$ spectra by observing time $t$
in order to reveal the time during which the bulk of the energy is
radiated. For these parameters, roughly equal energy is radiated per
decade of time, except at the very highest energies and during early
times. From the analytic results for the temporal index in the
afterglow phase, we find that $\chi\cong -0.32$ and $\chi \cong -1.1$
at intermediate energies, in agreement with the calculations. The
lower energy regime with $\chi \cong -0.975$ is not encountered
here. The abrupt cutoffs at early and late times are due to the
definite ranges of particle energies implied by the analysis.

\section{Neutrinos from GRBs}

The detailed calculations provide a lower limit to the neutrino fluxes
if the UHECR/GRB hypothesis is correct. Even within the context of the
external shock model, other effects could enhance the neutrino
emissivity. For example, reverse shock emission provides additional
soft photons that would enhance photomeson production
\citep{dl01}. Larger neutrino fluxes could also be obtained if we
relax the assumption that the surrounding medium is uniform, which is
probably the case in many GRBs, in view of the short timescale
variability observed in their $\gamma$-ray emission \citep{dm99}. Also
important is the uncertainty in determining the rate density
of dirty and clean fireballs.

Before displaying calculations, it is useful to make an estimate of
the neutrino background expected from GRBs. The energy density of
high-energy neutrinos from GRBs is
\begin{equation}
u_\nu \simeq \;\eta_\nu\cdot\Sigma(1)\cdot \dot\e_{GRB}(z = 0) \cdot t_{\rm H} \simeq 6\times 10^{-19} \eta_\nu \; {\rm ergs~cm}^{-3}
\label{u_nu}
\end{equation}
where we use equations (\ref{dEdVdt}) and (\ref{SFR}) to give the mean
emissivity at $z \approx 1$, and let the Hubble time $t_{\rm H} =
10^{10}$ yr. The term $\eta_\nu$ represents the production efficiency
which, as we have seen, is $\sim 0.1$\%. For $p \sim 2$, the principal
behavior of the time-integrated GRB neutrino spectrum varies  $\propto
E_\nu^{p/2}\sim E_\nu^{-1}$ up to some maximum energy 
$E_{\nu,{\rm max}} \approx
10^{18}$-$10^{19}$ eV (see Fig.\ 2 and App.\ B), so that the diffuse neutrino
background flux $\Phi_\nu(E_\nu)$ also varies $ \propto
E_\nu^{-1}$. Using equation (\ref{u_nu}) to normalize this flux, we
find
\begin{equation}
{E_\nu \cdot\Phi_\nu(E_\nu)} \simeq {9\times
10^{-19}(\eta_\nu/10^{-3}) \over (E_{\nu,{\rm max}}/10^{18}
{\rm~eV})}\;{\rm~cm}^{-2}{\rm~s}^{-1}{\rm~sr}^{-1}\;.
\label{dNdlnE}
\end{equation}
 GRBs will therefore produce a diffuse neutrino flux at the level of
$10^{-9}$ GeV cm$^{-2}$ s$^{-1}$ sr$^{-1}$ at $E_\nu \gtrsim 10^{18}$ eV.
 This flux\footnote{A similar estimate for the diffuse extragalactic
$\gamma$-ray background, though now assuming a spectrum varying
$\propto \e_\gamma^{-2}$ as might be expected from an electromagnetic
cascade in the GRB blast wave, gives a diffuse flux of
$\e_\gamma^2\cdot \Phi_\gamma \sim 0.1\eta_\gamma$ keV cm$^{-2}$
s$^{-1}$ sr$^{-1}$.  The $\gamma$-ray production efficiency
$\eta_\gamma$ could reach 10\%, but this emission still falls well
below the observed diffuse extragalactic $\gamma$-ray background,
which has an intensity of $\sim 1$ keV cm$^{-2}$ s$^{-1}$ sr$^{-1}$
between $\sim 100$ MeV and $\sim 100$ GeV \citep{sre98}.} is
comparable to other estimates of cosmological neutrinos above $\sim
10^{16}$ eV \citep{ghs95,yt93,ste91}. 

\subsection{Diffuse Neutrino Background}
Each GRB produces a time-integrated neutrino spectrum ($dE/dE^{\rm
em}_\nu)$ $= $$E^{\rm em}_\nu $$(dN/dE^{\rm em}_\nu) $, where $E^{\rm
em}_\nu$ is the energy of the emitted neutrino. The luminosity
distance $d_L$ is defined so that the relationship $dE/dA dt = (4\pi
d_L^2)^{-1} (dE/dt_{\rm em}$) holds, where $dt_{\rm em}$ is the
differential element of time in the emitter frame. Because $dt = (1+z)
dt^{\rm em}$ and $E_\nu = E^{em}_\nu/(1+z)$, we find that
\begin{equation}
{dE\over dA dE_\nu} = {(1+z)^2\over 4\pi d_L^2} \; ({dE\over dE_\nu^{\rm em}})\;.
\label{dEdAde}
\end{equation}
The differential event rate observed from bursting sources with
comoving density $\dot n$ is $d\dot N = (1+z)^{-3}$$ \dot n c d_L^2
$$d\Omega $ $\times|dt_{\rm em}/dz | dz$ \citep{wei72,der92}. For a
Friedmann-Robertson-Walker universe, $|dt_{\rm em}/dz |^{-1}$ $=$ $
H_0 (1+z) \sqrt{(1+\Omega_m z)(1+z)^2 - \Omega_\Lambda(2z+z^2)}$
\citep{tot99}. The diffuse flux of neutrinos produced by the
superposition of GRBs throughout the universe is therefore
\begin{equation}
{dE\over dA dE_\nu dt d\Omega} = {c\over 4\pi H_0}\;\int_0^\infty dz
\int_1^\infty d\Gamma_0\int_0^\infty dE_0\;{\dot n_{\rm GRB}
(\Gamma_0, E_0;z)\cdot [dE(\Gamma_0,E_0)/dE_\nu^{\rm em}]\over (1+z)
\sqrt{(1+\Omega_m z)(1+z)^2 - \Omega_\Lambda(2z+z^2)} }\;.
\label{dEdAdedtdO}
\end{equation}

Calculations of the diffuse neutrino background, using equation
(\ref{ndot}) with standard parameter sets (A) and (B) in equation
(\ref{dEdAdedtdO}), are shown in Fig.\ 6. As before, we use a
cosmology with $(\Omega_0, \Omega_{\Lambda}) = (0.3, 0.7)$ and $h =
0.65$. The relevant units conversion is $c/(4\pi H_0$-Gpc$^{3}$-yr) $=
1.22\times 10^{-63}$ cm$^{-2}$ s$^{-1}$. The solid curve is for
parameter set (A), and the dotted curve is for parameter set (B). The
calculations are at the level of $E_\nu\cdot \Phi_\nu(E_\nu) \sim
10^{-18}$ cm$^{-2}$ s$^{-1}$ sr$^{-1}$ at $E_\nu \ll 10^{18}$ eV. Thus
the estimate of the diffuse neutrino background (\ref{dNdlnE}) is in
accord with these results. Parameter set (A) produces a more luminous
neutrino flux at lower energies because there are greater number of
high energy soft photons, due to the smaller magnetic field used in
this parameter set.  It is not clear in this representation, but there
is approximately equal energy fluxes in parameter sets (A) and (B),
but most of the energy for set (B) is carried by neutrinos with
energies between $10^{18}$ and $10^{19}$ eV.

\subsection{GRB Neutrino Event Rate}

The number of neutrino events that would be detected per year by a
muon detector with an effective area of $A$(km$^2$) due to
upward-going neutrinos is
\begin{equation}
N_{\rm events/yr} \cong 3.16\times 10^7 \cdot 10^{10} A\cdot 2\pi \int
_0^\infty {dE_\nu \over E_\nu} \cdot {dE\over dA dE_\nu dt
d\Omega}\cdot P_{\nu\rightarrow \mu}(E_\nu)\;.
\label{Nevents/yr}
\end{equation}
In this expression, $P_{\nu\rightarrow \mu}(E_\nu)$ is the probability
that a neutrino with energy $E_\nu$, on a trajetory passing through a
detector, produces a muon above threshold. From the work of
\citet{gg87} and \citet{ls91} as summarized in \citet{ghs95}, we use
the following approximation to calculate neutrino event rates:
\begin{equation}
P_{\nu\rightarrow \mu}(E_\nu) \approx \cases{5.2\times
 10^{-33}[E_\nu({\rm eV})]^{2.2} \; ,& for $10^9 \leq E_\nu({\rm eV})
 < 10^{12}$ \cr 3.3\times 10^{-16}[E_\nu({\rm eV})]^{0.8} \; ,& for
 $10^{12} \leq E_\nu({\rm eV}) < 1.2\times 10^{15}$ \cr 1.1\times
 10^{-11}[E_\nu({\rm eV})]^{0.5} \; ,& for $1.2\times 10^{15} \leq
 E_\nu({\rm eV})$ \cr }
\label{Pnumu}
\end{equation}
(see also \citet{dl01}).

The number of events detected from a GRB at redshift $z$ is
\begin{equation}
N_{\rm events} \cong {(1+z)^2 10^{10} A\over 4\pi d_L^2} \int
_0^\infty {d E_\nu \over E_\nu}\; ({dE\over dE_\nu })\;
P_{\nu\rightarrow \mu}(E_\nu)\;.
\label{Nevents}
\end{equation}
Here, in contrast to equation (\ref{dEdAdedtdO}), it is necessary to
evaluate $d_L$ explicitly. It is given by
\begin{equation}
d_{L} = {(1+z)c \over H_0} \int_0^z {dz\pr \over \sqrt{(1+\Omega_m
z\pr)(1+z\pr)^2 - \Omega_\Lambda(2 z\pr + z^{\prime 2})}}\;.
\label{dL}
\end{equation}
The size distribution of neutrino events can be obtained by evaluating
the quantity
\begin{equation}
\dot N_\nu (> N_{\rm events}) = {4\pi c\over H_0}\;\int_0^\infty dz \int_1^\infty d\Gamma_0\int_0^\infty dE_0\;{d_L^2 \;\dot n_{\rm GRB} (\Gamma_0, E_0; z)\over (1+z)^3 \sqrt{(1+\Omega_m z)(1+z)^2 - \Omega_\Lambda(2z+z^2)} }\;,
\label{N>Nevents}
\end{equation}
where a contribution to the integral occurs only if the number of
neutrino events, calculated through equation (\ref{Nevents}), exceeds
$N_{\rm events}$.

This calculation is displayed in Fig.\ 7. Neutrino detectors at
energies $\gg 10^{12}$ eV are more sensitive to neutrino number flux
rather than energy flux, so these neutrino spectra do not regrettably
yield large numbers of neutrino events per year. Very weak neutrino
fluxes are predicted in the external shock model of GRBs {\ebf for
smooth profile GRBs, and we predict that no neutrinos will be detected
in coincidence with such GRBs. Detection by km$^2$ detectors of
multiple neutrino events from smooth-profile GRBs would probably rule
out the external shock model. Obversely, the lack of detection of
neutrino events from GRBs with smooth profiles
 is fully consistent with the underlying
assumptions of this study.  The predicted neutrino signal from
highly variable GRBs will require more study. Detectable neutrino emission
with km$^3$  neutrino telescopes is predicted within an internal shell model
\citep{wb97}, though it is not clear if these predictions also apply 
to smooth-profile GRBs.}

\section{Radiation Halos from Neutrons Produced by GRBs}

After production, neutrons with Lorentz factor $\gamma_n =
10^{10}\gamma_{10}$ will travel a characteristic distance $\lambda_n
\cong c\gamma_n t_n \cong 100 \gamma_{10}$ kpc before their numbers
are depleted by $\beta$-decay. Figs.\ 2a and 2b show that for strong
GRBs, $\gtrsim 1\%$ of the explosion energy is carried by neutrons
with $E\sim 10^{18}$-$10^{20}$ eV, or $0.1 \lesssim \gamma_{10}
\lesssim 10$. Thus a halo of neutron-decay electrons, protons, and
neutrinos will be formed around the site of a GRB that extends over a
size scale of $\sim 10$ kpc - 1 Mpc. The characteristic neutron-decay
lifetime is $t_n\gamma_n \sim 3\times 10^5 \gamma_{10}$ yrs. If the
$\beta$-decay electrons radiate on a timescale that is much shorter
than the neutron-decay timescale, then the peak power of a single
energetic GRB explosion in the extended nonthermal radiation halo
could reach
\begin{equation}
{dE_{\rm halo}\over dt} \simeq 0.01{\cal F}\;{m_e\over m_p} {10^{54}
E_{54}\over \gamma_n t_n}\sim 10^{36}{\cal F}\;{E_{54}\over
\gamma_{10}}\; {\rm~ergs~s}^{-1}\; , \label{dEhalodt}
\end{equation}
where ${\cal F}\cong 0.1$ is a temporal correction factor. This
assumes a neutron-production efficiency of 1\%, as applies to GRBs
with $E_0 \gtrsim 2\times 10^{53}$ ergs (see Fig.\ 3). The power in
the outflowing neutron-decay protons is $\sim m_p/m_e$ larger than
that given by equation (\ref{dEhalodt}), but will only be detected if
the protons can radiate this energy or deposit it in a galaxy halo to
be radiated through other processes.  Given the rarity of powerful GRB
events, it is unlikely that the nonthermal halo from neutron-decay
electrons surrounding a galaxy will consist of the superposition from
several GRBs at the frequency of greatest luminosity, but will instead
be formed by a single event. If GRBs are stongly beamed, however, the
radiation halos will consist of a superposition of emissions from many
GRBs.  In Section 5.1 we derive the radiation halo from a single
powerful GRB ``neutron bomb." Highly beamed GRBs will produce halos
that are time averages of the emission spectra formed by a single GRB,
weighted by the energy of events. Section 5.2 outlines energy
deposition into the halo from neutron-decay protons. We also note that
this process will make a very weak, long-lived afterglow of very
energetic neutron-decay neutrinos. Section 5.3 describes
multiwavelength prospects for detecting these halos.

\subsection{Radiation Halo from Neutron $\beta$-Decay Electrons}

The neutron flux (neutrons cm$^{-2}$ s$^{-1}$) at location $x$ in the
case of a spherically symmetric explosion is simply
\begin{equation}
\Phi_n(\gamma_n,t_*;x) = {\dot N_n(\gamma_n;t_*-x/c)\over 4\pi x^2}\; 
\exp(-x/ c\gamma_n t_n)\;,
\label{Phi_n}
\end{equation}
where $t_*$ is the time measured in the rest frame of GRB source and
$\dot N(\gamma_n;t_*)d\gamma_n$ is the differential number of neutrons
produced at time $t_*$ with Lorentz factors in the range between
$\gamma_n$ and $\gamma_n + d\gamma_n$.  The $\beta$-decay electron and
antineutrino each receive on average $\sim $ 0.6 MeV from the decay
(the neutron-proton mass difference is $\approx 1.3$ MeV). It is
sufficiently accurate for the purposes here to let the proton and
$\beta$-decay electron each receive the same Lorentz factor as the
neutron originally had. Thus the differential emissivity of either
neutron-decay protons or electrons is simply
\begin{equation}
\dot n(\gamma,t_*;x) = x^{-2}\;{\partial [x^2\Phi_n(\gamma;x,t_*)]\over
 \partial x}
\label{ndot_e}
\end{equation}
(see also \citet{gk90,ck95}).

If we consider only the neutrons with $\gamma_n \gg 10^4$ which decay
on timescales $\gg 10^7$ s, then the GRB explosion and afterglow can
be approximated as a $\delta$-function in time. If we also approximate
the neutron production spectrum as a single power law, then the
neutron source spectrum can be represented by
\begin{equation}
\dot N(\gamma_n,t_*) = K_n \gamma_n^{-q} \delta(t_* - \bar t_*) 
H[\gamma_n; 1, \gamma_{\rm n, max}]\; ,
\label{Ndot_n}
\end{equation}
where $q$ is the spectral index of the neutron number spectrum, and
$\gamma_{\rm n, max} \sim 10^{10}$ is the maximum neutron Lorentz
factor. Normalizing this spectrum to the total energy $E_n$ in
neutrons, we have
\begin{equation}
K_n = {E_n\over m_n c^2}\;({2-q\over \gamma_{\rm n, max}^{2-q} - 1})\;.
\label{K_n}
\end{equation}
Because $q \simeq 1$, it is irrelevant whether the minimum value of
$\gamma_n$ is 1, as used here, or $10^4$. Without loss of generality,
we let the explosion time $\bar t_* = 0$. Substituting equations
(\ref{Ndot_n}) and (\ref{Phi_n}) into equation (\ref{ndot_e}) gives
\begin{equation}
\dot n(\gamma,t_*;x) = {K_n\over c t_n}\; {\gamma^{-(q+1)}\over 4\pi
x^2}\;H[\gamma;1,\gamma_{\rm n, max}] \exp(-x/c\gamma
t_n)\delta(t_*-x/c)\;.
\label{ndot_e1}
\end{equation}

We now consider the radiation signature of the neutron-decay
electrons. Subsequent transport of the electrons can be neglected if
an on-the-spot approximation is valid, which holds if the electron
Larmor radius is much less than $\lambda_n$. This requires that the
halo magnetic field $B \gg m_e c/(e t_n) \cong 6\times 10^{-11}$
G. Faraday rotation measures of galaxy clusters and inferences from
synchrotron radio halos indicate that cluster fields are $\sim \mu$G
(see, e.g., the review by \citet{eil99}). It seems likely that galaxy
halos would also be this strong. The solution to the electron
continuity equation
\begin{equation}
{\partial n(\gamma;t)\over \partial t} + {\partial [\dot \gamma n(\gamma;t)]\over \partial \gamma} = \dot n(\gamma,t)\;,
\label{e_cont}
\end{equation}
is
\begin{equation}
n(\gamma;t) = |\dot\gamma |^{-1}\;\int_\gamma^\infty d\gamma\pr \;
\dot n(\gamma\pr,t\pr)\;,{\rm ~where~} t\pr = t -
\int_\gamma^{\gamma^\prime}\; {d\gamma^{\prime\prime}\over
|{d\gamma(\gamma^{\prime\prime})\over dt }|}\;.
\label{e_con_sol}
\end{equation}
Synchrotron radiation and Compton scattering of the cosmic microwave
background radiation will dominate the energy losses of the electrons,
although a more detailed treatment must treat Klein-Nishina effects
which become important for electrons with $\gamma \gtrsim (m_e
c^2/k_{\rm B}T_{bb}) \approx 2\times 10^9$, where $T_{bb} = 2.7$
K. These loss rates can be written as $-\dot\gamma = \nu_0\gamma^2$
(see discussion following equation (\ref{dotN(e)})).  Substituting
equation (\ref{ndot_e1}) into equation (\ref{e_con_sol}) and solving
gives the result
\begin{equation}
n(\gamma;x,t_*) = {K_n\over 4\pi x^2 c t_n \gamma^2}\;
\bar\gamma^{\prime (1-q)}\exp(-x/ct_n\bar\gamma\pr)\;,
\label{n_solution}
\end{equation}
where $\bar\gamma\pr \equiv [\gamma^{-1} - \nu_0(t_* - x/c)]^{-1}$ and 
\begin{equation}
{x\over c}\leq t_* \leq {x\over c} + \nu_0^{-1}(\gamma^{-1} -
\gamma_{n,{\rm max}}^{-1})\;.
\label{x/c}
\end{equation}

Thomson and synchrotron losses can be treated on equal footing by
rewriting equation (\ref{dotN(e)}) in terms of the photon emissivity
\begin{equation}
\dot n_{ph}(\e;x,t_* ) = {3\over 8} \nu_0\e^{\prime-1/2}\bar\e^{-3/2} n(\sqrt{{\e\over\bar\e}};x,t_*)\;.
\label{nphe}
\end{equation}
The quantity $\bar\e = \e_H = B/B_{\rm cr}$ for synchrotron emission,
and $\bar\e = \e_i$ for Thomson scattering, where $\e_i$ is the
characteristic dimensionless photon energy of the radiation field.
The frequency $\nu_0 = 4c\sigma_{\rm T}u_i/3$, where the dimensionless
field energy density $u_i = u_B$ for synchrotron emission and $u_i =
\int_0^\infty d\e_i \e_i n_{{\rm soft}}(\e_i)$ for Thomson scattering,
where $n_{{\rm soft}}(\e_i)$ is the spectral density of the soft
photon field. These equations are valid when $\gamma \bar\e \ll
1$. The restriction to the classical synchrotron regime always holds
in this system, but the restriction to the Thomson regime may not
apply, as already noted.

It is elementary to substitute equation (\ref{n_solution}) into
equation (\ref{nphe}) to obtain the photon emissivity $\dot
n_{ph}(\e;x,t_* )$ at location $x$ and time $t_*$. The spectrum
observed at time $t$ requires an integration over volume. Because the
neutrons are flowing out at speeds very close to the speed of light,
the expression $t = t_* + x(1-\mu)/c$ accurately relates the explosion
frame time and the observer time. Taking this relationship into
account finally gives the synchrotron or Thomson spectrum observed at
time $t$ after a GRB explosion. It is
\begin{eqnarray}
\nu L_\nu(t){\rm~(ergs~s}^{-1}) = m_ec^2 \e^2 \int dV \;\dot n_{ph}[\e;x,t_*(t)] =  {3 K_n \nu_0 m_ec^2\over 16 c t_n}\;\sqrt{{\e\over\bar\e}}\nonumber\\\times\;\int_{-1}^1 d\mu \int_{\max\{ 0,{c\over (2-\mu)}[t - \nu_0^{-1}(\sqrt{{\bar\e\over\e}}-\gamma_{n,{\rm max}}^{-1}) ]\}}^{ct/(2-\mu)} dx\; [\bar\gamma(t)]^{1-q}\; \exp[-x/ct_n\bar\gamma(t)]\;,
\label{nuLnu}
\end{eqnarray}
where
\begin{equation}
\bar\gamma(t) \equiv \{\sqrt{{\bar\e\over\e}} - \nu_0[t - {x\over c}(2-\mu)]\}^{-1}\;.
\label{bargamma}
\end{equation}
The flux density $S(\nu)$(Jy) $= 10^{23}(\nu L_\nu)/(4\pi d_L^2\nu)$ where $\nu$ is the observing frequency.

Fig.\ 8 shows calculations of the synchrotron and Thomson spectra
emitted by neutron $\beta$-decay electrons using equation
(\ref{nuLnu}).  Here it assumed that $10^{52}$ ergs in neutrons are
emitted with a spectrum $q = 1$ up to a maximum Lorentz factor
$\gamma_{n,{\rm max}}=10^9$ in Fig.\ 8a, and up to $\gamma_{n,{\rm
max}}=10^{11}$ in Fig.\ 8b. In both calculations, we use a magnetic
field $B = 1~\mu$G and approximate the cosmic microwave background
radiation as a $\delta$-function soft photon source with dimensionless
photon energy $\bar \e = 4.6\times 10^{-10}$ and energy density
$4.1\times 10^{-13}$ ergs cm$^{-3}$. Note that although the
synchrotron and Thomson spectra are plotted in the same graph, they
are independently calculated.

The peak luminosities reach $\sim 10^{36}$ ergs s$^{-1}$ in Fig.\ 8a
and $\sim 10^{34}$ ergs s$^{-1}$ in Fig.\ 8b. The discrepancy with
equation (\ref{dEhalodt}) implies that ${\cal F} \simeq 0.1$. This
value is understood when one considers temporal smearing due to the
finite energy-loss timescale, light travel-time effects and, most
importantly, the contribution of late time ($t \gg \gamma_{n,{\rm
max}}t_n$) radiation. The bandwidth correction factor should also be
considered. For a 1 $\mu$G field, the energy-loss timescale is $\sim
7.7\times 10^{20}\gamma^{-1}$ s, which for $\gamma_{n,{\rm max}} =
10^9$ is comparable to the $10^{12}$ s neutron decay timescale. The
temporal smearing due to light travel-time effects arising from
emission produced on the far side of the explosion produces the
high-energy features in the spectra observed at late times,
particularly in Fig.\ 2b. In synchrotron and Thomson processes, the
integrated luminosity decays $\propto t^{-1}$ at late times when most
of the energy is injected in the form of high-energy electrons, as is
the case here. Thus there is comparable energy radiated per decade of
time at late times. A value of ${\cal F} \sim 0.1$ in equation
(\ref{dEhalodt}) is therefore reasonable.

A notable feature in Fig.\ 8 is the appearance of sharp emission peaks
at late times. These are the pileups that appear when electrons are
injected with number indices harder than $2$ and lose energy through
synchrotron and Compton processes \citep{pin80}. The synchrotron
pileup features might be considerably broadened by magnetic-field
gradients in the halos of galaxies. The situation regarding the pileup
features in the Thomson peaks at ultra-high $\gamma$-ray energies is
more complicated and will require further study. Besides the
Klein-Nishina effects that are not considered here and are crucially
important in Fig.\ 2b, $\gamma$ rays with energies $\gtrsim 100$ TeV
$\sim 10^{28}$ Hz will materialize into e$^+$-e$^-$ pairs through
$\gamma$-$\gamma$ interactions with the cosmic microwave background
radiation to form a pair halo surrounding the galaxy
\citep{aha94}. Photons with energies $\gtrsim 10$ TeV will not be
observed due to pair-production attenuation on the diffuse infrared
radiation field. The energy processed by this electromagnetic cascade
will be transferred, in most cases, from the Thomson to the
synchrotron components (consider, however, \citet{km92}). Given
detailed modeling and sensitive observations, the relative powers in
the X-ray/soft $\gamma$-ray synchrotron component and the high-energy
$\gamma$-component could, in principle, be used to infer the halo
magnetic field. 

Effects of different magnetic field geometries and spatial 
variations of neutron injection due to beaming in GRBs should
be considered in future work. Fig.\
8 represents the simplest field geometry, but is representative
of the integrated emission spectrum from radiation halos due to 
UHECR production in GRBs.

\subsection{Radiation Halos from Neutron-Decay Protons}

The neutron-decay protons carry three orders of magnitude more energy
than the neutron-decay electrons, but this energy is also more
difficult to extract. The proton Larmor radius is $r_{\rm L} \cong
10\gamma_{10}$ B$^{-1}(\mu$G) kpc, so that much of the energy will be
carried directly into intergalactic space when $\gamma_{n,{\rm max}}
\cong 10^{11}$, even in the optimistic case of an extended ($\gtrsim
100$ kpc) galaxy plasma halo with a mean magnetic field of $\sim 1~
\mu$G. Such neutrons, together with the ultra-high energy protons and
ions that diffusively escape from the blast wave, are of course
postulated here to constitute the UHECRs. The Larmor timescale $t_{\rm
L} = r_{\rm L}/c \approx 10^{12}\gamma_{10}$B$^{-1}(\mu$G) s, so that
a $10^{19}$ eV proton might random walk for $\sim 10^{14}$ s before
diffusively escaping from a 1 $\mu$G, 100 kpc halo into intergalactic
space. The timescale for energy loss through secondary production is
$(n_{\rm halo}\sigma_{pp}c)^{-1} \cong 10^{15}/n_{\rm
halo}$(cm$^{-3}$) s, where the mean halo particle density $n_{\rm
halo}\ll 1$ cm$^{-3}$. Even though secondary production is very
inefficient, it could however compete with the energy deposition by
neutron $\beta$-decay electrons if $n_{\rm halo} \gtrsim 10^{-2}$
cm$^{-3}$, $\gamma_{n,{\rm max}} \lesssim 10^{10}$, and $B \gtrsim
1~\mu$G. Although the distinctive signature of secondary production is
the $\pi^0\rightarrow 2\gamma$ decay bump at $\sim 70$ MeV, it would
be severely broadened due to the large Lorentz factors involved and
would probably either not be detectable, or would form a low plateau
to the diffuse galactic and intergalactic $\gamma$ radiation fields.

Streaming instabilities excited by the outflowing neutron-decay
protons could convert a large fraction of the available energy into
long wavelength magnetohydrodynamic turbulence. Subsequent cascades of
the turbulence energy to  shorter wavelengths could
accelerate electrons through gyroresonant interactions  with
whistler and Alfven waves. Such processes have been invoked to
explain the formation of diffuse radio halos in rich clusters (for a
recent review, see articles in the collection edited by
\citet{bfs99}). It would be difficult, however,
 to distinguish neutron-decay halo nonthermal radio emission from
electrons accelerated by cluster merger shocks \citep{lw00}. Searches
for neutron-decay halos from field galaxies would therefore be more
definitive than searching for such halos around galaxies within or
near the peripheries of a galaxy cluster.

\subsection{Prospects for Detecting GRB Neutron-Decay Radiation Halos}

Three neutron-decay radiation halos are distinguished.  In the {\it
Type-$\beta$~halo}, the power of the halo radiation field comes from
$\beta$-decay electrons.  The previous section outlined in sufficient
detail the principal radiative properties of a $\beta$ halo. The most
important uncertainty, besides the ever-present question of GRB source
collimation, is the ratio of the magnetic-field energy density in the
halo to that in the cosmic background radiation. The $\beta$-decay
electrons in a {\it synchrotron $\beta$ halo} place most of the
radiated power in the synchrotron component. In contrast, microwave or
ambient photons are Compton-scattered to ultra-high $\gamma$-ray
energies to precipitate a pair shower in a {\it Compton $\beta$ halo}.

In the {\it Type-p} (for proton) {\it halo}, the power of the halo
radiation field comes from $\beta$-decay protons.  A {\it p}-halo can
be much brighter than a $\beta$ halo, because it has a factor $\sim
2000$ more energy available, but the extraction and subsequent
reradiation of this energy is far less easily quantified than for the
$\beta$ halo.  Depending on the radiation transfer and environmental
effects there are, as for the $\beta$ halos, {\it synchrotron p-halos}
and {\it Compton p-halos}.

The third type of halo is the {\it Type-$\nu$} (for neutrino)
halo. The instantaneous neutrino energy spectra received at different
times after the GRB can be obtained by following the approach of
Section 5.1. The detection of a $\nu$ halo is not technically feasible
at present.

\subsubsection{Statistics of Neutron-Decay Halos}

Our starting point was the differential source rate density, equation
(\ref{ndot}). There we noted that in the power-law approximations for
the $E_0$- and $\Gamma_0$-dependences of the differential rate
density, one-half of the energy generated by the sources of GRBs comes
from cosmic sources with apparent isotropic energy releases $>
2.4\times 10^{53}$ ergs. Fig.\ 3 shows that the neutron production
efficiency increases monotonically with energy; therefore most of the
neutron energy comes from GRBs with $E_0\gtrsim 3\times 10^{53}$
ergs. These very energetic GRBs are, of course, much less
frequent. Our study of GRB statistics \citep{bd00} shows that in the
universe on small ($z\lesssim 0.1$) scale, the rate density of GRB
sources is $\sim 3.6[E_{52}^{-0.52}-(100)^{-0.52}]$ Gpc$^{-3}$
yr$^{-1}$. Thus on average there are 0.43 GRB-type explosions per
Gpc$^{3}$ per year with energy $\gtrsim 2\times 10^{53}$ ergs.

There are, speaking crudely, $\simeq \int_{L^*/2}^{2L^*} dL \;\Phi(L)
\simeq 0.52\Phi^*$ ($L^*$ galaxies)/Mpc$^{3}$, so that the density of
$L^*$ galaxies in the local universe is $n_{L^*} \simeq 2.3\times
10^{-3}$ Mpc$^{-3}$ (compare Section 2.1). If all the mass of galaxies
were wrapped up in $L^*$ galaxies, then each galaxy would see on
average a GRB-type explosion with energy $\gtrsim 2\times 10^{53}$
ergs every $\sim 5$ Myrs,  assuming that GRB explosions are 
uncollimated. Let $t_\nu$ represent the characteristic
FWHM duration when the emission at frequency $\nu$ from a
neutron-decay halo reaches its peak luminosity $L_0(\nu) =\nu
L_\nu$. If $t_\nu \ll 5$ Myr, then the fraction of $L^*$ galaxies
displaying emission at this level is $\sim t_\nu/5$ Myrs. If, on the
other hand, $t_\nu \gg 5$ Myrs, then the galaxy will exhibit a
superposition of the emissions from many GRB neutron-decay halos, with
the total halo brightness reaching $\sim L_0(\nu) (t_\nu /5$ Myr).
 If the beaming fraction is $\sim 1/500$ of the full sky \citep{fra01},
then GRBs will take place about once every $\sim 10,000$ years in an $L^*$ galaxy,
and will display neutron-decay halos at a
level corresponding to the average GRB power multiplied by 
the neutron $\beta$-decay production efficiency. Under these 
circumstances, the average bolometric power of an $L^*$ galaxy
 from $\beta$ halos is at a level of $2.5\times 10^{39}$ ergs s$^{-1} 
\times 4\times (2/1836)\times$ 1\% $\times k_{cl} \cong 10^{35}k_{cl}$
ergs s$^{-1}$, where the factor of 4 accounts for clean and dirty fireballs,
and the factor $k_{cl}> 1$ is a correction factor due to the enhancement
of neutrino production in a clumpy medium.

\subsubsection{$\beta$ Halos and $p$-Halos: Essential Features}

The essential features of a $\beta$ halo produced by a single
uncollimated GRB are given by the peak photon frequency $\nu_{pk}$,
the duration $t_{\rm dur}$ of peak luminosity $L_{pk}$, and radial
extent $r_h$ of the halo. For a synchrotron $\beta$ halo,
$\nu_{pk}\approx 3\times 10^{20}B(\mu{\rm G~})\gamma_{10}^2(1+z)^{-1}$
Hz, and $t_{\rm dur} \sim (1+z)\gamma_{n,{\rm max}}t_n \cong 3\times
10^5 \gamma_{10}$ yrs. Setting ${\cal F} \simeq 0.1$ in equation
(\ref{dEhalodt}), the peak luminosity $\sim
10^{35}E_{54}\gamma_{10}^{-1}$ ergs s$^{-1}$. The radial extent of the
halo is $r_h \sim 100 \gamma_{10}$ kpc.

A Compton $\beta$-halo will be formed if the mean halo magnetic field
$\langle B \rangle \ll 3 (1+z)^2$ $\mu$G. In this case, cosmic
microwave background photons are Thomson scattered to energies $\sim
\min ( 5\times 10^3,5\times 10^4 \gamma_{10})\gamma_{10}$ TeV. Many of
these photons will materialize into electron-positron pairs through
interactions with the cosmic diffuse background radiation field
\citep{gs67} to initiate an electromagnetic cascade that channels the
radiant power into lower energy $\gamma$ rays and into a radially
extended synchrotron component. The cascade ends when the photons
penetrate the optical depth of the universe to $\gamma\gamma$
attenuation. This quantity is not well known, but \citet{sdj98}
calculate that $\tau_{\gamma\gamma} \cong 0.5-1$ for $\sim$ TeV
photons from sources at $z = 0.075$ due to absorption by the diffuse
intergalactic infrared radiation field.

The $p$-halo will be brightest if the neutron-decay protons transfer
and radiate their energy on a timescale shorter than the
light-crossing time $\gamma_{n,{\rm max}}t_n$. Given the
model-dependent uncertainty of the emergent photon spectrum from a
$p$-halo, we approximate it with a $\nu L_\nu $ spectrum that has
constant value $L_p$ between $10^6$ Hz and $10^{26}$ Hz. The $\nu
L_\nu$ power radiated from the $p$-halo formed by a single strong GRB
is, in this crude approximation for the spectrum, therefore at best
\begin{equation}
\nu L_\nu ({\rm ergs~s}^{-1})\cong {2\times 10^{37}\over \ln(10^{20})}\;{E_{54}\over \gamma_{10}} \simeq 4\times 10^{35}\;{E_{54}\over \gamma_{10}} \;,\;{\rm ~for~} 10^6 \leq \nu({\rm Hz})< 10^{26}\;,
\label{vL_v}
\end{equation}
where we use ${\cal F} = 0.1$ (compare eq.[\ref{dEhalodt}]). If the
radiation is emitted in a narrow bandwidth, the $p$-halo could be 2-3
orders of magnitude brighter. Thus there is emission in all observable
wavebands at the level given by equation (\ref{vL_v}) during a period
of $\sim 10^{13} \gamma_{10}$ s. The first-generation emission is
distributed over a region of size $r_h \sim 100 \gamma_{10}$ kpc, but
the cascade radiation from the pair halo can occupy a much larger
volume.

\subsubsection{Halo Detection: Observational Issues}

GRBs were first detected with soft $\gamma$-ray instruments (see
\citet{dcb99}). To survey prospects for detecting neutron-decay halos,
we begin at soft $\gamma$-ray energies and move to lower frequencies,
returning at the end to the high-energy $\gamma$-ray domain.

\subsubsubsection{Soft $\gamma$-ray and X-ray Detection}

The sensitivity limit of a detector such as BATSE is $\sim 0.2\times
100$ keV cm$^{-2}$ s$^{-1} \sim 3\times 10^{-8}$ ergs cm$^{-2}$
s$^{-1}$ for a $\sim 10$-100 s observation, and that of OSSE is $\sim
10^{-11}$ ergs cm$^{-2}$ s$^{-1}$ for a two-week observation. Even
with many orders of magnitudes improvement in sensitivity as provided
by pointed instruments or position-sensitive technology, the detection
of a neutron-decay halo is not easy with available X-ray detectors,
much less $\gamma$-ray detectors. We estimate the limiting detection
distance $d_{lim}$ for a telescope with $\nu F_\nu$ sensitivity $S =
10^{-15}S_{-15}$ ergs s$^{-1}$ over its nominal point-source observing
time and bandpass. The peak luminosity of a neutron-decay $\beta$ halo
is given by equation (\ref{dEhalodt}), so that $d_{lim} = \sqrt{L
/4\pi S} = (E_{54}/\gamma_{10}S_{-15})^{1/2}$ Mpc. This criterion
eliminates all $\gamma$-ray instruments and all but the best X-ray
detectors, such as {\it Chandra}\footnote{Unfortunately, the fact that
the neutron-decay halos are spread over a region greatly exceeding the
extent of the galaxy makes them more difficult to detect, because they
they will be harder to resolve from the diffuse background. The
strength of instruments such as {\it Chandra} is that it focuses all
photons from a point source onto one or a few pixels, so that the
back-ground is greatly reduced (M. B\"ottcher, private communication,
2000).} with $S_{-15} \sim 1$. Within a few Mpc, the Milky Way and M31
are the closest $L^*$-type galaxies.  The rough odds are that a
detectable halo could be observed from $\sim 8\gamma_{10}$\% of nearby
$L^*$ galaxies if GRBs releases their energy isotropically, leaving
only the two $L^*$ galaxy candidates if $d_{lim} \sim 1$ Mpc. A
neutron-decay halo from M31 would cover a half-angle extent of
$\theta_{1/2} \sim 8\gamma_{10}^\circ$.  Even for galaxies at $\sim
10$ Mpc, the challenge of background subtraction to reveal a cleaned
X-ray image is severe, but would be assisted with model templates. It
is worth recalling that beaming can increase the chance odds of
sighting a galaxy that harbors a neutron-decay halo, but the halo
itself would be at a proportionately smaller flux.

\subsubsubsection{Optical Detection}

In the spherical region that surrounds us to a depth of 100 Mpc, or
within $z \cong 0.022$ for $h = 0.65$, there are, according to the
earlier statistic, $\sim 10^4 L^*$ galaxies. At $100d_{100~\rm Mpc}$
Mpc, the half-angular extent of a neutron halo is $\sim 3.4
(\gamma_{10}/d_{100~\rm Mpc}$) arc minutes. At a sampling distance
between $\sim 10$ and 100 Mpc, there are therefore abundant candidates
with galaxy disk sizes of $\sim (2$-$20) \gamma_{10}$ arc seconds and
a halo angular extent appropriate for an optical CCD. In the
following, we sketch some basic considerations that enter optical halo
detection.

\begin{itemize}
\item The predicted $\beta$-halo optical luminosity is $\approx 10^{35}$
ergs s$^{-1}$, but a neutron-decay halo could be as bright as
$10^{37}-10^{38}$ ergs s$^{-1}$ if the parameters in the model are
most optimistically tuned in favor of detecting a synchrotron
$p$-halo. Compared to the typical $L^*$ galaxy optical luminosity of
$\sim 2\times 10^{11} L_\odot \sim 6\times 10^{44}$ ergs s$^{-1}$, the
halo luminosity provides a very weak flux. On the other hand, the
emission is spread over a region that is far outside the optical
radius of the galaxy.
 
\item The relative brightnesses of the central source and halo is $\sim 6$-9
orders of magnitude, or $\sim 15$-22 magnitudes. If the limiting
magnitude is $m_V = 25$ for a good ground-based telescope, then a halo
could only be seen for galaxies with $m_V<10$. Noting that $m_V
\approx 5$ for M31 implies that the limiting distance to detect a
neutron-decay halo is $\sim 10$ Mpc for 2-3 meter class ground-based
telescopes. In this case, the advantage of a halo that fills the CCD
is lost, and the sensitivity of most large-aperture telescopes may not
be good enough to detect the halo above background sources, the sky
brightness and detector noise.

\item The limiting magnitude of the {\it Hubble Space Telescope} for point
sources is $m_V \cong 30$. We could then potentially see neutron-decay
halos to $d_{lim}\approx 100$ Mpc. Even at 100 Mpc, the halo subtends
much of the CCD and the central bright source emission would have to
be subtracted. For comparison, when subtracting central source flux
from galaxy-disk flux in HST images, the contrast between the optical
power of the AGN and that of the extended disk might have been $\sim
10^2$-$10^3$ (this estimate is made by comparing optical luminosities
of typical galaxies and QSOs, though the ratio could be even larger in
studies where blazar light is
subtracted from the host galaxy.) This still does not compare with the
extreme contrast between the surface brightnesses of the optical disk
of a galaxy and the surrounding diffuse halo.  It seems that a
blocking crystal for ground and space-based optical telescopes could
be developed to eliminate the intense flux of the much brighter galaxy
disks. The instruments on the {\it Solar and Heliospheric Observatory}
probably achieve the greatest technical feat to detect faint objects
in the field of a bright source ($|m_\odot - m_{\rm stars}|$ or
$|m_\odot - m_{\rm comets}|$ implies $>30$ orders of magnitude
blockage of the Sun), but the detection of halos around distant
galaxies will clearly pose different problems.

\item Optical central-source luminosity is suppressed in certain classes of
galaxies, most remarkably, those that are likely to harbor active star
formation.  Here we are thinking of edge-on starbursts (M82 or NGC
253-types) and dusty spirals, tidally-disturbed systems (e.g. Mrk 421
and its satellite galaxies \citep{gor00}), and infrared luminous
mergers such as Arp 220, Mrk 273, and other non-quasar members in
Arp's atlas of peculiar galaxies. The search for neutron-decay halos
also introduces a new avenue to examine the relative power of ULIGs
(ultra-luminous IR galaxies) in stellar formation and black hole
activity. The {\it Infrared Space Observatory} results on PHA/infrared
line tracers of the starburst and AGN activity \citep{lutz96} showed a
separation of different galaxy types in a way that can be tested,
because the strength of the neutron-decay halo is proportional to
star-formation activity. The magnitude of either a $\beta$ halo or a
$p$-halo is, in this picture, directly proportional to the rate at
which high-mass stars are formed, and is a basic assumption of the GRB
statistics treatment of \citet{bd00}.

\item AGNs and quasars introduce greater background subtraction problems, and
pose the added difficulty of an interfering zodiacal light from
high-latitude dust or gas that scatters the optical emission from the
galaxy's AGN and stellar radiation fields. The existence of rather
dense high latitude ($\sim 10$-100 kpc) dust seems quite likely in an
AGN environment due, for example, to tidal activity, disk winds, AGN
radiation pressure on surrounding gas, and gravitational effects from
distorted dark-matter halos and galaxy bars. Diffuse scattering plasma
might also, unfortunately, be found in ULIGs for the same reasons.

\end{itemize}

Technical considerations for detecting neutron-decay halos with
optical telescopes will require an examination beyond the scope of
this paper. A central insight is that even though point-source
(``light-bucket") fluxes dim with source distance according to $\phi
\propto d^{-2}$ in the Newtonian limit, the surface brightness of an
optically thin source is constant (again, in the Newtonian
limit). This effect has fundamental implications for observations
against a source-confused and sky-limited background.

\subsubsubsection{Radio Detection}

The radio regime has the best $\nu F_\nu$ sensitivity, with $S_r\sim
10^9$ Hz$\times 0.1$ mJy $\sim 10^{-18}$ ergs cm$^{-2}$ s$^{-1}$,
combined with excellent angular resolution. Improved resolution (VLBI)
must trade off with better limiting sensitivity (VLA), both of which
additionally depend on observing frequency. For an optimistic radio
halo power of $\sim 10^{35}$ ergs s$^{-1}$, the limiting sampling
distance is only $\sim 30(E_{54}/\gamma_{10})^{1/2}$ Mpc. Two effects
determine the actual radio luminosity of a $\beta$ halo.  The first,
as seen in Fig.\ 8, is that the radio luminosity is $\sim 10^8$ times
dimmer than the peak nonthermal synchrotron power from a synchrotron
$\beta$ halo for our standard halo with a randomly oriented $\sim 1$
$\mu$G mean magnetic field. This reduction is partially offset by the
fact that the synchrotron decay timescale from the radio-emitting
electrons and positrons is larger by a factor of $\sim
10^{16.5}/1.2\times 10^{14} \approx 260$ than the burst timescale (see
Fig.\ 8) in the case of a 1 $\mu$G halo. The net result is to reduce
the sampling distance so that detecting the radio emission from a halo
turns out again to be difficult. In the event of a very weak ($\langle
B \rangle \ll 0.1~\mu$G) halo magnetic field,  the reprocessing of the
Compton power into the synchrotron component could however improve
radio detectability by moving $\nu_{pk}$ to lower frequencies (see
Section 5.3.2). To take advantage of the good radio resolution (the
size of the radio halo for sources at $z \sim 1$ ($cz/H_0 = 4600$ Mpc)
is on the order of a few arc-seconds) would require detection of
$(\sim 10^{30}{\rm~ergs~s}^{-1}\times 10^{23}/ 10^{57}$ cm$^2$-$10^9$
Hz$)\lesssim 10^{-7} \mu$Jy fields spread over a surface area of this
extent. This is not yet feasible. The tradeoff between angular extent and
sensitivity will be helped if radio techniques can yield cleaned images
that are $\sim O(^\circ)$ in extent. Wherever the radio range proves
to have the greatest capability (probably for galaxies at a few tens
of Mpc), structure in the neutron-decay radio halo should be carefully
sought. The advantage here is that to test the UHECR/GRB hypothesis,
{\it every} $L^*$ galaxy should have a diffuse neutron-decay radio
halo,  whether or not GRB outflows are collimated.

The very low-frequency ($\lesssim 100$ MHz) emission from
neutron-decay halos persists around all $L^*$ galaxies, and forms part
of the diffuse low-frequency radio background. Whether detection of
such halos is technically feasible with new-generation radio arrays
(e.g., the planned low-frequency array LOFAR) will require more study.

\subsubsubsection{High Energy $\gamma$-ray Detection}

Returning now to the ultra-high energy gamma-ray regime, neither GLAST
nor the ground-based air and water Cherenkov telescopes operating or
in development (e.g., Whipple, Milagro, HESS, VERITAS) can be expected
to detect a Compton $\beta$ halo. Only under the most optimistic
conditions of a highly luminous Compton $p$-halo at $\sim 10^{38}$
ergs s$^{-1}$ is detection feasible. GLAST is $\sim 50$ times more
sensitive than EGRET, which had a limiting sensitivity of $\approx$
few$~\times 10^{-11}$ ergs cm$^{-2}$ s$^{-1}$, as does Whipple. This
gives a sampling distance of $\sim 300 E_{54}/\gamma_{10}$ kpc. With
the seven-fold increase in limiting distance for GLAST, and with the
improvement that will be achieved with the VERITAS array, there
remains a chance of detecting highly luminous Compton $p$-halos from
nearby galaxies.

In summary, the search for direct synchrotron, and both direct and
cascade $\gamma$ radiation from neutron-decay halos predicted by the
external shock model are at limits that challenge current radio,
optical, X-ray, and $\gamma$-ray detector technology. The predicted
weakness of the $\beta$-decay halos reflects the low neutron
production efficiency in this model. The stronger neutrino and neutron
production within the context of an internal shock scenario
\citep{wb00,dl01} would imply neutron decay halos $\sim 100$ times
brighter. However, because of the uncertainty in the brightness of a
$p$-halo, the detection of neutron-decay halos at the level of $\sim
10^{38}$ ergs s$^{-1}$ around $L^*$ galaxies would not discriminate
between internal and external shock models. Radiation halos detected
at the level of $\sim 10^{39}$-$10^{40}$ ergs s$^{-1}$ around $L^*$
galaxies would be inconsistent with an external shock scenario, but
would require stong damping of the very energy neutron-decay proton
energy in an internal shell model. The much stronger neutrino
production in the internal shock model will also provide a clear
discriminant between internal and external shock scenarios.

\section{Cosmic Ray Production by GRBs}

As summarized in the Introduction, observations indicate that GRBs are
associated with supernovae taking place in star-forming
galaxies. Statistical analyses within the external shock model show
that progenitor sources of GRBs inject a time- and space-averaged
power $\gtrsim 2.5\times 10^{39}$ ergs s$^{-1}$ into an $L^*$
galaxy. More general arguments also show that the power injected into
the Milky Way by fireball transient sources with relativstic outflows
is at the level of $\sim 10^{40\pm 1}$ ergs s$^{-1}$
\citep{der00b}. This is much smaller than the $\sim 10^{42}$ ergs
s$^{-1}$ thought to be injected into the Galaxy by SNe of all
types. It is also somewhat smaller than the galactic cosmic-ray
luminosity of $\sim 5\times 10^{40}$ ergs s$^{-1}$ that is estimated
to be required to power hadronic cosmic rays, depending on the assumed
efficiency for accelerating hadronic cosmic rays. But the global
cosmic-ray power estimate assumes that the locally observed cosmic-ray
energy density is typical throughout the galaxy, and that temporal
stochastic variations are not large. Both of these assumptions could
be wrong \citep{hun97,pe98}. The derived FT power is sufficiently
close to suggest that the progenitor sources of GRBs could power a
significant fraction of the hadronic cosmic rays in the Galaxy.

We additionally note several difficulties for the conventional view
that cosmic rays are accelerated by supernovae in the galaxy (see
\citet{der00a} for further detail): (i) Spectral signatures of the
hadronic cosmic-ray component associated with $\pi^0$ emission
features, which carries $\sim 30$-100 times as much energy as the
leptonic cosmic-ray component, have not been detected unambiguously in
the vicinity of SNRs \citep{esp96}.  (ii) The unidentified EGRET
sources have not been firmly associated with SNRs \citep{rbt99}, and
several candidate SNRs are more likely to be associated with pulsars
\citep{mir01}. (ii) TeV gamma rays are not detected from SNRs at the
level expected from hadronic acceleration in SNR shocks
\citep{buc98,aha01}. (iii) The measured spectrum of the diffuse
galactic $\gamma$-ray background is harder than predicted if the
locally measured cosmic-ray proton spectrum is typical of other places
in the Milky Way \citep{hun97}. (iv) The origin of cosmic rays at and
above the knee of the cosmic-ray spectum and the smooth transition at
the knee are difficult to explain with a SN shock model
\citep{lc83,der01}.

We therefore suggest that cosmic rays originate from the subclass of
SNe that are progenitors of GRBs.  (\citet{mu96} and \citet{dp99} have
also suggested that cosmic rays might be accelerated by the sources of
GRBs.) How frequent are these events compared to other types of SNe?
We can evaluate this rate from the statistical study of \citet{bd00},
or by modifying this study in view of the constant energy reservoir
result of \citet{fra01}.

The ``supernova unit" 
\begin{equation}
{\rm SNu}\; [{{\rm events} \over 10^{10} L_{\odot,B}{\rm -}10^2{\rm
~yr}}] \cong {1.3\times 10^{-5}\over h_{70}}\;\dot n({\rm
Gpc}^{-3}{\rm~yr^{-1}})\;
\label{SNu}
\end{equation}
is defined in terms of the number of events of a given type per
$10^{10}$ Solar luminosities in the blue band per century, recalling
from Section 2.1 that $J_{gal,B} \cong 7.6\times 10^{16}
h_{70}L_{\odot,B}$ Gpc$^{-3}$, where $h_{70} \equiv 0.7h$. For
reference, note also that 1 SNu $ \cong 7\times 10^{-5}$ GEM, where
the conversion factor to galactic events per Myr [GEM $= \#$ of events
$/{\rm (MW~galaxy}$ -$ 10^6$ yr); \citet{wij98}] uses a Milky Way
blue-band luminosity $L_{MW,B} \cong \pi (15{\rm~kpc})^2\cdot 20
L_\odot$ pc$^{-2} \cong 1.4\times 10^{10} L_{\odot,B}$
\citep{bm98,sw02}.

The local rate density of FTs calculated by \citet{bd00} when
$E_{52}^{min} = 10^{-4}$ in equation (\ref{dNdVdt}) is $\dot n_{bd} =
440$ Gpc$^{-3}$ yr$^{-1}$, implying a rate of $\sim 80$ GEM, or
$SNu(FT) \cong 6\times 10^{-3}$ in supernova units. In galaxies of
type Sbc-Sd, $SNu(II) \cong 0.7(\pm 0.35) h_{70}^2$ and $SNu(Ib/c)
\cong 0.14(\pm 0.07) h_{70}^2$ for Type II and Type Ib/c SNe,
respectively \citep{cap97,sw02}.  Hence the ratio $SNu(II)/SNu(FT)
\cong 120\pm 60$, and $SNu(Ib/c)/SNu(FT) \cong 24\pm 12$. The FT rate
density is, however, very sensitive to the number of low energy GRBs,
and so could be $\approx 30$ times less frequent if $E_{52}^{min}
\cong 0.1$.
   
If we accept the standard energy reservoir result of \citet{fra01},
then the rate of the brightest GRBs will be $\gtrsim 500$ times the
observed rate. For the FT rate of $\dot n_{GRB}(z = 0) \cong 3.6$
Gpc$^{-3}$ yr$^{-1}$, which is the minimum rate necessary to fit the
BATSE statistics within the external shock model, this implies a FT
within the Milky Way $\sim 4\times 10^{-4}$-$10^{-3}$ yr$^{-1}$,
depending on the $L^*$ density.\footnote{Two measures of the density
of galaxies like the Milky Way, which is assumed to be L$^*$-like, are
used in this paper. The $2.3\times 10^6$ Gpc$^{-3}$ figure refers to
the number density of galaxies with luminosities between $0.5 L^*$ and
$2L^*$ (Section 5.3.1). The method of energy-weighting that was
applied to the Schechter luminosity function in equation
(\ref{dE(L)dt}) implies an $L^*$ density of $4.6\times 10^6$
Gpc$^{-3}$.} Given the uncertainties associated with the rate and
beaming estimations, we can expect that a GRB or FT will occur about
once every $10^3$-$10^4$ yr in the Galaxy. Consequently $\sim 1$ in
every 10-100 SNe will display relativistic outflows including, in
$\sim 10$-50\% of the cases, a GRB. GRBs are probably associated with
the rarer Type Ib/c SNe that have lost their hydrogen envelopes. This
prediction can be tested with a sample of many dozens of SNe Ib/c,
where follow-on radio monitoring is used to identify relativistic
outflows \citep{kul98,wei00}. A small fraction of SNe in our Galaxy
should exhibit strong hadronic signatures associated with cosmic ray
production.

There are $\sim 200$-1000 black holes formed per Myr in the Milky Way
if a black hole is formed by every GRB and FT.  Over the $1.2\times
10^{10}$ yr age of our galaxy, $\sim 2\times 10^6$-$10^7$ black holes
are thus formed. Gravitational deflection of the black holes off other
stars and molecular clouds would increase the scale height of the
older black holes to values exhibited by the older K and M stellar
populations \citep{bs80}. A population of isolated black holes that
accrete matter from the ISM will thus be formed by GRBs
\citep{an99,der00b}. Some of the EGRET unidentified sources could
originate from accreting, isolated black holes.


\section{Summary and Conclusions}

We considered implications of the hypothesis that UHECRs are
accelerated by the sources of GRBs in this paper. Here are the main
points deduced from this study:

\begin{enumerate}

\item The statistical study of GRBs in the external shock model
\citep{bd00} gives the rate density and emissivity of GRBs, and the
number of GRBs with different Lorentz factor $\Gamma_0$ and total
apparent isotropic energy $E_0$.  The event rate is very uncertain for
weak GRBs, but most of the energy comes from the rare strong GRBs with
$E_0 \gtrsim 2\times 10^{53}$ ergs. If GRBs are highly beamed, as
suggested by \citet{fra01} and \citet{pk01}, then GRBs and fireball
transients with $E_0\gtrsim 10^{51}$ ergs could occur as frequently as
once every $10^3$-$10^4$ yrs in the Galaxy.

\item Formulae for the nonthermal synchrotron emission spectra produced by
an external shock were used to derive the comoving nonthermal
synchrotron photon spectra in the blast wave. This radiation provides
target photons for the very energetic nonthermal particles. Neutrons,
neutrinos, positrons, and pairs are formed as byproducts of $\gamma$-p
and $\gamma$-ion interactions.

\item Neutron and neutrino production spectra and light curves formed in
photomeson interactions are readily derived in the external shock
model. The neutrino flux from individual GRBs is far too weak to be
detected by km$^2$ neutrino detectors, because most of the energy is
carried by relatively few, very energetic neutrinos. GRBs might still
contribute the major fraction of the diffuse neutrino background for
neutrinos with energies $\gtrsim 10^{16}$ eV. The energy carried away
by neutrons from very energetic GRBs with $E_{54} \equiv
E_0/10^{54}{\rm~ergs} \gtrsim 0.2$ will exceed 1\% of the total
energy; reverse shock emission giving enhanced target photons could
make the neutron-production efficiency even larger, though SSC
processes might reduce it.

\item Galaxies with GRB activity will be surrounded by neutron-decay halos
formed by emissions from $\beta$-electrons and neutron-decay
protons. The halo size is $\sim 100\gamma_{10}$ kpc, where
$\gamma_{10}$ is the Lorentz factor of the neutrons that carry most of
the energy from the GRB. The shortest halo emission lifetime from a
single GRB is $\sim 3\times 10^5$ yrs. The peak luminosity of a
$\beta$ halo from a single  smooth-profile
GRB is $\sim 10^{35}E_{54}/\gamma_{10}$
ergs s$^{-1}$. Depending on the magnetic field strength in halos of
galaxies, the neutron $\beta$-decay electrons will produce a
nonthermal synchrotron $\beta$ halo with peak luminosities at
optical/X-ray/soft $\gamma$-ray energies, and a Compton $\beta$ halo
at very high (GeV-TeV) $\gamma$-ray energies when the high-energy
electrons Compton-scatter photons of the cosmic microwave radiation
and induce a cascade.  The $p$-halo formed by neutron-decay protons is
more difficult to quantify and could be much brighter than the
emission from a $\beta$ halo.

\item Because of sensitivity and imaging capabilities, prospects for
detecting neutron $\beta$-decay halo emission are best at optical and
radio frequencies. For optimistic model parameters, it might also be
technically feasible to detect these halos at X-ray and $\gamma$-ray
energies. The subtraction of the light from the bright central galaxy
is a major obstacle to halo detection at optical
frequencies. Approximately $8\gamma_{10}$\% of $L^*$ galaxies should
display a $\beta$ halo from a single GRB near the peak of its
luminosity output for unbeamed GRBs. If GRBs are highly beamed, then
essentially all $L^*$ galaxies will be surrounded by such halos, but
at a weaker average flux.  The average bolometric neutron-decay
$\beta$ halo emission
surrounding an $L^*$ galaxy is $\approx 10^{35}$ ergs s$^{-1}$
from smooth-profile fireball transients, but could be greater for
GRBs occurring in inhomogeneous surroundings.

\item The emissivity of the progenitor sources of GRBs in our Galaxy
potentially provides a large fraction of the luminosity required to
power the galactic cosmic rays. If cosmic rays are accelerated by the
supernovae that collapse to form a GRB, about 1 out of $\sim 20$-100
supernova remnants in the Galaxy should have harbored a GRB, so that
detection of a strong hadronic signature from a subset of supernovae
should support this model for the origin of cosmic rays.

\end{enumerate}

In addition to the search for GRB neutron-decay radiation halos around
galaxies, further progress on these problems will be achieved by
searching for neutrino emission from GRBs.  Detection of
high-energy neutrinos from smooth profile GRBs would probably rule out
the external shock model for the prompt phase of GRBs. Another
important study is to search for the sister classes of GRBs that do
not trigger GRB detectors. Present GRB telescopes have strong
triggering biases against dirty and clean fireballs transients
predicted by the external shock model that can be remedied with
appropriate slewing strategies and new detector designs
\citep{dcb99}. Most important is to identify hadronic emission from
SNRs through gamma-ray observations, and to determine if cosmic ray
production is typical of all SNRs or, as suggested here, only a small
subset of SNe that are associated with GRBs.

\acknowledgments

I thank T. K. Gaisser, M. B\"ottcher, K. Weiler, S. Woosley, M. Leising,
C. Kouveliotou, R. Berrington, K. Wood, T. Galama, and R. Schlickeiser
for comments, and the referees for their reports.
 The work of CD is supported by the Office of Naval
Research and the NASA Astrophysics Theory Program (DPR S-13756G).

\appendix

\section{Synchrotron Radiation Limit on Maximum Proton Energy}

We derive an expression for the maximum proton energy that results
from a competition between the acceleration energy-gain rate and the
synchrotron energy-loss rate. For the acceleration rate, we assume
that a particle cannot gain a large fraction of its energy at a rate
more rapid than the gyrofrequency \citep{gfr83,kir91,jag96,rm98}. This
condition follows from general considerations for either first- or
second-order acceleration. In either case, $\dot \pp \cong f_{\rm L}
(eB/2\pi m_p c)$, where $\pp$ is the particle momentum in the comoving
frame and $f_{\rm L}\lesssim 1$ is an unknown parameter. For
first-order Fermi acceleration in the Bohm diffusion approximation,
$f_{\rm L} \cong 2\pi \beta_+^2$, where $\beta_+ c$ is the upstream
speed. For second-order Fermi acceleration, $f_{\rm L} \cong
(8\pi/3)\beta_{\rm A}^2 (\delta B/B)^2$, where $\beta_{\rm A}c$ is the
Alfven speed of the scattering centers and $(\delta B/B)^2$ is the
fractional energy density of resonant waves \citep{der01}.

The synchrotron radiation limit is obtained by balancing the
acceleration rate with the synchrotron energy-loss rate given through
equation (\ref{t_psyn}). It easily follows that the maximum observable
proton Lorentz factor is
\begin{equation}
\gamma_{{\rm L}, max} = \Gamma \; ({m_p\over m_e})\; \sqrt{{3 f_{\rm
L} e\over \sigma_T B}} \cong 1.4\times 10^{11}\; {(f_{\rm L}
\Gamma)^{1/2}\over [e_B n({\rm cm}^{-3})]^{1/4}}\simeq {2.4\times
10^{12}\over 1 + (4\tau)^{3/16} }\; {(f_{\rm
L}\Gamma_{300})^{1/2}\over [e_B n({\rm cm}^{-3})]^{1/4}} \; .
\label{gLmax}
\end{equation}
The synchrotron radiation limit is implemented in the calculations
with $f_{\rm L} = 1$. Thus, for example, the timescales of the highest
energy protons shown in Figs. 1a and 1b derive from this limit. Other
effects that limit maximum particle energy through second-order Fermi
acceleration in a GRB blast wave, such as available time and the
requirement that the Larmor radius of the accelerated particle be less
than the blast-wave width, are considered elsewhere \citep{dh01}.

\section{Analytic Estimate of Photomeson Production Efficiency}

 The comoving differential photon energy density in a blast-wave geometry is
\begin{equation}
\ep u^\prime(\ep ) \cong ({d_L\over 2x})^2\; 
{f_\e \over c\Gamma^2}\;,\;{\rm and}\;\e \cong {2\Gamma \ep\over 1+z}\;.
\label{epkupk}
\end{equation}
 The comoving differential photon density is $n^\prime(\ep ) = \ep u^\prime(\ep
)/m_e c^2 \e^{\prime 2}$, and the synchrotron emission is assumed to
be isotropic in the comoving frame, so that $n^\prime(\ep , \mu^\prime
)\cong {1\over 2} n^\prime (\ep )$.

The time scale for significant energy loss by photohadronic reactions is
given, following equation (\ref{tpr_pg}), by
\begin{equation}
t^{\prime~-1}_{p\gamma\rightarrow \pi}\simeq {c\over 5}
\int_0^\infty d\ep \int_{-1}^1 d\mu^\prime 
(1-\mu^\prime ) n^\prime (\ep ,\mu^\prime)
\sigma_{p\gamma\rightarrow \pi}(\e^{\prime\prime})\; , \;
\label{nprime}
\end{equation}
where $\sigma_{p\gamma\rightarrow \pi}(\e^{\prime\prime})\cong
\sigma_0\delta[\mu^\prime - (1-\e_\Delta/\gamma_p^\prime\ep )]$,
$\sigma_0\approx 2\times 10^{-28}$ cm$^2$, $\epsilon_\Delta \approx
640$, $\e^{\prime\prime} = \gamma_p^\prime\ep (1-\mu^\prime )$, and we
now use primes to refer particle Lorentz factors to their proper
frame. The comoving time available to undergo hadronic reactions is
$t^\prime\cong x/\Gamma c$. The quantity
\begin{equation}
\eta \equiv {t^\prime\over t^\prime_{p\gamma\rightarrow \pi}}
\cong {\sigma_0d_L^2\e_\Delta t^\prime\over 40 \gamma_p^\prime m_ec^2\Gamma^2
x^2}\;\int_{\epsilon_\Delta/2\gamma_p^\prime}^\infty
d\ep\;\e^{\prime-3}\; f_\e\;
\label{eta}
\end{equation}
represents the efficiency to lose energy through photohadronic processes.

In a fast cooling scenario, the photon energy at the peak of the $\nu F_\nu$
spectrum is
\begin{equation}
\e_{pk} \cong {2\Gamma\over (1+z)}\e_B\gamma_{min}^2\cong
{500\over 1+z}\;k_p^2 e_e^2 (e_Bn_0)^{1/2}\Gamma_{300}^4\;.
\label{epknew}
\end{equation}
The threshold condition $\gamma_p^\prime \ep \cong
\e_\Delta$ requires acceleration of protons with observer-frame
energies $\gamma_p \cong \Gamma^2
\epsilon_\Delta/[(1+z)\e_{pk}]$ when
scattering photons with $\e\approx\e_{pk}$. Because $\gamma_p\propto
\Gamma^{-2}\psim t^{3/4}$ following the prompt phase,
 demands on particle acceleration to scatter photons with
$\e \cong \e_{pk}$ are more easily satisfied during the
early episodes ($t\lesssim t_d$) of a GRB.

The $\nu F_\nu$ synchrotron flux can therefore be written as
\begin{equation}
f_\epsilon \cong  f_{\epsilon_{pk}}(t)
[{\epsilon\over \epsilon_{pk}(t)}]^{\alpha_\nu}\;.
\label{fe}
\end{equation}
We evaluate equation (\ref{eta}) at $t^\prime = t^\prime_d$, and
 denote quantities at this time with hats. From equations (\ref{N_e}) 
and (\ref{dotN(e)}) at $\e = \e_{pk}$ corresponding to $\gamma = \gamma_1$
in a fast cooling scenario,
\begin{eqnarray}
\nonumber
\hat f_{\e_{pk}} \simeq {2\Gamma^2\over 4\pi d_L^2}\;
({B^2\over 8\pi}c\sigma_{\rm T})\;N_e^0\gamma_c\gamma_{min}
\simeq {3\over 4}\,{k_p e_eE_0\over 4\pi d_L^2t_d}\simeq 10^{-6}\;
{E_{52}^{2/3}\Gamma_{300}^{8/3}n_0^{1/3}k_p e_e\over d_{28}^2 (1+z)}\;
{{\rm ergs}\over{\rm ~cm}^{2}{\rm s}}\;,
\label{epk17}
\end{eqnarray}
where $\hat y =
 \Gamma_0\epsilon_\Delta/[(1+z)\hat\e_{pk}\gamma^\prime_p$]. Hence
 $\hat f_\e = \hat f_{\e_{pk}}(\e/\hat \e_{pk})^{\alpha_\nu}$ and
 $\alpha_\nu = 1/2$ for $\e\leq \hat \e_{pk}$, and $\alpha_\nu = (2-p)/2$
 for $\e > \hat \e_{pk}$ in the fast cooling limit (at smaller photon
 energies, though still above the synchrotron self-absorption
 frequency, $\alpha_\nu =4/3$).  From eq.\ (\ref{eta}),
\begin{equation}
\hat\eta  \simeq {3\over 20}\;{k_pe_e E_0 \sigma_0 \hat y^{\alpha_\nu-1}\over 
 m_ec^2(1+z)^2 (2-\alpha_\nu) 4\pi x_d^2 \hat \e_{pk}}
\;.
\label{etahat}
\end{equation}
Thus 
\begin{equation}
\hat\eta  \simeq 7\times 10^{-5}k_pe_e \;{E_{52}^{1/3}n_0^{2/3}
(\Gamma_0/300)^{4/3}\over 
 (1+z)^2 (2-\alpha_\nu) \hat \e_{pk}} \;\hat y^{\alpha_\nu-1}
\;.
\label{etahat1}
\end{equation}

We \citep{dcb99,dbc99,bd00,der00b} have previously shown that joint
consideration of blast wave temporal and spectral characteristics of
external shock emission, levels of the diffuse background radiation,
and triggering characteristics of $\gamma$-ray detectors cause burst
telescopes to be biased in favor of the detection of GRBs with the
prompt phase $\nu F_\nu$ peak frequency $\hat\e_{pk} \cong \e_{det}$,
where $\e_{det}$ is the photon energy of the telescope's largest
effective area.  For the BATSE telescope, $\e_{det}\cong
0.1$-1. Parameter sets A and B imply that
 $\hat \e_{pk} = 0.33/(1+z)$ and $\hat \eta \cong
10^{-3}$, and $\hat \e_{pk} = 0.42/(1+z)$ and $\hat
\eta \cong 2\times 10^{-4}$, respectively. 
The estimate for $\hat \e_{pk}$ compares favorably with the results of
Figs.\ 1a and b. This estimates apply to protons with energy $E_p \cong
\Gamma_0^2 \epsilon_\Delta m_pc^2/[(1+z)\hat\e_{pk}]\cong 5\times
10^{17} (\Gamma_0/300)^2/[(1+z)(\hat \e_{pk}/0.1)$ eV, which follows from
the condition $\hat y = 1$.

From equation (\ref{etahat}) it is easy to show that in the fast cooling
limit, $\hat\eta \propto E_p^{p/2}, E_p^{(p-1)/2},$ and $E_p^{-1/3}$
at progressively higher proton energies.

The spectral model used to fit the BATSE statistics is degenerate in
the quantity  $n_0\Gamma_0^8$ \citep{bd00}, and $\e_{pk}$ is also degenerate
in this quantity (eq.\ [\ref{epknew}]). But the photomeson production
efficiency $\eta \propto n_0^{2/3}\Gamma_0^{4/3}$ which, when
the invariance $n_0\Gamma_0^8$ is removed, indicates that 
\begin{equation}
\hat\eta \propto \sqrt{n_0}\;.
\label{etan0}
\end{equation}
Based on the measured directional energy releases and the duration
distribution of GRBs, $\Gamma_0 \sim 100$-$1000$ and uniform CBM
densities $n_0\sim 10^{-2}$-$10^2$ cm$^{-3}$ are implied.  The lower
range of the densities are similar to values deduced from GRB
afterglow fits \citep{pk01,wg99,pk01a}.  Clouds or clumps in the
external medium with densities $n_0 \gg 10^2$ cm$^{-3}$ are argued in
this model to produce shorter duration spikes in GRB light curves, and
so from equation (\ref{etan0}) could be more neutrino and neutron
luminous. The conclusions of neutrino and neutron power calculated in
this paper apply to smooth profile GRBs, and we predict no coincident
neutrino fluxes from this type of GRB.

\clearpage

\clearpage

\begin{deluxetable}{crrrrrrrrrrr}
\tabletypesize{\scriptsize}
\tablecaption{Standard Parameter Sets \label{tbl-1}}
\tablewidth{0pt}
\tablehead{
\colhead{Variable} & \colhead{Quantity} & \colhead{Parameter Set A\tablenotemark{a}}   & \colhead{Parameter Set B\tablenotemark{a}}   
}
\startdata
$E_0$ & Total Energy\tablenotemark{c} ~~~(ergs) & $2\times 10^{53}$ & $2\times10^{53}$   \\
$\Gamma_0$ & Initial Lorentz factor & 300  & 300   \\
$e_e$ & Electron Energy Transfer Parameter & 0.5 & 0.1   \\
$e_B$ & Magnetic Field Parameter & $10^{-4}$ & $10^{-1}$   \\
$e_{\rm max}$ & Maximum Particle Energy Parameter & 1 & 1  \\
$n_0$ & Density of Surrounding Medium\tablenotemark{d}~~(cm$^{-3}$) & $10^2$  & $10^2$   \\
$p$ & Nonthermal Particle Injection Index & 2.2 & 2.2   \\
$\xi$ & Nonthermal Proton Energy Fraction & 0.5 & 0.5   \\

\enddata

\tablenotetext{a}{Parameter set giving good spectral fits to $\gamma$-ray luminous phase of GRBs}
\tablenotetext{b}{Parameter set giving good spectral fits to afterglow phase of GRBs}
\tablenotetext{c}{Apparent isotropic energy release}
\tablenotetext{d}{Surrounding medium is assumed to be uniform density}

\end{deluxetable}



\clearpage

\begin{figure}
\figurenum{1}
\epsscale{1.1}
\plottwo{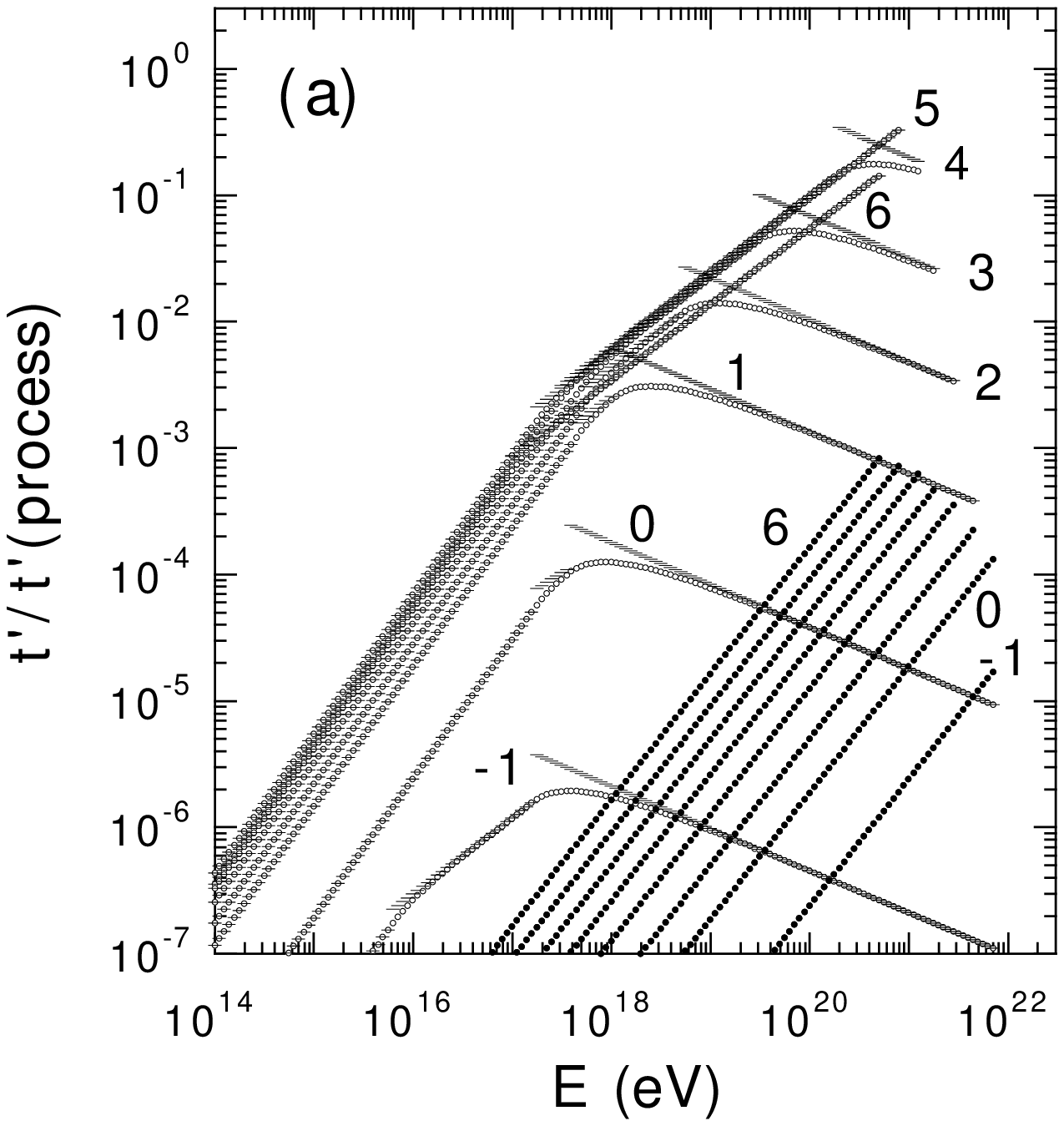}{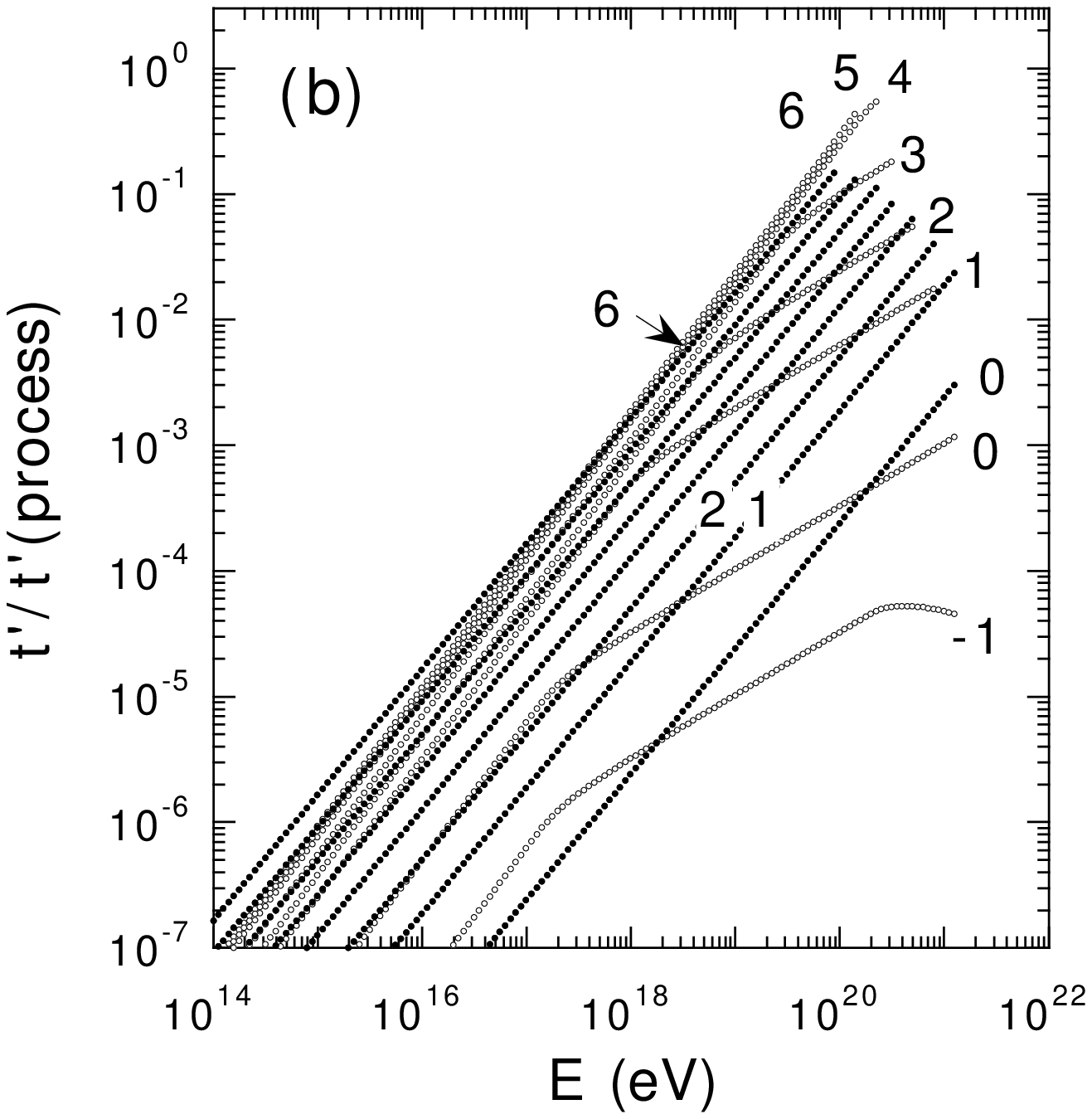}
\caption{Ratio of comoving time to the energy-loss timescales for the processes
$p + \gamma \rightarrow p + \pi^0, n + \pi^+$ (open circles) and
proton synchrotron radiation (filled circles) as a function of the
proton energy $E$ measured by an observer. Curves are denoted by the
base 10 logarithm of the observing time in seconds. (a) Parameters
giving good fits to GRBs during the prompt phase with $e_e = 0.5$ and
$e_B = 10^{-4}$.  Dotted lines show approximate expressions for the
photomeson process given by equation (\ref{Iap(gamma)}).(b) Parameters
giving good fits to GRBs during the afterglow phase with $e_e = 0.1$
and $e_B = 0.1$. Other parameters are given in Table 1.  }
\end{figure}

\begin{figure}
\figurenum{2}
\epsscale{1.0}
\plottwo{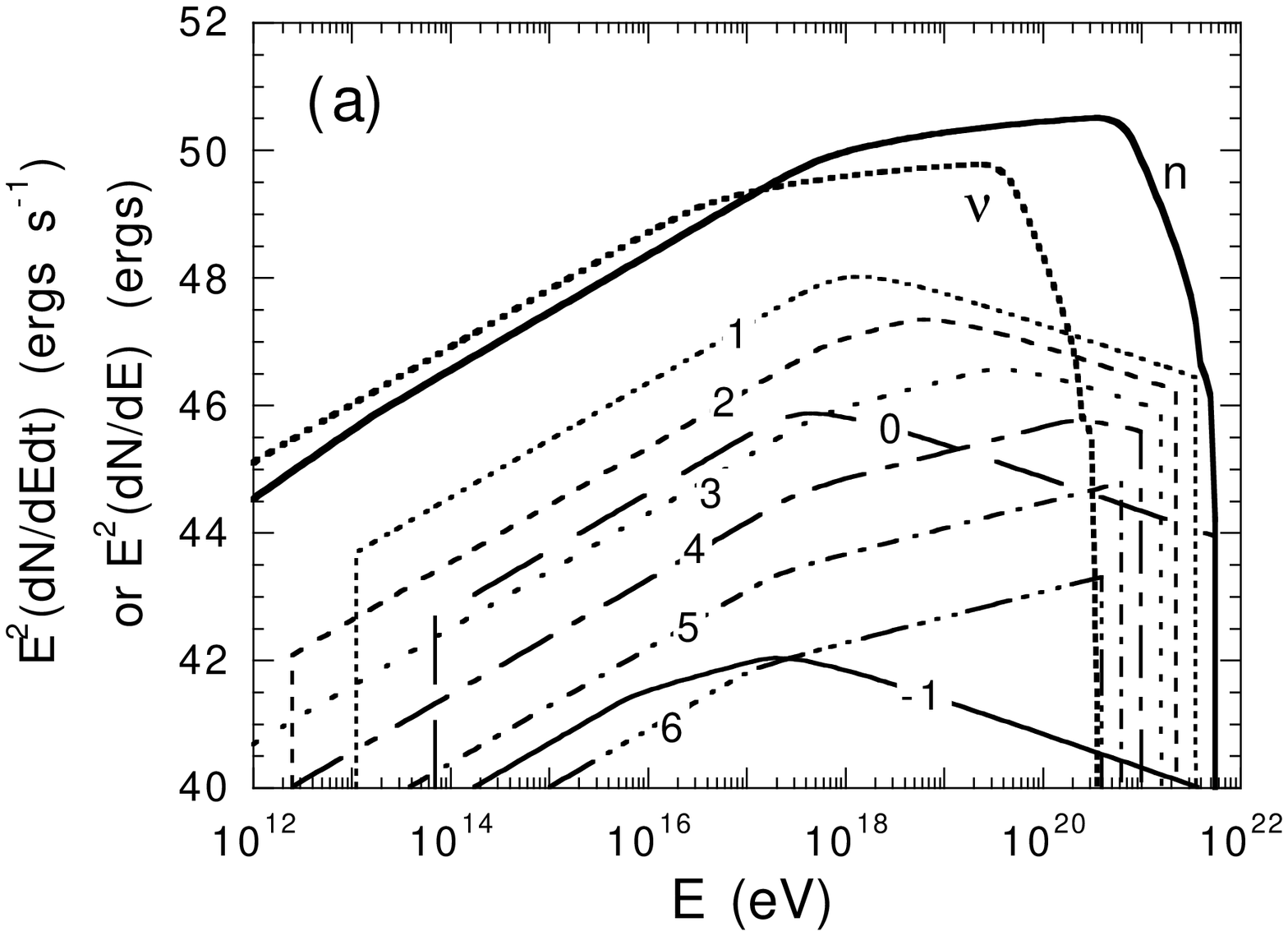}{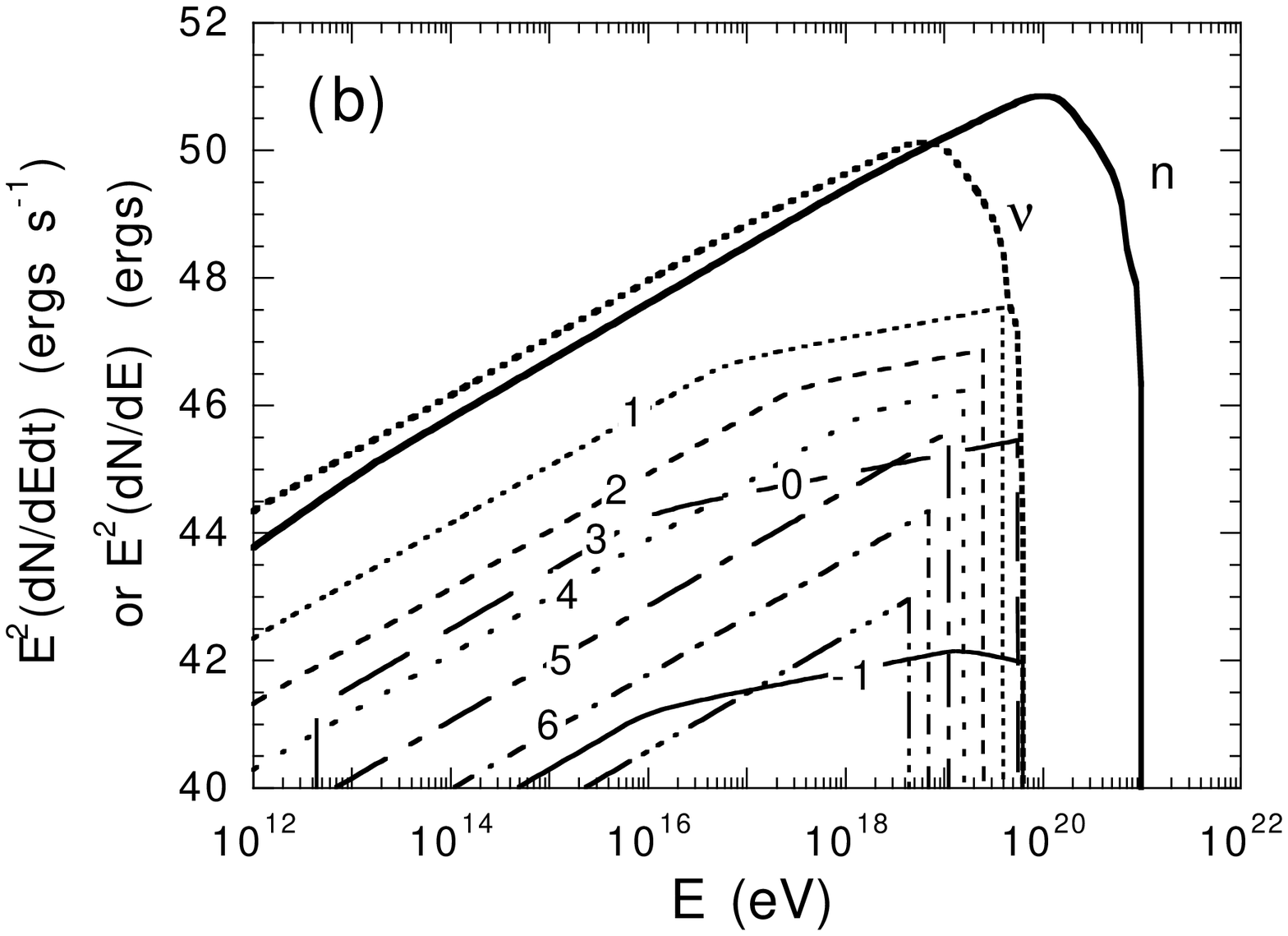}
\caption{Instantaneous production spectra for neutrons (a) and neutrinos (b),
labeled by the base 10 logarithm of the observing time in seconds.
The prompt phase parameter set (A) is used in (a) and the afterglow
parameter set (B) in (b). The thick solid and dotted curves give the
time-integrated spectra for neutrons and neutrinos, respectively.  }
\end{figure}

\begin{figure}
\figurenum{3}
\epsscale{0.9}
\plotone{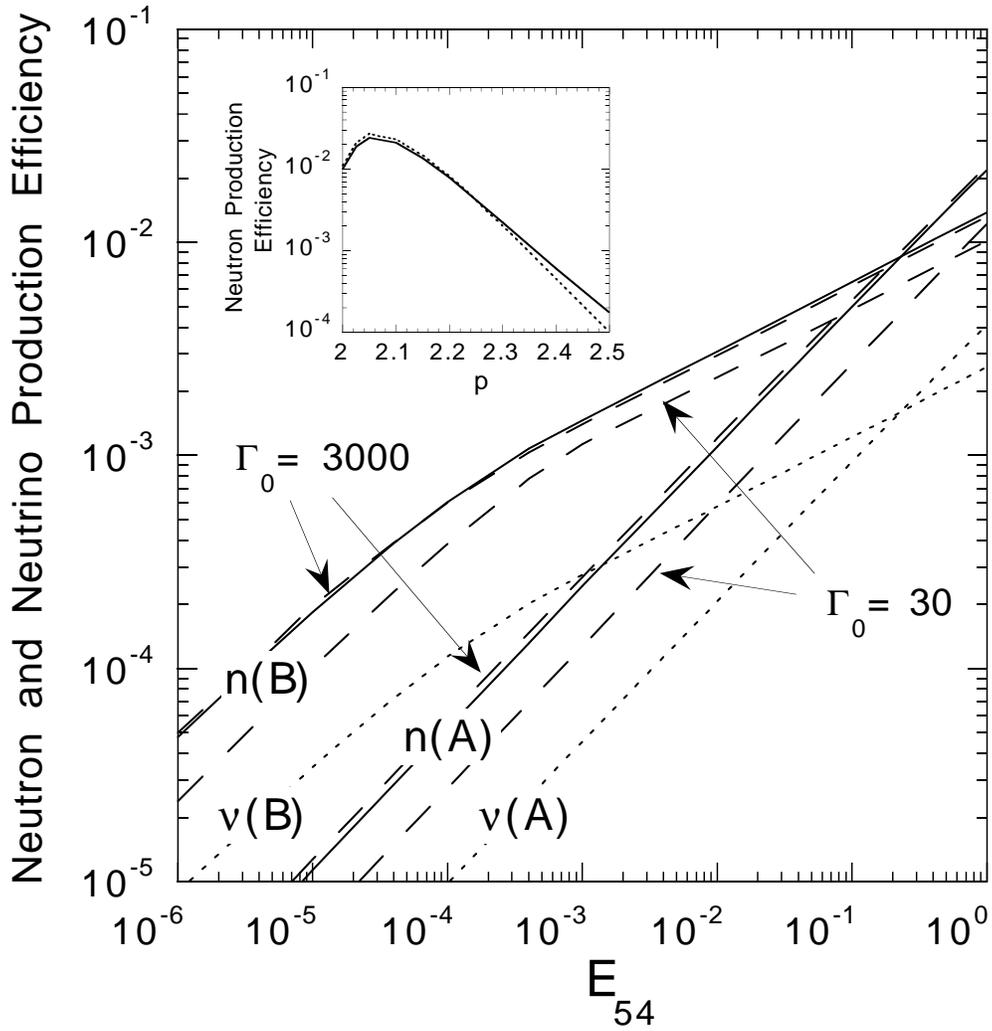}
\vskip-0.5in
\caption{Production efficiencies for neutrons, denoted by ``n," and neutrinos,
denoted by ``$\nu$," as a function of apparent isotropic explosion
energy $E_{54}$ in units of $10^{54}$ ergs. Prompt and afterglow
parameter sets are denoted by labels (A) and (B),
respectively. Neutron production efficiencies are also shown for the
standard parameter sets, except now with $\Gamma_0 = 30$ and $\Gamma_0
= 3000$.  Inset shows the dependence of the neutron production
efficiency on the injection index $p$ of the nonthermal protons for
parameter sets (A) and (B), shown by the solid curve and dotted curve,
respectively.  }
\end{figure}

\begin{figure}
\figurenum{4}
\epsscale{1.0}
\plotone{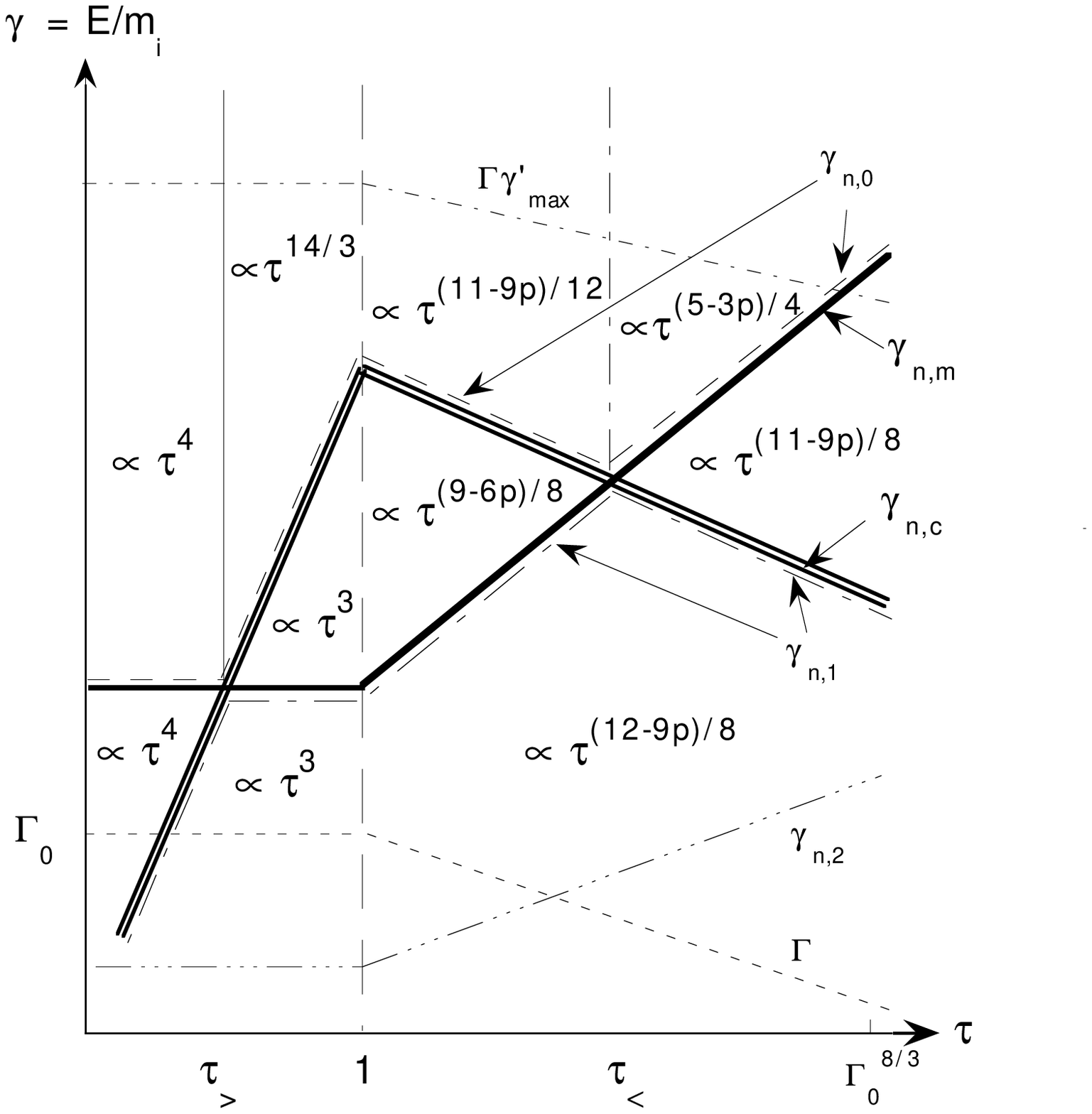}
\vskip-1.5in
\caption{Diagram illustrating the temporal behavior of the rate at which
neutrons and neutrinos with energies $E = m_i \gamma$ are produced by
photomeson production in a GRB blast wave. The dimensionless time
$\tau = t/t_d$, where $t_d$ is the deceleration time.  The lower
triple-dotted dashed lines and the upper dot-dashed lines bound the
production of particles due to the lack of available photons or the
synchrotron radiation limit, respectively. The other regimes are a
consequence of the spectral shape and evolution of the synchrotron
soft photons, and the evolution of the energy in nonthermal
protons. The interior tetragon defines the time during which the blast
wave evolves in the strong cooling regime.  }
\end{figure}

\begin{figure}
\figurenum{5}
\epsscale{1.00}
\plotone{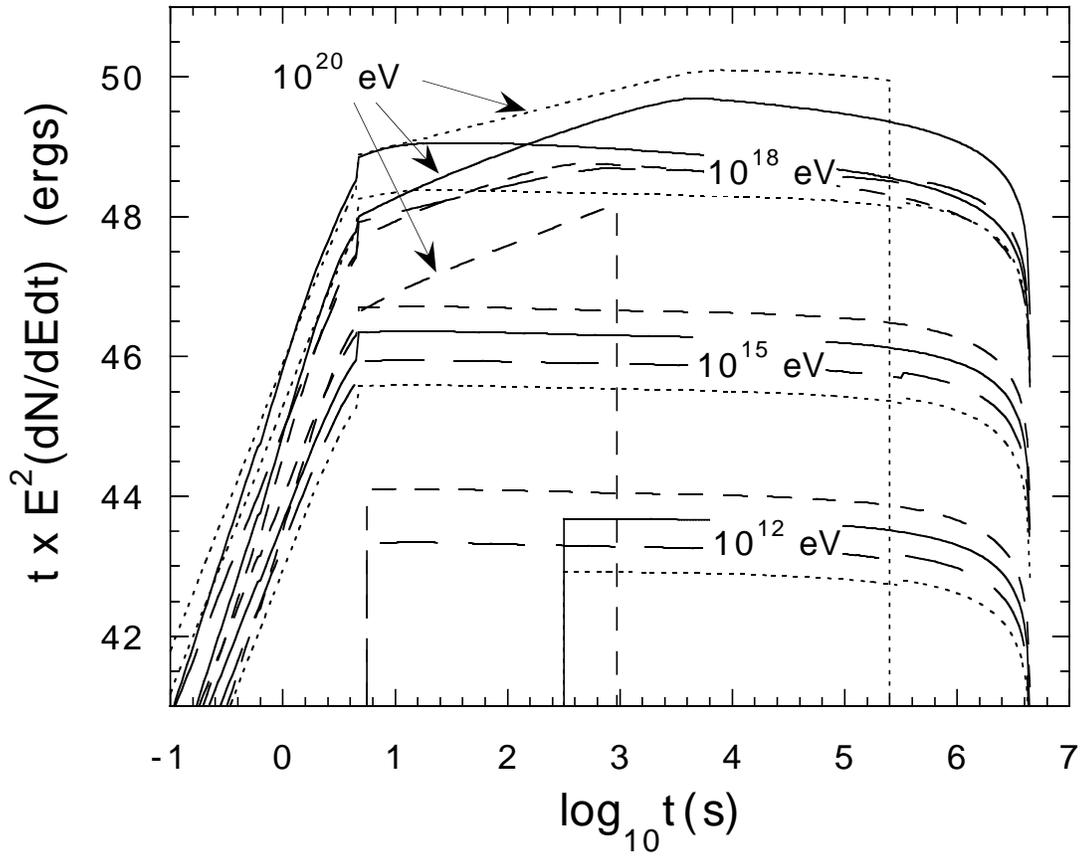}
\vskip-1.5in
\caption{Calculations of neutron and neutrino production time profiles at
$10^{12}$, $10^{15}$, $10^{18}$, and $10^{20}$ eV energies. Neutron
production time profiles are given by the solid and dotted curves, and
neutrino time profiles are given by the short-dashed and long-dashed
curves for parameter sets (A) and (B), respectively. No neutrinos are
produced at $10^{20}$ eV for parameter set (B), as can be seen in
Fig.\ 2b.  }
\end{figure}

\begin{figure}
\vskip-1.0in
\figurenum{6}
\epsscale{0.8}
\plotone{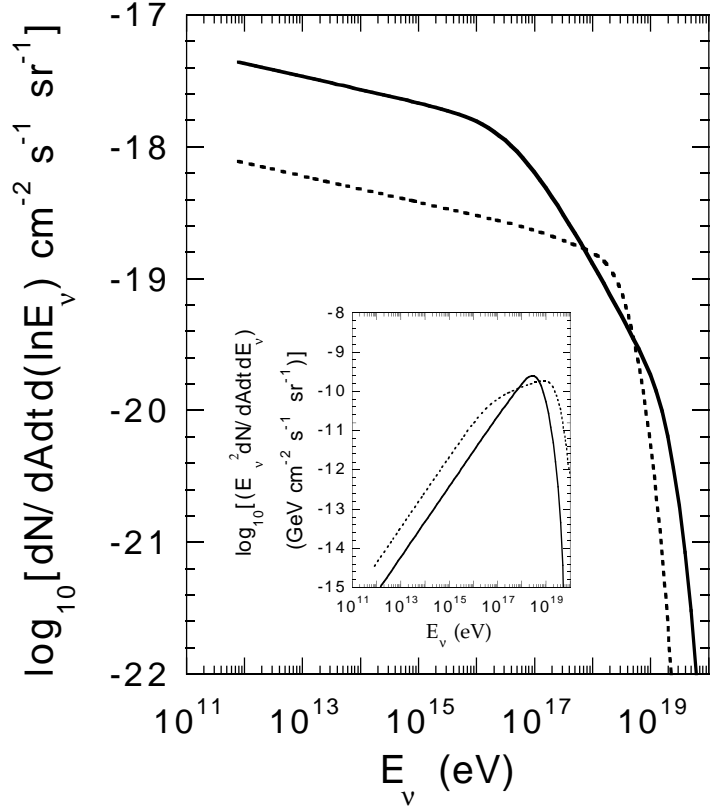}
\caption{Calculations of the diffuse neutrino background flux produced by 
 smooth-profile GRBs
with parameters given by sets (A) (solid curve) and (B) (dotted
curve), except that integrations are performed over $E_0$ and
$\Gamma_0$ using equation (\ref{ndot}) for the differential GRB
emissivity. Inset shows the calculated GRB neutrino background
for the smooth-profile GRBs.  }
\end{figure}

\begin{figure}
\figurenum{7}
\epsscale{1.0}
\plotone{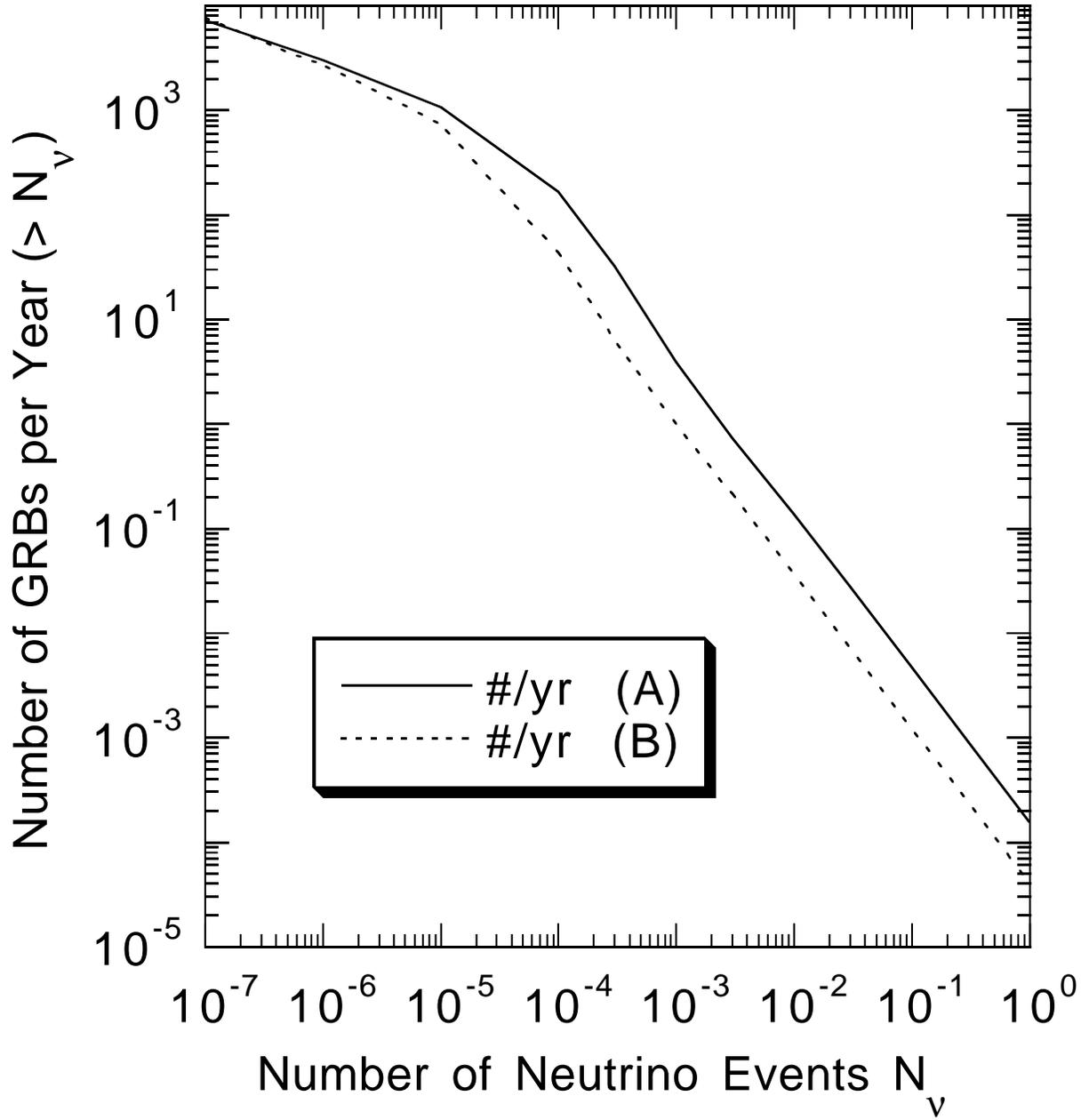}
\vskip0.5in
\caption{Calculations of expected event rate of neutrinos from 
smooth-profile GRBs for a 1 km$^2$ detector for parameter sets
(A) and (B), shown by the solid and dotted curves, respectively.  }
\end{figure}

\begin{figure}
\figurenum{8}
\epsscale{1.0}
\plottwo{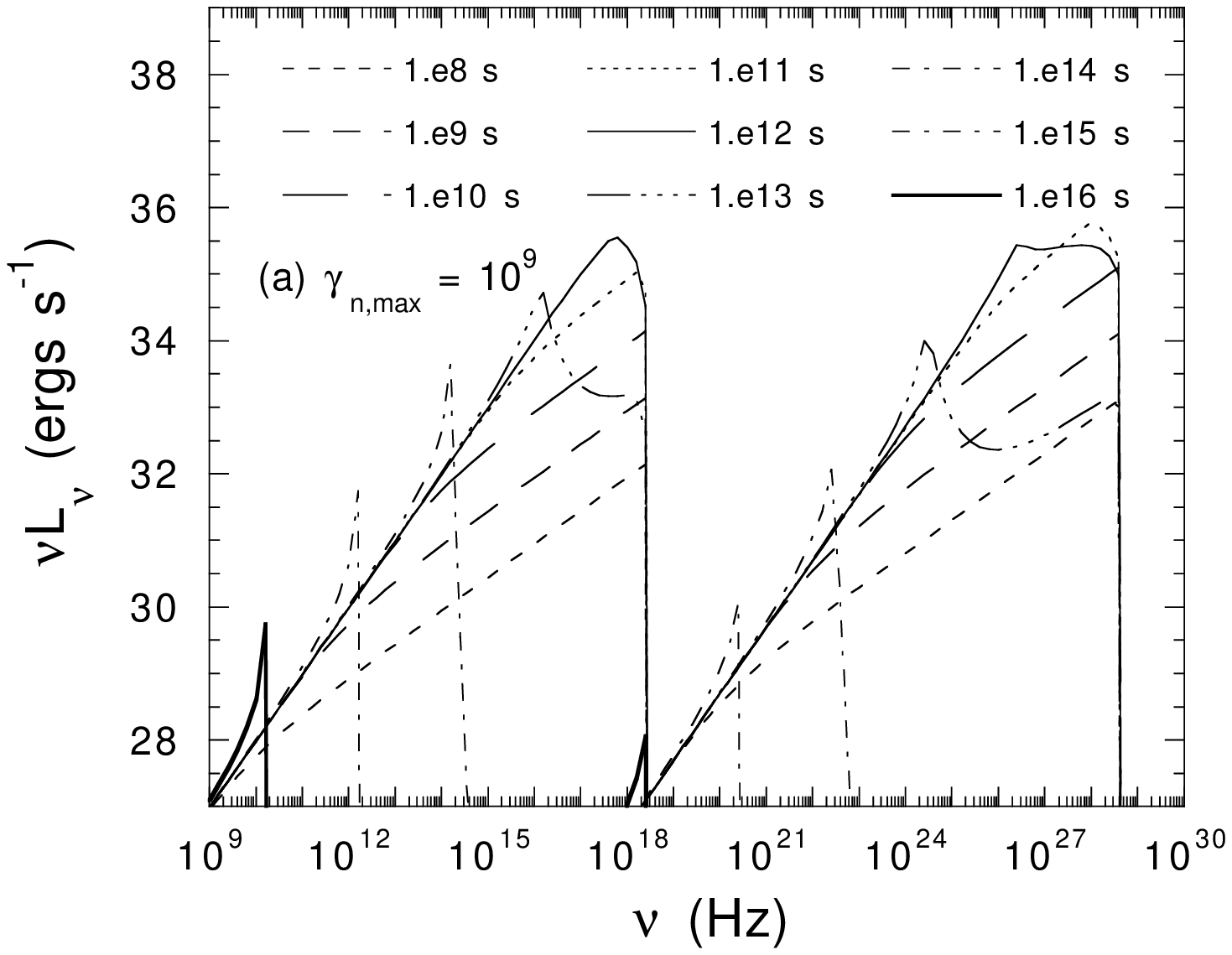}{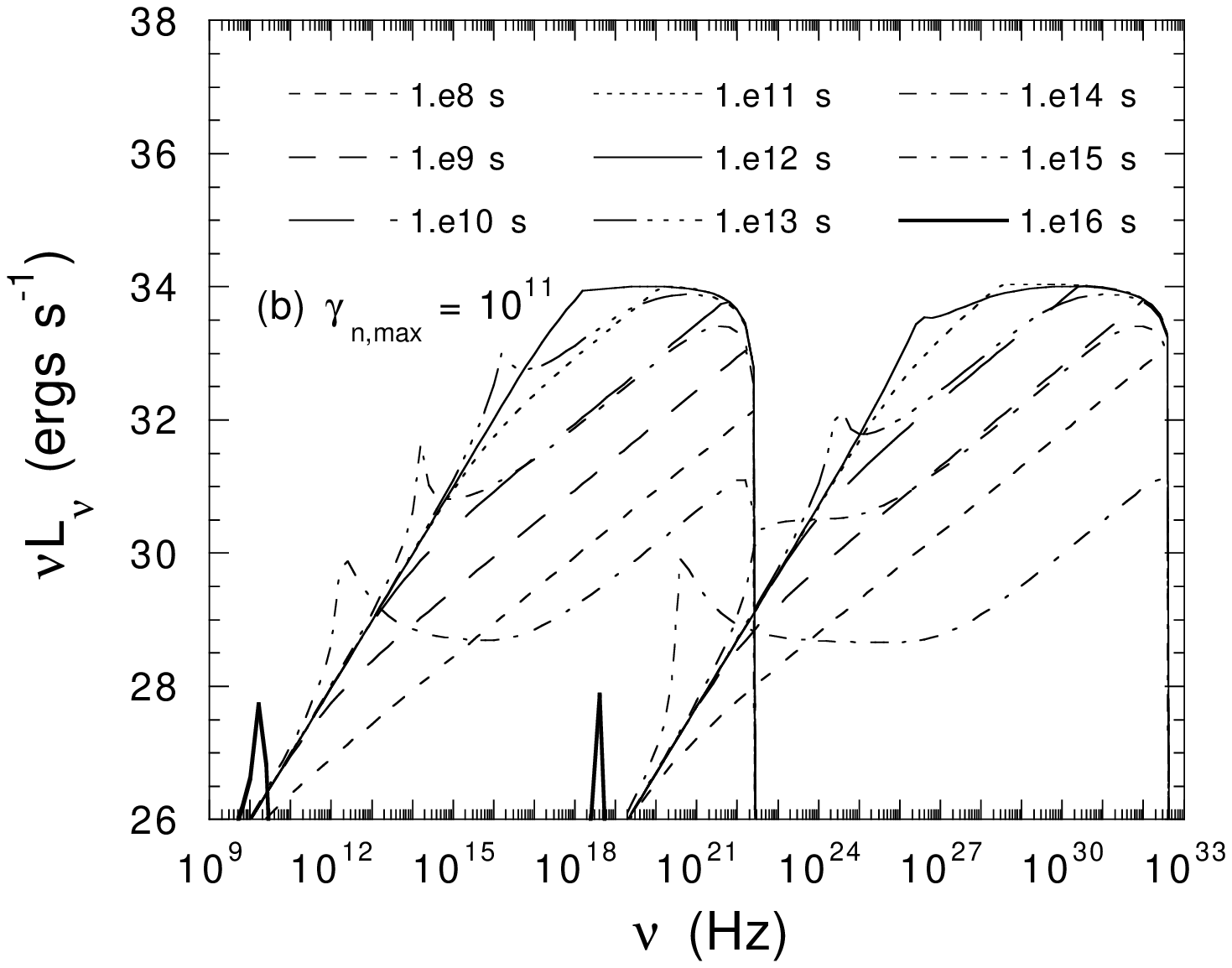}
\vskip-1.2in
\caption{Calculations of spatially integrated synchrotron and Thomson spectra
radiated by neutron $\beta$-decay electrons emitted from a single GRB
that radiates $10^{52}$ ergs in neutrons. The neutron decay spectra is
assumed to be represented by a power-law spectrum with
$dN/d\gamma_n\propto \gamma_n^{-1}$ up to the maximum Lorentz factors
of $\gamma_{n,{\rm max}} = 10^9$ in panel (a) and $\gamma_{n,{\rm
max}} = 10^{11}$ in panel (b).  }
\end{figure}


\begin{thebibliography}{}
\bibitem[Aharonian et al.(2001)]{aha01} Aharonian, F.~et al.\ 2001, \aap, 370, 112 
\bibitem[Aharonian et al.(1994)]{aha94} Aharonian, F. A., Coppi, P. S., and V\"olk, H. J. 1994, \apj, 423, L5
\bibitem[Armitage and Natarajan(1999)]{an99} Armitage, P. J., and Natarajan, P. 1999, \apj, 523, L7
\bibitem[e.g., Bahcall and Soneira(1980)]{bs80} Bahcall, J. N., and Soneira, R. M. 1980, \apjs, 44, 73
\bibitem[however, see Beloborodov(2000)]{bel00}Beloborodov, A. M. 2000, \apj, 539, L25
\bibitem[Binney and Merrifield(1998)]{bm98} Binney, J., and Merrifield, M. 1998, Galactic Astronomy (Princeton: Princeton University Press)
\bibitem[Blain et al.(1999)]{bla99} Blain, A.~W., Jameson, A., Smail, I., Longair, M.~S., Kneib, J.-P., and Ivison, R.~J.\ 1999, \mnras, 309, 715
\bibitem[Blandford and McKee(1976)]{bm76} Blandford, R. D., and McKee, C. F. 1976, Phys. Fluids, 19, 1130
\bibitem[Bloom et al.(1999)]{blo99}Bloom, J. S., et al.\ 1999, \nat, 401,453
\bibitem[Bloom et al.(1999a)]{blo99a} Bloom, J.  S., et al.\ 1999a, \apj, 518, L1
\bibitem[B\"ohringer et al.(1999)]{bfs99} B\"ohringer, H., Feretti, L., and Schuecker, P., eds., 1999, Diffuse Thermal and Relativistic Plasma in Galaxy Clusters, MPE Report 271
\bibitem[B\"ottcher(2000)]{boe00} B\"ottcher, M. 2000, \apj, 539, 102
\bibitem[B\"ottcher and Dermer(1998)]{bd98} B\"ottcher, M., and Dermer, C. D. 1998, \apj, 499, L131
\bibitem[B\"ottcher and Dermer(2000)]{bd00} B\"ottcher, M., and Dermer, C. D. 2000, \apj, 529, 635; (e) 2000, \apj, 536, 513
\bibitem[B\"ottcher and Dermer(2000a)]{bd00a} B\"ottcher, M., and Dermer, C. D. 2000a, \apj, 532, 281
\bibitem[Briggs et al.(1999)]{bri99} Briggs, M.~S.~et al.\ 
1999, \apj, 524, 82
\bibitem[Buckley et al.(1998)]{buc98}Buckley, J. H., et al.\ 1998, A\&A, 329, 639
\bibitem[Cappellaro et al.(1997)]{cap97}Cappellaro, E., Turatto, M., Tsvetkov, D. Y., Bartunov, O. S., Pollas, C., Evans, R., and Hamuy, M. 1997, A\&A, 322, 431
\bibitem[Castander and Lamb(1999a)]{cl99a}Castander, F. J., and Lamb, D. Q. 1999a, \apj, 523, 593
\bibitem[Castander and Lamb(1999b)]{cl99b}Castander, F. J., and Lamb, D. Q. 1999a, \apj, 523, 602
\bibitem[Chiang and Dermer(1999)]{cd99} Chiang, J., and Dermer, C. D. 1999, \apj, 512, 699
\bibitem[Cohen et al.(1997)]{coh97}Cohen, E., Katz, J. I., Piran, T., Sari, R., Preece, R. D., and Band, D. L. 1997, \apj, 488, 330
\bibitem[Contopoulos and Kazanas(1995)]{ck95}Contopoulos, J., and Kazanas, D. 1995, \apj, 441, 521
\bibitem[Costa et al.(1997)]{cos97} Costa, E., et al.\ 1997,  Nature, 387, 783
\bibitem[Dai and Lu(2001)]{dl01} Dai, Z. G., and Lu, T. 2001, \apj, 551, 249
\bibitem[Dar and Plaga(1999)]{dp99} Dar, A., and Plaga, R. 1999, A\&A, 349, 259
\bibitem[de Jager et al.(1996)]{jag96}de Jager, O. C., Harding, A. K., Michelson, P. F., Nel, H. I., Nolan, P. L., Sreekumar, P., and Thompson, D. J. 1996, ApJ, 457, 253
\bibitem[Dermer(1992)]{der92} Dermer, C. D. 1992, \prl, 68, 1799
\bibitem[Dermer(2001)]{der01} Dermer, C. D., 2001, in Proc. 27th ICRC, Hamburg, Germany, 6, 2039
\bibitem[Dermer and Mitman(1999)]{dm99} Dermer, C. D., and Mitman, K. E. 1999, ApJ, 513, L5 
\bibitem[Dermer et al.(1999a)]{dbc99} Dermer, C. D., B\"ottcher, M., and Chiang, J. 1999a, \apj,  515, L49
\bibitem[Dermer et al.(1999)]{dcb99} Dermer, C. D., Chiang, J. and B\"ottcher, M. 1999, \apj, 513, 656
\bibitem[Dermer et al.(2000b)]{dcm00}Dermer, C. D., Chiang, J., and Mitman, K. E. 2000b, ApJ, 537, 785
\bibitem[Dermer and Humi(2001)]{dh01} Dermer, C. D., and Humi, M. 2001, ApJ, 556, 479
\bibitem[Dermer and B\"ottcher(2000)]{db00} Dermer, C. D., and B\"ottcher, M. 2000, ApJ, 534, L155
\bibitem[Dermer and B\"ottcher(2002)]{db02} Dermer, C. D., and B\"ottcher, M. 2002, to be submitted to Reviews of Modern Physics
\bibitem[Dermer et al.(2000a)]{dbc00} Dermer, C. D., B\"ottcher, M., and Chiang, J. 2000a, \apj, 537, 255
\bibitem[Dermer(2000a)]{der00a} Dermer, C.D., astro-ph/0005440, v.\ 1
\bibitem[Dermer(2000b)]{der00b} Dermer, C.D., Proc.\ Heidelberg 2000 High-Energy Gamma-Ray Workshop, ed. F. A. Aharonian and H. V\"olk (AIP: New York), p. 202 (astro-ph/0010564)
\bibitem[Djorgovski et al.(2001)]{djo01} Djorgovski, S.\ G., et al.\ 2001, in Gamma Ray Bursts in the Afterglow Era, ed. E. Costa, F. Frontera, and J. Hjorth (Springer: Berlin), 218
\bibitem[Efremov et al.(1998)]{eeh98} Efremov, Y. N., Elmegreen, B. G., and Hodges, P. W. 1998, \apj, 501, L163
\bibitem[Eilek(1999)]{eil99} Eilek, J. 1999, in Diffuse Thermal and Relativistic Plasma in Galaxy Clusters, ed. H. B\"ohringer, L. Feretti, and P. Schuecker, MPE Report 271, 71
\bibitem[Esposito et al.(1996)] {esp96} Esposito, J. A., Hunter, S. D., Kanbach, G., and Sreekumar, P. 1996, ApJ, 461, 820
\bibitem[Fenimore and Ramirez-Ruiz(1999)]{frr99} Fenimore, E. E., and Ramirez-Ruiz, E. 1999, \apj, submitted (astro-ph/9909299)
\bibitem[Frail et al.(1999)]{fra99} Frail, D. A., et al.\ 1999, ApJ, 525, L81
\bibitem[Frail et al.(2001)]{fra01} Frail, D. A., et al.\ 2001 (astro-ph/0102282)
\bibitem[Fruchter et al.(1999)]{fru99} Fruchter, A. S., et al.\ 1999, \apj, 519, L13
\bibitem[Furlanetto and Loeb(2002)]{fl02}Furlanetto, S., and Loeb, A.\ 2002, \apj, submitted (astro-ph/0203044)
\bibitem[Gaisser(1990)]{gai90} Gaisser, T. K. 1990, Cosmic Rays and Particle Physics (New York: Cambridge University Press), 148
\bibitem[Gaisser and Grillo(1987)]{gg87}Gaisser, T. K., and Grillo, A. F. 1987, \prd, 36, 2752
\bibitem[Gaisser et al.(1995)]{ghs95}Gaisser, T. K., Halzen, F., and Stanev, T. 1995, Phys. Repts., 258(3), 173 
\bibitem[Galama et al.(1998)]{gal98} Galama, T. J., et al.\ 1998, Nature, 395, 670
\bibitem[Galama et al.(2000)]{gal00} Galama, T. J., et al.\ 2000, ApJ, 536, 185
\bibitem[Gallant and Achterberg(1999)]{gal99} Gallant, Y. A., and Achterberg, A., 1999, \mnras, 305, L6
\bibitem[Gallant et al.(1999)]{gak99} Gallant, Y. A., Achterberg, A., and Kirk, J. G. 1999, A\&AS, 138, 549
\bibitem[Giovanoni and Kazanas(1990)]{gk90} Giovanoni, P. M., and Kazanas, D. 1990, Nature, 345, 319
\bibitem[Gorham et al.(2000)]{gor00} Gorham, P. W., van Zee, L., Unwin, S. C., and Jacobs, C. S. 2000, AJ, 119, 1677
\bibitem[Gould and Schr\'eder(1967)]{gs67} Gould, R. J., and Schr\'eder, G. P. 1967, Phys. Rev., 155, 1404
\bibitem[Grindlay(1999)]{gri99}Grindlay, J. E. 1999, \apj, 510, 710
\bibitem[Guilbert et al.(1983)]{gfr83} Guilbert, P. W., Fabian, A. C., and Rees, M. J. 1983, MNRAS, 205, 593
\bibitem[Halzen and Hooper(1999)]{hh99} Halzen, F., and Hooper, D. W. 1999, \apj, 527, L93
\bibitem[Harrison et al.(1999)]{har99} Harrison, F. A., et al.\ 1999, \apj, 523, L121
\bibitem[Hogg and Fruchter(1999)]{hf99} Hogg, D. W., and Fruchter, A. S. 1999, \apj, 520, 54
\bibitem[Hunter et al.(1997)]{hun97} Hunter, S. D., et al.\ 1997, \apj, 481, 205
\bibitem[Kirk(1991)] {kir91} Kirk, J.G., in Relativistic Hadrons in Cosmic Compact Objects, ed. A. A. Zdziarski and M. Sikora (New York: Springer-Verlag), 91
\bibitem[Kirk and Mastichiadis(1992)]{km92} Kirk, J. G., and Mastichiadis, A. 1992, Nature, 360, 135
\bibitem[Krumholz et al.(1998)]{kth98}Krumholz, M., Thorsett, S. E., and Harrison, F. A., 1998, ApJ, 506, L81
\bibitem[Kulkarni et al.(1999)]{kul99} Kulkarni, S. R., et al.\ 1999, Nature 398, 389
\bibitem[Kulkarni et al.(1998)]{kul98} Kulkarni, S. R., et al.\ 1998, Nature 395, 663
\bibitem[Kumar(1999)]{kum99} Kumar, P. 1999, \apj, 523, L113
\bibitem[Lagage and Cesarsky(1983)]{lc83} Lagage, P. O., and Cesarsky, C. J. 1983, A\&A, 118, 223
\bibitem[e.g., Lamb(1999)]{lam99} Lamb, D.  Q. 1999, A\&AS, 138, 607 
\bibitem[Lipari and Stanev(1991)]{ls91}Lipari, P., and Stanev, T., 1991, \prd, 44, 3543
\bibitem[Loeb and Perna(1998)]{lp98} Loeb, A., and Perna, R. 1998, \apj, 503, L35
\bibitem[e.g., Loeb and Waxman(2000)]{lw00} Loeb, A., and Waxman, E.\ 2000, Nature, 405, 156
\bibitem[Loveday et al.(1992)]{lov92} Loveday, J., Peterson, B. A., Efstathiou, G., and Maddox, S. J. 1992, \apj, 390, 338
\bibitem[Lutz et al.(1996)]{lutz96} Lutz, D., et al.\ 1996, A\&A, 315, L137
\bibitem[Madau et al.(1998)]{mpd98} Madau, P., Pozzetti, L., and Dickinson, M. 1998, \apj, 498, 106
\bibitem[Mallozzi et al.(1996)]{mpp96}Mallozzi, R. S., Pendleton, G. N., and Paciesas, W. S., 1996, ApJ, 471, 636
\bibitem[Mao and Mo(1998)]{mm98} Mao, S., and Mo, H. J. 1998, A\&A, 339, L1
\bibitem[McBreen and Hanlon(1999)]{mh99} McBreen, B., and Hanlon, L. 1999, A\&A, 351, 759
\bibitem[M\'esz\'aros and Rees(1993)]{mr93}M\'esz\'aros, P., and Rees, M. J. 1993, \apj, 405, 278 
\bibitem[Milgrom and Usov(1995)]{mil95} Milgrom, M., and Usov, V. 1995, \apj, 449, L37
\bibitem[Milgrom and Usov(1996)]{mu96} Milgrom, M., and Usov, V. 1995, Astroparticle Physics, 4, 365
\bibitem[e.g., Mirabal et al.(2001)]{mir01} Mirabal, N.,  Halpern, J. P., Eracleous, M.,  and Becker, R. H. 2000, \apj, 547, L137
\bibitem[Narayan et al.(1992)]{npp92} Narayan, R., Paczy\'nski, B., and Piran, T. 1992,, \apj, 395, L83
\bibitem[Odewahn et al.(1998)]{ode98} Odewahn, S. C., et al.\ 1998, \apj, 509, L5
\bibitem[e.g., Owens et al.(1998)]{owe98} Owens, A., et al.\ 1998, A\&A, 339, L37
\bibitem[Paciesas et al.(1999)]{pac99} Paciesas, W.~S.~et 
al.\ 1999, \apjs, 122, 465
\bibitem[Panaitescu et al.(1998)]{pmr98}Panaitescu, A., M\'esz\'aros, P., and Rees, M. J. 1998, Astrophys.\ J., 503, 314
\bibitem[Panaitescu and Kumar(2001)]{pk01} Panaitescu, A., and Kumar, P.\ 2001, \apj, 554, 667 
\bibitem[Panaitescu and Kumar(2001a)]{pk01a} Panaitescu, A.\ and 
Kumar, P.\ 2001, \apjl, 560, L49
\bibitem[Panaitescu et al.(1999)]{psm99} Panaitescu, A., Spada, M., and M\'esz\'aros, P. 1999, \apj, 522, L105
\bibitem[Pian(2000)] {pia00} Pian, E. 2000, in Supernovae and Gamma Ray Bursts, ed. M. Livio, K. Sahu, and N. Panagia, in press (astro-ph/9910236)
\bibitem[Pinkau(1980)]{pin80} Pinkau, K. 1980, A\&A, 87, 192
\bibitem[Piran(1999)]{pir99} Piran, T. 1999, Phys.\ Rpts., 314, 575
\bibitem[Piro et al.(2000)]{piro00} Piro, L., et al.\ 2000, Science, 290, 955
\bibitem[Pohl and Esposito(1998)]{pe98} Pohl, M., and Esposito, J. A. 1999, \apj, 507, 32
\bibitem[Pohl and Schlickeiser(2000)]{ps00} Pohl, M., and Schlickeiser, R. 2000, A\&A, 354, 395
\bibitem[Rachen and M\'es\'azaros(1998)]{rm98} Rachen, J., and M\'esz\'aros, P. 1998, \prd, 58, 123005
\bibitem[Rachen and M\'es\'azaros(1998a)]{rm98a} Rachen, J., and M\'esz\'aros, P. 1998, in Fourth Huntsville Symposium on Gamma Ray Bursts, ed. C. A. Meegan, R. D. Preece, and T. M. Koshut (AIP: New York), 776
\bibitem[Rees and  M\'esz\'aros (1992)]{rm92} Rees, M. J., and  M\'esz\'aros, P. 1992, MNRAS, 258, 41P 
\bibitem[Reichart(1999)]{rei99}Reichart, D. E. 1999, \apj, 521, L111
\bibitem[Romero et al.(1999)]{rbt99} Romero, G. E., Benaglia, P., and Torres, D. F. 1999, A\&A, 348, 868
\bibitem[Sari et al.(1998)]{spn98} Sari, R., Piran, T., and Narayan, R.  1998, \apj, 497, L17
\bibitem[Scalo and Wheeler(2002)]{sw02} Scalo, J., and Wheeler, J. C. 2002, \apj, in press (astro-ph/9912564)
\bibitem[Schaefer(2000)]{sch00}Schaefer, B. E. 2000, \apj, 532, L21
\bibitem[Schlickeiser and Dermer(2000)]{sd00} Schlickeiser, R., and Dermer, C. D. 2000, A\&A, 360, 789
\bibitem[Schmidt(1999)]{sch99} Schmidt, M. 1999, \apj, 523, L117
\bibitem[Schuster, Pohl, \& Schlickeiser(2002)]{sps02}
Schuster, C., Pohl, M., \& Schlickeiser, R.\ 2002, \aap, 382, 829.
\bibitem[Sommers(2002)]{som02}Sommers, P., oral contribution at Aspen Workshop on UHE Particles from Space (January 27 - February 2, 2002)
\bibitem[Sreekumar et al.(1998)]{sre98} Sreekumar, P., et al.\ 1998, \apj, 494, 523
\bibitem[Stanev et al.(2000)]{sta00} Stanev, T., Engel, R., M\"ucke, A., Protheroe, R. J., and Rachen, J. P. 2000, \prd, 62, 093005
\bibitem[Stecker(1979)]{ste79}Stecker, F. W., 1979, \apj, 228, 919 
\bibitem[Stecker(2000)]{ste00}Stecker, F. W., 2000, Astroparticle Phys., 14, 207
\bibitem[Stecker and de Jager(1998)]{sdj98} Stecker, F. W., and de Jager, O. C. 1998, A\&A, 334, L85
\bibitem[see also Stecker et al.(1991)]{ste91} Stecker, F. W., Done, C., Salamon, M. H., and Sommers, P. 1991, \prl, 66, 2697; (e) 1992, \prl, 69, 2738
\bibitem[Takeda et al.(1998)]{tak98} Takeda, M., et al.\ 1998, \prl, 81, 1163
\bibitem[Totani(1997)]{tot97} Totani, T. 1997, \apj, 486, L71
\bibitem[Totani(1999)]{tot99} Totani, T. 1999, \apj, 511, 41
\bibitem[van Paradijs et al.(1997)]{vpa97} van Paradijs, J., et al.\ 1997, Nature, 386, 686  
\bibitem[van Paradijs et al.(2000)]{vpa00} van Paradijs, J., Kouveliotou, C., and Wijers, R. A. M. J. 2000, \araa, 38, 379
\bibitem[Vietri(1995)]{vie95} Vietri, M. 1995, \apj, 453, 883
\bibitem[Vietri(1997)]{vie97} Vietri, M. 1997, \prl, 78, 4328
\bibitem[Vietri(1998a)]{vie98a} Vietri, M. 1998a, \prl, 80, 3690
\bibitem[Vietri(1998b)]{vie98b} Vietri, M. 1998b, \apj, 507, 40
\bibitem[Waxman(1995)]{wax95} Waxman, E. 1995, \prl, 75, 386
\bibitem[Waxman and Bahcall(1997)]{wb97} Waxman, E., and Bahcall, J. N. 1997, \prl, 78, 229
\bibitem[Waxman and Bahcall(2000)]{wb00} Waxman, E., and Bahcall, J. N. 2000, \apj, 541, 707
\bibitem[Waxman and Coppi(1996)]{wax96} Waxman, E., and Coppi, P. 1996, \apj, 464, L75
\bibitem[e.g., Weinberg(1972)]{wei72} Weinberg, S., 1972, Gravitation and Cosmology (New York: Wiley), chpt. 14
\bibitem[Weiler et al.(2000)]{wei00} Weiler, K. W., Panagia, N., Sramek, R. A., van Dyk, S. D., Montes, M. J., and Lacey, C. K. 2000, in Supernovae and Gamma-Ray Bursts, ed. M. Livio, N. Panagia, and K. Sahu, in press (astro-ph/0002501)
\bibitem[Wijers et al.(1998)]{wij98}Wijers, R. A. M. J., Bloom, J. S., Bagla, J. S., and Natarajan, P. 1998, MNRAS, 294, L13
\bibitem[Wijers and Galama(1999)]{wg99}Wijers, R. A. M. J., and Galama, T. J. 1999, \apj, 523, 177
\bibitem[Yoshida and Teshima(1993)]{yt93}Yoshida, S., and Teshima, M. 1993, Progr. Theor. Phys., 89, 833


\end{thebibliography}
\end{document}